# Physical Characterization of Warm Spitzer-observed Near-Earth Objects

Cristina A. Thomas[a,b,c,h], Joshua P. Emery[d,h], David E. Trilling[a,h], Marco Delbó[e,h], Joseph L. Hora[f,h], Michael Mueller[g,h]

[a]*Northern Arizona University, Department of Physics and Astronomy,*
*PO Box 6010 Flagstaff, AZ 86011*
[b]*NASA Goddard Space Flight Center, 8800 Greenbelt Rd., Greenbelt, MD 20771*
[c]*NASA Postdoctoral Program, Oak Ridge Associated Universities*
[d]*University of Tennessee, Department of Earth and Planetary Sciences,*
*1412 Circle Dr., Knoxville, TN 37996*
[e]*Université de Nice Antipolis, CNRS, Observatoire de la Côte d'Azur,*
*BP4229, 06304 Nice Cedex 4, France*
[f]*Harvard-Smithsonian Center for Astrophysics, 60 Garden Street, MS-65, Cambridge, MA 02138*
[g]*SRON LEA, Netherlands Institute for Space Research*
*Postbus 800, 9700AV Groningen, Netherlands*
[h]*Visiting Astronomer at the Infrared Telescope Facility, which is operated by the University of Hawaii under Cooperative Agreement no. NNX-08AE38A with the National Aeronautics and Space Administration, Science Mission Directorate, Planetary Astronomy Program.*

**Abstract**

Near-infrared spectroscopy of Near-Earth Objects (NEOs) connects diagnostic spectral features to specific surface mineralogies. The combination of spectroscopy with albedos and diameters derived from thermal infrared observations can increase the scientific return beyond that of the individual datasets. For instance, some taxonomic classes can be separated into distinct compositional groupings with albedo and different mineralogies with similar albedos can be distinguished with spectroscopy. To that end, we have completed a spectroscopic observing campaign to complement the ExploreNEOs Warm Spitzer program that obtained albedos and diameters of nearly 600 NEOs (Trilling et al., 2010). The spectroscopy campaign included visible and near-infrared observations of ExploreNEOs targets from various observatories. Here we present the results of observations using the low-resolution prism mode ($\sim$0.7-2.5 $\mu$m) of the SpeX instrument on the NASA Infrared Telescope Facility (IRTF). We also include near-infrared observations of ExploreNEOs targets from the MIT-UH-IRTF Joint Campaign for Spectral Reconnaissance. Our dataset includes near-infrared spectra of 187 ExploreNEOs targets (125 observations of 92 objects from our survey and 213 observations of 154 objects from the MIT survey). We identify a taxonomic class for each spectrum and use band parameter analysis to investigate the mineralogies for the S-, Q-, and V-complex objects. Our analysis suggests that for spectra that contain near-infrared data but lack the visible wavelength region, the Bus-DeMeo system misidentifies some S-types as Q-types. We find no correlation between spectral band parameters and ExploreNEOs albedos and diameters. We investigate the correlations of phase angle with band area ratio and near-infrared spectral slope. We find slightly negative Band Area Ratio (BAR) correlations with phase angle for Eros and Ivar, but a positive BAR correlation with phase angle for Ganymed. The results of our phase angle study are consistent with those of (Sanchez et al., 2012). We find evidence for spectral phase reddening for Eros, Ganymed, and Ivar. We identify the likely ordinary chondrite type analog for an appropriate subset of our sample. Our resulting proportions of H, L, and LL ordinary chondrites differ from those calculated for meteorite falls and in previous studies of ordinary chondrite-like NEOs.

## 1. Introduction

Since many asteroids have diagnostic spectral features in near-infrared wavelengths, we can investigate their mineralogies remotely. These connections to mineralogy allow for detailed investigations of individual objects and can be applied to questions regarding the compositional distributions of asteroids throughout the Solar System. In near-Earth space, one puzzle is the relative distribution of ordinary chondrite-like material. Previous work (e.g Vernazza et al., 2008; Dunn et al., 2013; de León et al., 2010) suggests that a majority of near-Earth objects (NEOs) have compositions similar to LL ordinary chondrites. This result was unexpected since LL ordinary chondrites are only ∼7.9% of all meteorite falls (Burbine et al., 2002). The gap between these two populations will likely be bridged as we start to observe smaller objects, the direct precursors to our meteorite collection, and as our spectroscopic observing sample sizes grow.

There are several quantitative approaches to near-infrared spectral analysis. Taxonomic classification based on measurements of the spectral slope and absorption features is often used as a first step. Basic compositional properties can inferred if the average compositions of each of the taxonomic classes are known. Taxonomy can suggest the presence of certain minerals, such as mafic silicates, but in many cases the mineralogical interpretation can be ambiguous. For those objects with broad 1-$\mu$m and 2-$\mu$m absorption features we can study the composition more thoroughly using band parameter analysis. Band parameter analysis has been used in many studies (e.g. Gaffey et al., 1993; Thomas and Binzel, 2010; de León et al., 2010; Dunn et al., 2013; Burbine et al., 2009; Moskovitz et al., 2010; Reddy et al., 2011a) to determine the compositions and relative abundances of the olivines and pyroxenes present on the surface of various asteroids. For olivine-pyroxene mixtures, the band center wavelength of the broad 1-$\mu$m absorption feature is a function of the relative abundance and composition of olivine and pyroxene, and the band center of the broad 2-$\mu$m absorption is a function of the pyroxene composition. Additionally, the band area ratio (the ratio of the area of the 2-$\mu$m band to the area of the 1-$\mu$m band) is a function of the relative olivine and pyroxene abundances. Using the derived compositions, the diversity and distribution of material in near-Earth space can be investigated remotely.

The regions within the parameter space defined by calculated band centers and band area ratios provide insight into meteorite analogs and the diversity of compositions among the asteroid sample. These regions are correlated to potential meteorite types using laboratory derived mineralogies of various meteorite analogs. The relationship between band parameters and mineralogical composition is particularly well studied for ordinary chondrites and basaltic achondrites (e.g. Dunn et al., 2010; Burbine et al., 2009, 2007; Reddy et al., 2011a,b, 2012b).

The combination of spectroscopy and albedos and diameters derived from thermal infrared observations can increase the scientific return beyond that of the individual datasets. To that end, we have completed a ground-based observing campaign to measure visible and near-infrared spectra of near-Earth objects in support of the Warm Spitzer project entitled "The Warm Spitzer NEO Survey: Exploring the history of the inner Solar System and near Earth space" or ExploreNEOs (Trilling et al. 2010 and Trilling et al. submitted). This two year program (2009-2011) was allocated 500 hours to determine the albedos and diameters of nearly 600 NEOs using the Warm Spitzer capabilities of the IRAC (Infrared Array Camera, Fazio et al. 2004; Werner et al. 2004). In this paper, we present our near-infrared observations from the SpeX instrument at the NASA Infrared Telescope Facility (IRTF). We supplement our observations with spectra of ExploreNEOs targets from the MIT-UH-IRTF Joint Campaign for NEO Spectral Reconnaissance.

## 2. Data

### 2.1. Observations and Reduction

The observations presented in this paper were obtained using the low-resolution prism mode of the SpeX instrument (Rayner et al., 2003) on the NASA Infrared Telescope Facility (IRTF) between 1 September 2009 and 28 January 2012. Each object was observed with the dichroic out of the light path, which resulted in spectra covering ∼0.7-2.5$\mu$m. We used the 0.8 x 15 arcsec slit with the *abba* 7.5 arcsec dither pattern. Observations were taken at the parallactic angle. Each individual exposure was nominally limited to 120s due to variability of the atmosphere. Brighter object exposure times were calculated to avoid any non-linearity in the detector. Solar standards were selected using a list provided by S. J. Bus (personal communication) and SIMBAD searches for G-dwarfs with solar-like $B - V$ and $V - K$ colors near the position of the object. We used the internal SpeX calibration macro for flat-field frames and wavelength calibration argon lamp frames.

The spectra were reduced in one of two methods that employed standard near-infrared reduction techniques. One method used IDL tools (e.g. Emery et al., 2011) to perform the reduction, while the other used the IRAF `apextract` package with telluric atmosphere correction by the ATRAN model atmosphere (e.g. Lord, 1992; Rivkin et al., 2004). Both techniques included the creation of bad pixel maps, division by the flat field, subtraction of the dithered pairs to remove atmospheric emission, wavelength calibration, and the extraction of two-dimensional spectra to one-dimensional spectra. We chose solar-like stars as telluric calibrators in order to also correct for the spectroscopic signature of reflected sunlight from the targets. These techniques were tested against each other on a small number of objects and the reductions were nearly identical. Some reduced spectra show evidence of telluric absorption in the ∼1.4 and ∼1.8-1.9 $\mu$m regions, which should not be confused with features that present compositional information. The one exception to these described reduction techniques is the analysis of (3552) Don Quixote. Mommert et al. (submitted) discovered cometary activity on Don Quixote at the time of our ExploreNEOs observation. Therefore, a small reduction artifact found in the spectrum needed to be addressed. The final spectrum presented in this paper and in Mommert et al. (submitted) was reduced using Spextool (Cushing et al., 2004) and the ATRAN atmospheric correction.

We present 125 observations of 92 ExploreNEOs targets (see Table 1 & Figure 9). Some of these spectra were used to taxonomically classify objects for the study of average albedo by taxonomic complex presented in Thomas et al. (2011). Additional visible and near-infrared spectra from the campaign will be presented in future papers.

Table 1: Table of Observations

## 2.2. MIT-UH-IRTF Joint Campaign for NEO Spectral Reconnaissance

A number of ExploreNEOs targets were also observed with SpeX as part of the MIT-UH-IRTF Joint Campaign for NEO Spectral Reconnaissance[1]. The objects were identified using the MIT catalog of asteroid spectra and include observations from the MIT Joint Campaign, Binzel et al. (2001), Binzel et al. (2004a), and Rivkin et al. (2004). These objects were also observed with the low-resolution prism mode on SpeX with the 0.8 x 15 arcsec slit and the 7.5 arcsec dither pattern. These objects were reduced using the IRAF `apextract` package and the ATRAN model atmosphere. Data taken prior to mid-2007 (prior to and including `sp61`) used the dichroic to remove light short ward 0.82$\mu$m. The blue end of the data for subsequent observations ranges between ∼0.7-0.8$\mu$m. Visible wavelength data from the Small Main-Belt Asteroid Spectroscopic Survey (SMASS) are included when available (Xu et al., 1995; Bus and Binzel, 2002).

We include 213 observations of 154 ExploreNEOs targets. Table 2 shows all the spectra used in this study, including the file name from the MIT-UH-IRTF Joint Campaign. The file names are marked with a footnote if the file does not include data at or below 0.75$\mu$m, which affects our band parameter analysis (§2.4).

The combination of our survey and the MIT-UH-IRTF Joint Campaign for NEO Spectral Reconnaissance (hereafter "MIT Joint Campaign") yields a total of 340 observations of 187 ExploreNEOs targets.

Table 2: Table of Observations from MIT Joint Campaign

## 2.3. Taxonomic Classification

Taxonomic classifications of all 340 observations from both our ExploreNEOs survey and the MIT Joint Campaign determined from the Bus-DeMeo online taxonomic classifier[2] (DeMeo et al., 2009) are presented in Tables 1 & 2. We use the definitions of the complexes adopted by Thomas et al. (2011) and defined in Binzel et al. (2004c).

The online classification system only returns a definitive classification when a spectrum contains visible and near-infrared data. This is the case for a subset of the MIT Joint Campaign data. For all near-infrared only observations (those without visible spectra), the classification system returns a variety of possible subclasses. The additional absorption features and slope information contained in the visible wavelength region are vital to non-ambiguous classification, especially for the relatively featureless C- and X-complexes. DeMeo et al. (2009) show that the lack of visible wavelength data allow certain classes to occupy the same principal component space. This prevents the data from being formally classified by principal component analysis, and all possible types are returned by the system. Some of these objects can be classified if the presence or absence of weak absorption features or small spectral variations typical to the potential subclasses are taken into account. These features are too small to be distinguished in principal component space and too small to affect the residuals between the input spectrum and the class average spectrum, but are indicative of certain classifications.

---

[1] http://smass.mit.edu/minus.html
[2] http://smass.mit.edu/cgi-bin/busdemeoclass-cgi

For all near-infrared only spectra, we compare the object's spectrum with the average spectra of the possible subclasses. Potential classifications are ruled out when the shapes, slopes, and/or residuals of the object in question are not consistent with the average subclass spectra. For most objects, this examination reveals the most likely or two most likely subclass classifications for the object. Two subclass classifications are given only if the two subclasses are within the same taxonomic complex. If several potential subclass classifications within a single complex are possible following examination, the object is classified as belonging to that complex and the classification is marked as the complex letter followed by a star (e.g., S*).

If the potential subclass classifications span more than one taxonomic complex, then the object is labeled as being of indeterminate complex and is marked as such. This is particularly prevalent for objects of C- or X-complex. These complexes are often difficult to separate using only near-infrared data. The subtle visible wavelength spectral features are often crucial to making definitive assignments.

We include two labels that indicate an object has not been classified. If the classification system finds no matches in principal component space it returns "Classifier Indeterminate." If the classifier returns several options in principal component space that are not good matches to the spectrum, we do not classify the object and label it as "No Tax Match". These two labels appear in Tables 1 & 2. The problems classifying the "No Tax Match" spectra are caused primarily by thermal tails in the observations. Objects with thermal tails are labeled with a footnote and will be addressed in a future paper.

We do not allow any past classifications to influence the classifications presented in this paper. We compare our classifications with the taxonomic classes reported by the European Asteroid Research Node (EARN) Near-Earth Asteroid Database[3] and find that most classifications from our survey are the same class or are within the same taxonomic complex as previously determined taxonomic classifications. Five objects have distinctly different classifications in our survey than appear in EARN. Three of these objects, (24761) Ahau, (154029) 2002 CY46, and (297418) 2000 SP43, have EARN classifications from broadband photometry observations (Hicks et al., 2010; Somers et al., 2010; Ye, 2011). In these cases, the classification differences are likely due to the low resolution of broadband photometry. We classify (297418) 2000 SP43 as V-type with our ExploreNEOs survey spectrum and with the MIT Joint Campaign spectrum, but the classification on EARN is Sr-type. This discrepancy is likely due to the fact that the previous classification comes from a visible wavelength spectrum with no near-infrared data. The final classification discrepancy is (138911) 2001 AE2. This object was previously classified as T-type (Binzel et al., 2004b). The primitive, carbonaceous-like, material implied by the T-type classification and the low delta-V orbit of the object made this object a candidate for spacecraft exploration. The classification for our observation is S-complex. The S-complex classification suggests the thermal history and evolution of the object is different from the primitive material suggested by the previous T-type classification. The T-type classification was determined with visible wavelength data and our near-infrared S-complex classification is unambiguous. This suggests that (138911) 2001 AE2 is not an appropriate target for missions to primitive objects.

The spectrum for (433) Eros provides an example of a classification mismatch seen in our survey. Eros is often considered a quintessential S-type object; however the near-infrared spectra from the ExploreNEOs survey are formally classified in the Q-complex. Most of the MIT Joint Campaign observations are classified as S-complex. The key difference between the ExploreNEOs and MIT data is the availability of visible wavelength data. Our observations have no visible wavelength component included, while the MIT data uses visible wavelength data from Binzel et al. (2004c). We tested the robustness of these classifications by removing the visible data from the MIT spectra and appending it to each of the ExploreNEOs observations. We reclassified each of these alternate spectra. All MIT spectra with no visible wavelength component classified as Q-complex (either Q/Sq or Q), while most ExploreNEOs spectra with the visible wavelengths added classified as Sw (one classified as Sqw). This suggests that for near-infrared only observations, the Bus-DeMeo system returns more apparent Q-types than truly exist in the population. To address this problem quantitatively, visible spectra should be obtained for objects that have been classified as Q-complex using only near-infrared data. The classifications of the combined visible and near-infrared spectra can be compared against the classifications that use only near-infrared data.

---

[3]http://earn.dlr.de/nea/

### 2.4. Band Parameters

For objects belonging to the S-, Q-, and V- complexes, which show large 1- and 2-$\mu$m absorption features, certain spectral band parameters can be used to infer composition (e.g. Burns et al., 1972; Adams, 1974; Gaffey et al., 1993; Dunn et al., 2010; Sanchez et al., 2012; Dunn et al., 2013). These diagnostic absorption features are indicative of crystalline olivine and/or pyroxene. These minerals are common in various meteorites and asteroids. Olivine characteristically shows a broad absorption feature centered near 1.04-1.1$\mu$m that is comprised of three overlapping bands, while pyroxenes show two broad absorption features centered at 0.9-1$\mu$m and 1.9-2$\mu$m. These features are caused by $Fe^{2+}$ in the crystalline structure (e.g. Burns, 1993). For olivine-pyroxene mixtures, the wavelength of the 1$\mu$m feature (Band I, BI) is a function of the relative abundances and compositions of olivine and pyroxene, while the wavelength of the 2$\mu$m feature (Band II, BII) is a function of the pyroxene composition (e.g., Cloutis et al., 1986). The ratio of the area of Band II to Band I, or Band Area Ratio, is also a measure of the relative abundances of olivine and pyroxene. The depths of the bands can be used to investigate phase angle and space weathering effects under the assumption that mafic mineral abundance and particle size are the same (Sanchez et al., 2012; Reddy et al., 2012c). The band parameters that work best for this analysis are 1-$\mu$m Band Depth (BID), 1-$\mu$m Band Center (BIC), 2-$\mu$m Band Center (BIIC), and Band Area Ratio (BAR). For a very small number of taxonomically classified objects, the spectra are too noisy to calculate specific band parameters (e.g., (7358) Oze from 17 Sep 2011 does not have a BAR calculation). Calculated band parameters are shown in Tables 3 and 4.

All spectral band parameters are calculated based on the procedures described in Gaffey et al. (2002). For consistency between our two reduction techniques, all spectra are normalized to unity at 1-micron. We find no change between the parameters calculated before and after the normalization. The spectra are displayed using this normalization in the appendix (Figure 9). All parameters are calculated using a spectrum that is smoothed with a boxcar average of a chosen width. This ensures that no individual point can influence the calculation. Most objects are smoothed with a width of 10 spectral points; however, some high signal-to-noise observations use a smaller value. To calculate band centers, a linear continuum between the two local maxima is calculated for each band. The linear continua are divided out of each band and the center is calculated by fitting a fourth-order polynomial to the bottom of the band and finding the local minimum. For some objects, the linear continuum that defines Band I will cut through the spectrum at slightly shorter wavelength than the local maximum at $\sim$1.5$\mu$m. This is defined as a "rollover" spectrum and in these cases a new local maximum is found from the continuum removed spectrum and the Band I Center calculation is repeated. For Band II, the red edge is nominally defined as the end of the spectrum. However, many spectra are noisy in this region and we remove some spectral points without changing the angle of the continuum in order to most accurately calculate the band centers and band areas.

Band center errors are calculated via a Monte Carlo method. We generate offsets to each individual point in a spectrum by multiplying the observational error by a normally distributed random number with a standard deviation of 1. Each spectrum is altered 20,000 times and a new band center is calculated for each trial. The band center error is defined as the standard deviation of these trials. We define a minimum band center error of 0.01 for all observations to account for uncertainties that have not been quantified. Calculated uncertainties are only reported if they exceed this minimum. Band II centers often have a much higher error than Band I centers. This is not only due to increased noise in Band II, but also to the shallowness and relative flatness of Band II, and therefore the polynomial fit, which allows for subtle polynomial shifts to result in relatively large shifts in the band center calculations.

Since some observations do not contain data that extend to ∼0.75μm on the blue end, they do not contain a complete Band I. These Band I Centers are calculated using a linear continuum that begins at the start of the data, not at the local maximum blue-ward of Band I. Dividing Band I by this data-start continuum will produce a differently sloped continuum-divided Band I than if it had been divided by an appropriate local maxima continuum. Therefore, there is an additional source of error on objects that do not include data short ward of 0.75μm. We quantify this error by removing the spectral data <0.82 μm for a few high signal-to-noise observations to simulate an observation taken with the SpeX dichroic. We find that the BICs tend to shift to higher wavelength by values smaller than the defined minimum band center error of 0.01.

Band Area Ratio (BAR) is defined as the ratio of the area of Band II to the area of Band I. Each band area is calculated by using the trapezoidal rule to determine the area between the linear continuum and the continuum-divided spectral band. The error for each band area is calculated using the same Monte Carlo method used for band center errors. The error is defined as the standard deviation of the 20,000 altered area trials. The BAR error uses the individual area errors and propagates them with standard error analysis techniques. We define a minimum BAR error of 0.1 for all observations to account for uncertainties that have not been quantified such as the definition of the continuum slope and band depth changes due to phase angle. Calculated uncertainties are only reported if they exceed this minimum. This minimum error was also used by Reddy et al. (2010). The Band Area Ratio is not calculated for observations with no data <0.75μm. Using a data-start continuum instead of a local maxima continuum removes a significant amount of area from the Band I area calculation. Therefore, the BARs calculated in this manner would not be directly comparable to the BARs calculated for asteroids using the complete Band I and the BARs calculated for laboratory meteorite spectra.

Band depth is a measure of the ratio of the spectrum to the continuum at the calculated band center. The depth of a band is diagnostic of mafic mineral abundance and grain size, but is less diagnostic of composition than the other presented band parameters. The nominal band depth for an object can be modified by factors such as phase angle and lunar-style space weathering. We only present Band I depth in this paper because the telluric absorption remnants in Band II often make depth difficult to constrain. We use the smoothed, continuum-divided spectrum to determine Band I depth using the following equation:

$$\text{Band I Depth}(\%) = [1 - \frac{R_b}{R_c}] * 100 \qquad (1)$$

where $R_b$ is the reflectance spectrum value at the Band I center location and $R_c$ is the linear continuum value at the Band I center location (Clark and Roush, 1984).

2.5. *Temperature Corrections of Band Parameters*

Surface temperature variations on asteroids shift the band centers and change the width of the absorption feature (e.g. Moroz et al., 2000; Hinrichs and Lucey, 2002; Sanchez et al., 2012; Dunn et al., 2013). These temperature changes are of little importance for Main-Belt asteroids, but can be significant for the higher eccentricity near-Earth objects. Most importantly, in order to directly compare room temperature (∼300K) laboratory spectra with observational spectra, temperature corrections are necessary.

In order to apply temperature corrections to our sample, we need to identify the best laboratory matches to the two types of compositions in our band parameter sample. The S- and Q- complex objects are olivine-pyroxene assemblages. For these objects, we apply corrections derived from ordinary chondrites by Sanchez et al. (2012) and Dunn et al. (2013) using data from Moroz et al. (2000) and Hinrichs and Lucey (2002). The near-infrared spectra of the V-complex objects are dominated by the pyroxene features. For these objects, we apply corrections derived by Reddy et al. (2012c) using laboratory spectra of howardite, eucrite, and diogenite (HED) meteorites measured at various temperatures. All band parameters are corrected to room temperature values. Tables 3 and 4 contain the temperature corrections and corrected band parameters.

For each object, we calculate the average surface temperature (T) in Kelvin using the equation defined in Burbine et al. (2009).

$$T = [(1 - A)L_0/16\eta\varepsilon\sigma\pi r^2]^{1/4} \qquad (2)$$

where $A$ is the asteroid albedo, $L_0$ is the solar luminosity ($3.827 \times 10^{26}$ W), $\eta$ is the beaming parameter (in this case, assumed to be unity, Cohen et al. 1998), $\varepsilon$ is the asteroid's infrared emissivity (assumed to be 0.9, Salisbury et al. 1991), $\sigma$ is the Stefan Boltzmann constant ($5.67 \times 10^{-8}$ J s$^{-1}$ m$^{-2}$ K$^{-4}$), and $r$ is the asteroid's distance from the Sun in meters. When available, we use ExploreNEOs geometric albedos. Due to a small number of target list changes over the course of the Warm Spitzer observations, there are some objects on the ExploreNEOs target list for which albedos and diameters have not been measured. For those objects, we use the average albedo of the taxonomic complex (S=0.26, Q=0.29, and V=0.42) derived by Thomas et al. (2011). All calculated corrections are added to the spectrally derived values. Corrections are on order of 0.001 for BIC, 0.01 for BIIC, and 0.1 for BAR. We use equation 2 to calculate the asteroid surface temperatures to be consistent with previous asteroid spectroscopy temperature corrections (e.g. Burbine et al., 2009; Moskovitz et al., 2010). The temperature calculated by this equation gives the temperature of a fast-rotating spherical asteroid with high thermal inertia and uniform temperature. This is known as the Fast-Rotating Model (FRM, Lebofsky and Spencer, 1989) and assumes that energy is absorbed over the side of the asteroid facing the sun, but is radiated from the entire sphere. The temperatures determined using equation 2 are lower than those calculated using the Standard Thermal Model (STM, Lebofsky et al., 1986). If we calculate the sub-solar temperature using the STM we get higher temperatures, which correspond to smaller corrections compared to the FRM temperature corrections. Both the FRM and STM corrections are on the order of our defined minimum band parameter errors and therefore the differences between them do not contribute significantly to the overall error budget. We note that the FRM is not consistent with asteroids in this size range with known rotation periods (e.g., Delbó et al., 2003; Müller et al., 2012; Emery et al., 2010).

The majority of our objects in the band parameter sample belong to the olivine-pyroxene-rich S- and Q-complexes. Sanchez et al. (2012) shows that Band II is the most affected by temperature variations and that Band II center shifts to longer wavelength as temperature increases. The equation derived by Sanchez et al. (2012) for the temperature correction of the Band II center is the following:

$$\Delta \text{BIIC}(\mu m) = 0.06 - 0.0002 \times T(K) \tag{3}$$

The calculated BIIC correction is added to the spectrally derived BIIC values.

Sanchez et al. (2012) find the correction for Band I center to be very small in general ($\sim 0.003 \mu$m) and do not calculate a temperature correction for it. Dunn et al. (2013) use different corrections from the LL chondrite Soko-Banja and the L chondrite Elenovka and also find the correction to be small. However, since the direction of movement of the Band I center depends on the amount of olivine in the sample, the temperature corrections derived from these two ordinary chondrites do not shift in the same directions. Since both corrections are small, Dunn et al. (2013) apply a correction based on the average of the two corrections. We apply this method from Dunn et al. (2013) to our sample using the average of the following corrections:

$$\Delta \text{BIC, LL correction}(\mu m) = 0.000078 \times T(K) - 0.02194 \tag{4}$$

$$\Delta \text{BIC, L correction}(\mu m) = -0.000048 \times T(K) + 0.01512 \tag{5}$$

We note that the constant in Eq. 4 is printed incorrectly as 0.2194 in Dunn et al. (2013). If we use 0.2194 instead of 0.02194, we calculate a temperature correction that is two orders of magnitude different from the corrections calculated using the complementary L chondrites BIC correction (Eq. 5). This would result in a correction of order 0.1 $\mu$m instead of <0.001 $\mu$m. Dunn et al. (2013) state that the corrections are consistent with Burbine et al. (2009) and are on order <0.001 $\mu$m. Therefore we use 0.02194 throughout our analysis.

Sanchez et al. (2012) determine the correction for Band Area Ratio (BAR) to be the following:

$$\Delta \text{BAR} = 0.00075 \times T(K) - 0.23 \tag{6}$$

Dunn et al. (2013) again use different corrections from the two meteorites and apply the average correction:

$$\Delta \text{BAR, LL correction} = 0.0009 \times T(K) - 0.2666 \tag{7}$$

$$\Delta \text{BAR, L correction} = 0.0005 \times T(K) - 0.1708. \tag{8}$$

We find the Sanchez and Dunn BAR corrections to be nearly identical and use the average of the two in our analysis.

Band I depth was not addressed by Dunn et al. (2013) and Sanchez et al. (2012) found no obvious correlation between Band I depth and temperature. No correction was applied.

For the few pyroxene-rich V-types in our sample, we apply temperature corrections derived by Reddy et al. (2012c) using laboratory spectra of howardite, eucrite, and diogenite (HED) meteorites measured at various temperatures. The band center corrections for howardites and eucrites were calculated using spectra from Hinrichs and Lucey (2002) and the corrections for diogenites were calculated using spectra from Schade and Wäsch (1999). We apply an offset to each nominal V-type band center equal to the average of the two calculated corrections for the appropriate band. The corrections for Band I center are defined as the following:

$$\Delta \text{BIC}, \text{Howardites \& Eucrites } (\mu m) = 0.01656 - 0.0000552 \times T(K) \tag{9}$$

$$\begin{aligned}\Delta \text{BIC}, \text{Diogenites } (\mu m) = &\ 0.0000000017 \times T^3(K) - 0.0000012602 \times T^2(K) \\ &+ 0.0002664351 \times T(K) - 0.0124\end{aligned} \tag{10}$$

while the corrections for Band II center are the following:

$$\Delta \text{BIIC}, \text{Howardites \& Eucrites } (\mu m) = 0.05067 - 0.00017 \times T(K) \tag{11}$$

$$\Delta \text{BIIC}, \text{Diogenites } (\mu m) = 0.038544 - 0.000128 \times T(K). \tag{12}$$

The resulting corrections are similar to those calculated with equations derived by Burbine et al. (2009) who used spectra of enstatites at multiple temperatures measured by Moroz et al. (2000). Temperature corrections for band area ratio were not addressed in Reddy et al. (2012c), Burbine et al. (2009) or Moskovitz et al. (2010). Therefore, we do not correct the BARs of V-types for temperature variations.

Table 3: Band parameters for Q-, S-, and V-complex in our survey

Table 4: Band parameters for Q-, S-, and V-complex in the MIT Joint Campaign

## 3. Results

### 3.1. Band Parameters

The results of our band parameter analysis are shown in Figure 1. Panel (a) shows all band parameters for observed Q-, S-, and V-complex ExploreNEO targets in our survey (black circles) and in the MIT Joint Campaign for NEO Spectral Reconnaissance (red triangles). The Gaffey S-subtypes are shown in panel (b). These subtypes are the result of the S-complex containing a progression of mixtures that range from pure olivine (S(I)) to pure pyroxene (S(VII) and the basaltic acondrites (BA)). The comparison of the Gaffey S-subclasses and our calculated band parameters is shown in panel (c). Our data closely match the region defined by the continuum of S-subtypes defined by Gaffey et al. (1993) and the olivine-orthopyroxene mixing line determined by Cloutis et al. (1986). However, it seems that a small offset to higher BAR could improve the fit. A calibration offset between laboratory derived data and telescopic data was proposed in Gaffey (1984) to explain a discrepancy between their results and calibrations in Adams (1974). It was not determined whether this correction was correct. We do not calculate an offset correction for our data, but do allow for an extra margin when investigating S(IV)-subclass members (panel (d), §3.5).

Only one outlying object is excluded from this figure, (16834) 1997 WU22 from `sp77` of the MIT Joint Campaign. Visual inspection of the spectrum shows an anomalous 2-micron band that is significantly different from other observations of 1997 WU22. We deem this spectrum to be anomalous and not indicative of the true composition and band parameters of the object.

Figure 1: Four-panels. Band I Center v. Band Area Ratio

*3.2. ExploreNEOs albedos and diameters*

We combine ExploreNEOs albedos with our calculated Band I Centers (BICs). For this analysis we use the albedos calculated via the method described in Mueller et al. (2011) and Trilling et al. (submitted). The analysis produces an albedo and error bars generated by a Monte Carlo procedure. The calculations include a robust error analysis that considers the various sources of uncertainty present in the ExploreNEOs data and asymmetric error bars to capture the true uncertainty. The band parameter sample in this paper is taken from the Q-, S-, and V-complexes which are the higher albedo objects in our NEO sample. Due to the relatively large reflected light component for these objects, the albedo and diameter errors are correspondingly larger than for the low albedo objects (Harris et al., 2011; Mueller et al., 2011).

To potentially identify correlations using the albedo data, we limit the sample to those with $\sigma_{pV+}$ and $\sigma_{pV-}$ < 0.20. This error cutoff was chosen to eliminate the albedos with the highest error. Of the band parameters, we choose Band I Centers because they are the most accurate and precise measurements within our sample and the values can be used to determine the olivine and pyroxene chemistries (Dunn et al., 2010).

As discussed in Trilling et al. (2010), the nominal albedo and diameter values for Eros are quite different from those reported by the NEAR mission and supporting ground-based observations. This discrepancy is due to the near pole-on geometry for the Spitzer observations. Using a beaming parameter ($\eta$) and absolute magnitude (H) that were more appropriate for the viewing geometry of the observation, Trilling et al. (2010) recalculated the albedo ($p_V$) and diameter (D) to be 0.23 and 22.3 km respectively. These recalculated values agree with previous observations of Eros (e.g. Harris and Davies, 1999; Li et al., 2004). We incorporate these recalculated values in our analysis. All other ExploreNEOs albedos and diameters are the nominal values (Trilling et al., submitted).

One object, (141052) 2001 XR1, from the MIT Joint Campaign observations has multiple Warm Spitzer observations. This object was part of a subset of ExploreNEOs objects that were observed multiple times in order to study phase angle effects. We use the average of the four albedos and diameters in our analysis. We propagate the errors using standard error propagation techniques. The individual albedo and diameter values are shown in Table 4. The average values are $p_V$=0.22$^{+0.08}_{-0.05}$ and $D$=0.91$^{+0.09}_{-0.08}$ km.

We add NEOWISE (Mainzer et al., 2011) albedos and diameters to our analysis. For objects that lacked ExploreNEOs albedo and diameter solutions, we include the NEOWISE solutions. When the object appeared in both the ExploreNEOs and NEOWISE catalog, we average the albedos and diameters and propagated the errors using standard techniques in order to reduce the errors for these multiply observed objects.

We find no correlation between our error-limited albedos or diameters (ExploreNEOs and NEOWISE) with respect to Band I Centers (Figure 2). Since object albedos can depend on object chemistry and olivine is more reflective on average than pyroxene (e.g. Reddy et al., 2011c, 2012a), one might expect to see a trend of increasing albedo with increasing Band I Center. We find no correlation between albedo and BIC down to error limits of 0.05 in albedo. This suggests that processes such as space weathering and photometric effects due to various factors including regolith grain size have a non-negligible contribution to the observed albedo.

Figure 2: Band I Center v. Warm Spitzer Albedo

*3.3. Phase Angle- Band Area Ratio correlations*

Band parameters can be influenced by factors not related to compositional variation. One important factor to consider is the phase angle of the observation. Phase angle ($\alpha$) is the angle between the light incident on the asteroid surface and the light reflected from the surface towards an observer. Increased phase angle can result in an increase of spectral slope known as phase reddening and changes in the strength of the absorption bands.

In §3.1 we note that a small offset to higher Band Area Ratio (BAR) could improve the match between our data and the laboratory-derived Gaffey S-subclasses and olivine-orthopyroxene mixing line. To investigate potential causes of this offset, we investigate the effect of phase angle on BAR. We examine the relationship between phase angle and BAR on individual objects, not the full population, to understand the effects of varying compositions and optical properties. Our data has not been corrected for phase angle effects although data presented in Sanchez et al. (2012) demonstrates that changes in phase angle will affect the spectral band depth. Ideally, we would correct BAR to a phase angle of 30 degrees to best match laboratory spectra. Sanchez et al. (2012) did not find any correlations between increasing phase angle and BAR values. Our observations demonstrate that individual objects can show different spectral responses to changing phase angle and that there is not a universal phase angle-BAR correlation that can be applied to our spectral sample (Figure 3).

Three objects in our sample were observed at multiple phase angles: (433) Eros, (1036) Ganymed, and (1627) Ivar. Eros was observed 11 times total by the ExploreNEOs and MIT Joint Campaign surveys. Using the phase angles ($\alpha$) of the observations and the temperature corrected BARs, we calculate a least-squares linear fit of BAR=$(-0.003 \pm 0.002) \times \alpha + (0.445 \pm 0.083)$ with $R^2$=0.180. $R^2$ is the coefficient of determination, which ranges from 0 to 1 and indicates how well data fits a line (1 indicates the line perfectly fits the data) . Ganymed also has 11 observations from the two surveys. We find a least-squares linear fit for Ganymed of BAR=$(0.014 \pm 0.005) \times \alpha + (0.773 \pm 0.131)$ with $R^2$=0.494. Ivar has 7 observations from the two surveys, which we use to find a least-squares linear fit of BAR=$(-0.005 \pm 0.004) \times \alpha + (0.454 \pm 0.098)$ with $R^2$=0.215.

Eros and Ivar have similar phase angle-Band Area Ratio correlations: for these objects, BAR is anti correlated with phase angle. However, for Ganymed the correlation is opposite with a trend of increasing BAR with increasing phase angle. If most S-type objects behaved in a manner similar to Eros and Ivar, a correction from high phase angle to 30 degrees could cause a shift in our BAR distribution to better match the Gaffey S-subclasses. Sanchez et al. (2012) saw similar correlations of BAR with respect to phase angle. Their two observations of Ganymed also displayed a correlation opposite to the other objects'.

We find no consistent correlation between phase angle and BAR in our data. The large differences between the phase angle-BAR relationships for our three objects suggest that no individual correction should be applied to all objects. Therefore, we do not correct our sample for phase angle variations.

Figure 3: Band Area Ratio (BAR)- Phase Angle correlations

*3.4. Phase Angle- near-infrared spectral slope correlations*

We also examine the relationship between phase angle and spectral slope in near-infrared wavelengths for (433) Eros, (1036) Ganymed, and (1627) Ivar. To investigate these correlations we use phase angles ($\alpha$) and the near-infrared spectral slope between 0.82-2.4 $\mu$m. We choose this wavelength range to be consistent for all spectra. Spectra from the ExploreNEOs survey do not include visible wavelength observations. Since we do not want to include data points from visible wavelength observations and several of the Joint Campaign near-infrared spectra were taken with the SpeX dichroic in the light path, we select a lower wavelength of 0.82 $\mu$m. We choose the upper limit of wavelength to be 2.4 $\mu$m to avoid contributions to the spectral slope from spectral noise at higher wavelengths.

For the 11 observations of Eros we calculate a nominal least-squares linear fit of Near-Infrared Spectral Slope (NISS)=$(0.005 \pm 0.001) \times \alpha + (0.180 \pm 0.053)$ with $R^2$=0.530. We find a least-squares linear fit for the 11 observations of Ganymed of NISS=$(0.002 \pm 0.001) \times \alpha + (0.096 \pm 0.024)$ with $R^2$=0.302. We find a least-squares linear fit for the 7 observations of Ivar of NISS=$(0.002 \pm 0.002) \times \alpha + (0.231 \pm 0.036)$ with $R^2$=0.186.

Each of the three datasets has a single datapoint that lies far from the nominal correlation (Figure 4). We also calculate each correlation without the single outlying datapoint. For Eros the outlying point is from the MIT Joint Campaign (`sp105` on 28 Dec 2011). Without that observation we calculate a least-squares linear fit of NISS=$(0.004 \pm 0.001) \times \alpha + (0.197 \pm 0.018)$ with $R^2$=0.867. For Ganymed the outlying point is from the ExploreNEOs survey on 17 Sep 2011. Without the outlying observation we calculate a least-squares linear fit of NISS=$(0.001 \pm 0.001) \times \alpha + (0.107 \pm 0.15)$ with $R^2$=0.099. The outlying point for Ivar is from the MIT Joint Campaign (`sp77` on 27 Jan 2009). Without this outlying observation we calculate a least-squares linear fit of NISS=$(0.002 \pm 0.001) \times \alpha (0.237 \pm 0.019)$ with $R^2$=0.661. The $R^2$ values improve for Eros and Ivar when the outlying points are removed.

We find evidence for phase reddening for all three objects. Figure 4 shows the correlations for Eros, Ganymed, and Ivar. The solid lines indicate the nominal correlations and the dashed lines indicate the correlations calculated without the outlying points.

Figure 4: Near-infrared spectral slope- phase angle correlations

3.5. Ordinary Chondrite-like mineralogies

A subset of the olivine-pyroxene mixtures represented in the Band I Center versus Band Area Ratio diagram represent mineralogies consistent with the ordinary chondrites. We identify the potential ordinary chondrites via two methods. First, we use the S(IV) region defined by Gaffey et al. (1993) to include the mineralogical assemblages typical of ordinary chondrites. Figure 1d shows the Gaffey S(IV) region and the S(IV) objects that were identified in our sample. This region, and similarly shaped regions, have been used by various other studies. For example, Dunn et al. (2010) used the S(IV) region defined by Gaffey et al. (1993) to determine the potential ordinary chondrites and de León et al. (2010) identified ordinary chondrite-like objects by calculating a similar, but slightly larger, region using the spectral band parameters of ordinary chondrite spectra. We identify S(IV) objects by first selecting all objects with data and error bars wholly within the S(IV) subclass. Objects in the nearby S(III) region and with lower Band I Centers than the S(IV) region are included if the datum is close to the boundary and at least half of the associated error bar falls within the S(IV) region. We include some objects with higher Band I Centers than the S(IV) region if the datum and error fall wholly outside of all other Gaffey regions.

We established these generous selection criteria to allow for the possible shift in BAR due to phase angle effects. All spectra of Eros and Itokawa that are not selected by this procedure are added to the sample due to the confirmed ordinary chondritic nature of these asteroids (e.g. Trombka et al., 2000; Nakamura et al., 2011; Foley et al., 2006). Some of the Eros and Itokawa spectra fall in the S(III) region to the left of the S(IV) region. For Eros this is consistent with the S(III) and S(IV) spectra observed with the near-infrared spectrometer on the Near Earth Asteroid Rendezvous (NEAR) mission (Bell et al., 2002).

The second method used to identify potential ordinary chondrite analogues is by calculating the BIC-BAR regions that correspond to the Dunn et al. (2010) ordinary chondrite mineralogies. Dunn et al. (2010) used laboratory derived values of Fayalite (Fa, the iron-rich end member of the olivine solid solution series), Ferrosilite (Fs, the iron-rich end member of the pyroxene solid solution series), and olivine/(olivine+pyroxene) in 48 equilibrated ordinary chondrites to define regions (with and without errors) for the H, L, and LL ordinary chondrites. The derived mineralogies and spectral band parameters for the same 48 ordinary chondrites enabled Dunn et al. (2010) to calculate the relationships between the mineralogical parameters (Fa, Fs, and ol/(ol+px)) and the spectral parameters (BIC and BAR). We use the defined H, L, and LL ordinary chondrite regions to determine the corresponding region in B1C-BAR space. Objects are selected as potential ordinary chondrites if the datum is inside one of the three regions or if at least half of the corresponding error bar enters one of the regions. Figure 5 shows the three ordinary chondrite type regions compared to the S(IV) region and includes all the objects determined to be ordinary chondrite-like via this method. All of the objects inside the Gaffey S(IV) region are included as a subset of the Dunn region objects. Using this method, all observations of Eros and Itokawa are automatically included in the sample. The Dunn et al. (2010) H and L regions are reasonably consistent with the Gaffey S(IV) region. However, the LL region is much larger than the S(IV) region and includes the majority of the S(III) region which Gaffey et al. (1993) suggests contains different meteorite analogs (urelites).

We calculate the mineralogies for all of our potential ordinary chondrites identified via both the S(IV) region and the Dunn regions using the following three equations from Dunn et al. (2010):

$$\text{ol/(ol+px)} = -0.242 \times BAR + 0.728 \tag{13}$$

$$\text{Fa (mol \%)} = -1284.9 \times (BIC)^2 + 2656.5 \times (BIC) - 1342.3 \tag{14}$$

$$\text{Fs (mol \%)} = -879.1 \times (BIC)^2 + 1824.9 \times (BIC) - 921.7 \tag{15}$$

ol/(ol+px) is a modal ratio of olivine to olivine and pyroxene. Fa is the amount (in mol %) of fayalite and Fs is the amount (in mol %) of ferrosilite. The root mean square errors presented by Dunn et al. (2010) are 0.03 for ol/(ol+px), 1.3 mol% for Fa, and 1.4 mol% for Fs.

We classify each of our potential ordinary chondrite analogs into a subtype using the Dunn et al. (2010) ordinary chondrite mineralogy regions with errors included (Figure 5). The regions available for analysis are Fa (mol%) with respect to ol/(ol+px) and Fs (mol%) with respect to ol/(ol+px). We classify each object in both of these regimes to determine our final classification. Each object is determined to be a type if it falls within a region that is unique to a specific ordinary chondrite type. Objects that fall in the overlap regions are classified as potentially belonging to either class. In both regimes, there are several outliers near the H and LL ordinary chondrite regions. These are potentially not ordinary chondrite analogues and are classified as "H?" and "LL?". These are the objects whose BIC-BAR data placed them outside of the defined regions, but whose error bars did not allow for exclusion from the sample. In order to be conservative, the final classification given is the least certain of the two classifications. For example, if one calculation is H, but another is H/L, then the final classification is H/L. Figure 6 displays the ordinary chondrite mineralogies of our selected objects. Objects in the Gaffey S(IV) sample are shown with black circles (this ExploreNEOs survey) or red triangles (MIT Joint Campaign). As mentioned previously, all S(IV) objects are a subset of the Dunn regions sample. The Dunn classified objects that are not part of the S(IV) sample are displayed as squares and use the same color scheme to differentiate between this survey and the MIT Joint Campaign. The ordinary chondrite mineralogies calculated often identify Eros and Itokawa as LL ordinary chondrites. This suggests that, at least for the high signal-to-noise objects, this method is a reliable indicator of object composition.

We have a total of 82 objects in the Gaffey S(IV) sample, which breaks down into 4.9% "H?", 29.3% H, 11.0% H/L, 13.4% L, 14.6% L/LL, and 26.8% LL. The sample determined using the Dunn ordinary chondrite regions contains 109 objects and breaks down into 3.7% "H?", 22.9% H, 8.3% H/L, 10.1% L, 12.8% L/LL, 40.4% LL, and 1.8% "LL?". We discuss the sample statistics and the comparison to the meteorite population in §4.

Table 5: Mineralogies and ordinary chondrite analogues for S(IV) ExploreNEOs objects
Table 6: Mineralogies and ordinary chondrite analogues for S(IV) MIT-UH-IRTF objects
Figure 5: Dunn ordinary chondrite region with potential OCs
Figure 6: Mineralogies of S(IV) asteroids and the compositional regions for the H, L, and LL ordinary chondrites

3.6. Spectral Variation: Sisyphus and Cuyo

As part of the survey, we obtained rotationally resolved spectra of two near-Earth asteroids: (1866) Sisyphus and (1917) Cuyo. Sisyphus was observed once on 1 May 2011, in six sequential observations on 10 June 2011, and in three sequential observations on 24 June 2011. Cuyo was observed in six sequential observations on 24 June 2011 and once on 14 August 2011. Figure 7 shows Band I Center for Sisyphus (top panel) and Cuyo (bottom panel) for our sequences of observations. The x-axis is not directly correlated with time. Each set of observations are connected and presented sequentially. Additional observations for each object are offset along the x-axis.

The Band I Centers of the Sisyphus spectra vary from 0.915 to 0.932 $\mu$m among the ten observations, implying that the amounts of fayalite and ferrosilite vary across the surface. While the variations are not greater than the errors associated with the BICs, it should be noted that the errors on BIC are artificially limited to be equal to or greater than 0.01. This error limit accounts for many unquantified sources of error, but this limit is much higher than the nominal Sisyphus BIC errors which are ∼0.001-0.002. The nominal errors are low due to the high signal-to-noise of the observations that also enable us to separate the observations into rotationally resolved components instead of requiring the co-addition of the frames over the rotating body. The low nominal errors suggest that Sisyphus likely contains compositional changes over its surface. The variation corresponds to the difference between the S(VII) region (for the 0.915 $\mu$m BIC) and the S(IV) region with H chondrite mineralogy (for the 0.932 $\mu$m BIC).

Cuyo shows a much smaller variation in Band I Center from 0.922 to 0.926 $\mu$m. These BICs are also presented with the error limit of 0.01. Even within the low nominal Cuyo BIC errors ($\sim$0.001-0.002, similar to those of Sisyphus), there is no evidence for compositional variation across the surface of Cuyo.

Figure 7: Band I Center variation of Sisyphus and Cuyo

## 4. Discussion

We found no correlation between albedo and Band I Center. The absence of a correlation could be due to the large uncertainties in the ExploreNEOs albedo calculations ($\pm$50%, Harris et al., 2011). Other datasets should be used to confirm the lack of a relationship between these parameters before it is concluded that another process, such as space weathering or regolith evolution, is masking the nominal albedo of the grains.

Our investigation of the relationship between phase angle and Band Area Ratio for (433) Eros, (1036) Ganymed, and (1627) Ivar confirms the phase angle-BAR correlations seen in Sanchez et al. (2012). The Sanchez et al. (2012) sample contained only two or three observations for each object and most correlations were flat or slightly negative in slope. Our results for Eros (11 spectra) and Ivar (7 spectra) show similarly negative slopes. Ganymed is an exception in both our ExploreNEOs sample and in the Sanchez et al. work with a positive slope that correlates increasing BAR with increasing phase angle. There is no *a priori* reason to expect the optical properties of Ganymed to be different than other objects. All three objects investigated in this work are classified as potential ordinary chondrites in our analysis. One difference of note that separates Ganymed from Eros and Ivar is the higher calculated BARs. This leads to Ganymed's classification as a potential H chondrite while Eros is often correctly identified as an LL chondrite (e.g. Trombka et al., 2000; McCoy et al., 2001) and Ivar has band parameters associated with a range of potential ordinary chondrite types. While the BARs calculated for Ganymed are much larger than those of Eros and Ivar, Sanchez et al. (2012) presents band parameters of one other object at similar BAR, (4954) Eric, which shows a nearly flat, slightly negative correlation. This object also has band parameters correlating to classification as a potential H chondrite. Therefore, it is unlikely that the differing slopes can be solely attributed to the compositions associated with the higher BARs calculated for Ganymed and Eric. Additional studies of the phase angle-BAR relationship should focus on examining objects with a variety of BARs. The Sanchez et al. (2012) sample had 12 different objects and 10 of those had BARs less than or equal to $\sim$0.6. Another possible explanation is that Ganymed is much larger than many other NEOs (D=35.01 km, Usui et al., 2011). The variations observed in the BARs for Ganymed could be partially attributed to compositional heterogeneities on the surface. The potential for large objects to display surface heterogeneity does not apply to Eros, which displays spectral homogeneity that was observed during the Near-Earth Asteroid Rendezvous (NEAR) mission (Bell et al., 2002).

Our 11 near-infrared spectra of Ganymed suggest compositional heterogeneity on the surface. Five of the Ganymed spectra are classified as potential H chondrites. The other six fall either in the region with lower BIC than the S(IV) region (outside of any S subclass region), in the S(VI) region, or in the S(VII) region. Fieber-Beyer et al. (2011) classified Ganymed as S(VI) and suggested that it was a mesosiderite analog containing HED pyroxenes and NiFe metal. Our analysis does not confirm this result. If the entire range of mineralogies suggested by our band parameters exist on Ganymed, it is possible that the Fieber-Beyer et al. (2011) spectrum observed only the S(VI)-type surface assemblages since their observations were taken in a time period (3.6 hrs) shorter than the rotational period of Ganymed (10.31 hrs, Skiff et al., 2012).

The Dunn et al. (2010) equations for olivine and pyroxene chemistry used in §3.5 are quadratic. Unfortunately, the local maxima for both the fayalite and ferrosilite equations are within the range of laboratory derived chemistries of the LL ordinary chondrite. This implies that there is a range of fayalite and ferrosilite values that exists in the ordinary chondrite population, but will never be predicted by the Dunn et al. (2010) equations. The maximum calculable value for fayalite is $\sim$30.76 which corresponds to a Band I Center of 1.034 and the maximum calculable value for ferrosilite is $\sim$25.37 which corresponds to a Band I Center of 1.038. This upper limit falls within the true LL ordinary chondrite region, not the larger region that includes the root square mean errors associated with spectrally derived values that is used in our previous analysis. The artificial limit caused by the turnover of the fayalite and ferrosilite equations is clear in Figure 6 and does not appear to affect our ordinary chondrite class assignments. However, we note that the limit causes an artificial enhancement of objects with similar fayalite and ferrosilite values near the limit and an artificial dearth of objects with fayalite and ferrosilite values above the limit.

One large unanswered question in the study of the compositions of near-Earth objects is the apparent mismatch between the relative proportions of the ordinary chondrite subclasses (H, L, LL) seen in meteorite falls compared to the relative proportions of the ordinary chondrite subclass analogues observed spectroscopically in near-Earth space. Burbine et al. (2002) calculate that the H, L, and LL ordinary chondrites are 34.1%, 38.0%, and 7.9% of all meteorite falls, respectively. This corresponds to H, L, and LL ordinary chondrites being 42.6%, 47.5%, and 9.8% of ordinary chondrite falls. However, the majority of near-Earth asteroids that are ordinary chondrite analogs have spectra consistent with LL chondrite mineralogies (e.g. Vernazza et al., 2008). The analysis of the 47 NEOs with ordinary chondrite compositions by Dunn et al. (2013) found that H, L, and LL ordinary chondrite analogs are 15%, 10%, and 60%, respectively, with objects that could not be distinguished between L and LL accounting for 15%. de León et al. (2010) showed that the concentration of BAR and BIC values in the NEO population is consistent with a more significant presence of olivine in NEOs compared to Main Belt objects (see Figure 5 of de León et al. (2010)). This concentration mostly occupies the LL ordinary chondrite region.

The 82 objects from our sample that fall in the Gaffey S(IV) sample break down into 4.9% "H?", 29.3% H, 11.0% H/L, 13.4% L, 14.6% L/LL, and 26.8% LL. If we include the "H?" objects in the H chondrite group, the Gaffey S(IV) sample contains 34.1% potential H chondrites. The 109 objects from our sample in the Dunn ordinary chondrite regions sample break down into 3.7% "H?", 22.9% H, 8.3% H/L, 10.1% L, 12.8% L/LL, 40.4% LL, and 1.8% "LL?". If we include the "H?" and "LL?" objects into their respective groups, the Dunn sample contains 26.6% potential H chondrites and 42.2% potential LL chondrites. The relative proportions of the ordinary chondrites in our samples (the S(IV) and Dunn regions) are between the H and L chondrite dominant meteorite statistics (Burbine et al., 2002) and the LL chondrite dominant NEO statistics calculated by Dunn et al. (2013) and seen in Vernazza et al. (2008) and de León et al. (2010). Our samples have elevated percentages of potential LL ordinary chondrites compared to the meteorite fall statistics, but do not have values as high as the Dunn et al. (2013) sample. Similarly, we do not show the same concentration of NEOs in BIC-BAR space (near the LL ordinary chondrites) as was seen in de León et al. (2010). Instead we show a broader distribution of NEOs in BIC-BAR space. This suggests that our objects are more distributed over the possible mineralogies than previous surveys.

Our sample could be a step forward in finding the link between the meteorite population and previous NEO observations. However, there are no clear physical distinctions between our sample and the Dunn et al. (2013) sample that would explain the discrepancies between them. There are 13 objects that are present in both our sample and the Dunn et al. (2013) sample. Most ordinary chondrite classifications are consistent between the samples (there are two exceptions). One initial possibility was the sizes of the objects in each sample: small asteroids are the immediate precursors to meteorite falls and therefore should show mineralogies similar to meteorite falls. We can not directly compare diameters since not all of the objects in the Dunn et al. (2013) sample have thermal diameters. In order to be consistent, we compared the samples using H magnitudes from the Minor Planet Center as a proxy for diameter. The H magnitude ranges for both samples are nearly identical with a peak in the distribution around H$\sim$16-17. Figure 8 shows the histograms with the ExploreNEOs sample in blue and the Dunn et al. (2013) sample in green. The ExploreNEOs sample is clearly not sampling a smaller size regime than the Dunn et al. (2013) sample. We also compared the orbital elements of the objects in our survey to the objects in the Dunn et al. (2013) sample. There were no clear differences in semi-major axis ($a$), eccentricity ($e$), inclination ($i$), perihelion distance ($q$), or aphelion distance ($Q$).

It is important to note that some observations of Eros and Itokawa fall outside of the S(IV) region, but are included in the Dunn regions. Since these spectra have been observed with high signal-to-noise, this suggests that their positions outside of the S(IV) region are not due to spectral errors and are an accurate representation of the Band I Centers, Band Area Ratios, and the resulting mineralogy. Additionally, our spectral samples have several objects with multiple spectral observations (e.g. Ganymed). In several cases, only a subset of the spectra for each object are classified as ordinary chondrite analogues using the S(IV) region. When the larger Dunn region is considered, many of the spectra from these multiply observed objects that did not fall into the S(IV) region are also classified as ordinary chondrite analogues. Ganymed, 2002

EZ16, and 2002 TE66 are among the objects whose full set of observations were split between ordinary chondrite analogues and non-ordinary chondrite-like objects prior to the inclusion of the Dunn regions. This suggests that classification using the larger Dunn regions (or at minimum, some subset of them that is larger than the S(IV) region) is a valid way to identify potential ordinary chondrites.

It is clear that ordinary chondrite-like spectra can fall outside of the S(IV) region. However, it does not necessarily follow that the area outside the S(IV) region that is covered by the Dunn ordinary chondrite regions contains only ordinary chondrites. Gaffey et al. (1993) suggest that other mineralogies could be present in the regions labeled as S(III). The spectral signatures quantified by the band parameters used in this work are those caused by FeO. Other pyroxene and olivine chemistries that are not as well suited to band parameter analysis can overlap the ordinary chondrite regions in BIC-BAR space. For example, previous spectral studies (e.g. Gaffey et al., 1993; Sunshine et al., 2004) have cautioned that high-calcium pyroxene, a pyroxene variation that is not being measured by the Dunn equations, can affect the calculated band area ratios. If this BIC-BAR overlap in compositions is primarily in the LL ordinary chondrite region then it could help explain the over abundance of NEOs with LL ordinary chondrite-like compositions compared to the meteorite fall population.

## 5. Conclusions and Future Work

We have completed a ground-based near-infrared spectral survey of near-Earth objects in support of the ExploreNEOs Warm Spitzer project and complement our observations with near-infrared spectra of ExploreNEOs targets from the MIT Joint Campaign for NEO Spectral Reconnaissance. We report taxonomic classifications, spectral band parameters, and ordinary chondrite mineralogies for objects within our combined dataset of 340 observations of 187 targets. We find no correlation between spectral band parameters and ExploreNEOs albedos and diameters. This suggests that various optical effects, such as space weathering, can change the nominal albedo expected from composition alone. Our analysis suggests that for spectra that contain near-infrared data but lack the visible wavelength region, the Bus-DeMeo system returns more apparent Q-types than exist in the NEO population. We find negative Band Area Ratio correlations with phase angle for (433) Eros and (1627) Ivar, but a positive Band Area Ratio correlation with phase angle for (1036) Ganymed, confirming the findings of Sanchez et al. (2012). We find evidence for spectral phase reddening for Eros, Ganymed, and Ivar. We identify ordinary chondrite subtypes for all potential ordinary chondrites in our sample using equations from Dunn et al. (2010). Our resulting proportions of H, L, and LL ordinary chondrites differ from those calculated for meteorite falls and in previous studies of ordinary chondrite-like NEOs.

One complementary investigation that has not been completed is a source region analysis for our various taxonomic types and ordinary chondrite types. Future work will use the source region model described in Bottke et al. (2002) to do a full source region analysis of the sample. We will also incorporate our visible wavelength observations that were not described in this paper.

### Acknowledgements

We thank Juan Sanchez and Vishnu Reddy for answers to various questions regarding the methodology in Sanchez et al. (2012) and for their helpful reviews. We thank Lucy Lim for helpful comments on the manuscript.

This work is based in part on observations made with the Spitzer Space Telescope, which is operated by JPL/Caltech under a contract with NASA. Support for this work was provided by NASA through an award issued by JPL/Caltech.

This research was supported by an appointment to the NASA Postdoctoral Program at Goddard Space Flight Center, administered by Oak Ridge Associated Universities through a contract with NASA.

Part of the data utilized in this publication were obtained and made available by the MIT-UH-IRTF Joint Campaign for NEO Reconnaissance. The IRTF is operated by the University of Hawaii under Cooperative Agreement no. NCC 5-538 with the National Aeronautics and Space Administration, Office of Space Science,

Planetary Astronomy Program. The MIT component of this work is supported by the National Science Foundation under Grant No. 0506716.

This research has made use of the SIMBAD database, operated at CDS, Strasbourg, France.

# 6. Appendix

We show our near-infrared spectroscopic observations from the SpeX component of our ExploreNEOs spectroscopy survey in Figure 9. All spectra have been normalized to unity at 1-$\mu$m.

Figure 9.

Table 1. Table of observations of ExploreNEOs targets from our spectroscopic survey. The table includes object number, name, the UT date and time of observation, the integration time ($T_{int}$) in minutes, airmass, the solar standard used in the reduction, and the taxonomic classification.

Table 2. Table of ExploreNEOs spectra from the MIT-UH-IRTF Joint Campaign for NEO Spectral Reconnaissance. The table includes object number, name, the file name from the MIT survey catalog (smass.mit.edu), the UT date of observation, and the taxonomic classification.

Table 3. Calculated band parameters for our ExploreNEOs observations. In order to distinguish between multiple observations of the same object, the objects appear in the same order as they appear in Table 1. The table includes object number, name, phase angle ($\alpha$), heliocentric distance ($r$), calculated temperature (in K), Band I Depth (BID), Band I Center (BIC), BIC error ($\sigma_{BIC}$), calculated BIC temperature correction ($\delta_{BIC}$), corrected BIC ($\Delta$BIC), Band II Center (BIIC), BIIC error ($\sigma_{BIIC}$), calculated BIIC temperature correction ($\delta_{BIIC}$), corrected BIIC ($\Delta$BIIC), Band Area Ratio (BAR), BAR error ($\sigma_{BAR}$), calculated BAR temperature correction ($\delta_{BAR}$), corrected BAR ($\Delta$BAR), Warm Spitzer diameter ($D$), positive diameter error ($\sigma_{D+}$), negative diameter error ($\sigma_{D-}$), Warm Spitzer albedo ($p_V$), positive albedo error ($\sigma_{pV+}$), and negative albedo error ($\sigma_{pV-}$).

Table 4. Calculated band parameters for ExploreNEOs objects in the MIT-UH-IRTF Joint Campaign. As with Table 3, in order to distinguish between multiple observations of the same object the objects appear in the same order as they appear in Table 2. The table includes the same parameters as Table 3.

Table 5. Calculated mineralogies and ordinary chondrite types for all potential ordinary chondrite types in the ExploreNEOs spectral survey. Potential ordinary chondrites were determined using the S(IV) region and the Dunn et al. (2010) ordinary chondrite regions. Since the objects in S(IV) region are a subset of the Dunn classified objects, they are listed first. The table includes object number, object name, spectral type, corrected Band I Center ($\Delta$BIC), corrected Band Area Ratio ($\Delta$BAR), olivine/(olivine+pyroxene), Mol % of Fayalite (Fa), Mol % of Ferrosilite (Fs), ordinary chondrite type using Fa-ol/(ol+px), ordinary chondrite type using Fs- ol/(ol+px), and the final ordinary chondrite type classification using the least certain of the two classifications.

Table 6. Calculated mineralogies and ordinary chondrite types for all potential ordinary chondrite types for ExploreNEOs objects in the MIT-UH-IRTF Joint Campaign. As with Table 5, the S(IV) objects are listed first. The table includes the same parameters as Table 5.

Figure 1. (a) Band I Centers (BICs) versus Band Area Ratios (BARs) for all S-, Q-, and V-complex objects in our full spectral sample. Objects observed as part of the ExploreNEOs survey are displayed using black circles and those from the MIT Joint Campaign are shown using red triangles. (b) The Gaffey et al. (1993) S-subclasses and basaltic achondrites (BA) in BIC-BAR space. (c) The comparison of our spectral band parameters with the Gaffey et al. (1993) S-subclasses. The shape of the distribution of S-subclasses matches the distribution of objects in our analysis. (d) We use the S(IV) region as one method of identifying potential ordinary chondrites in our sample. This panel shows the S(IV) region with objects classified as S(IV) in our analysis. We include Eros and Itokawa because of their confirmed ordinary chondritic nature.

Figure 2. We find no correlation between ExploreNEOs (E) and NEOWISE (N) albedos and Band I Center for our spectral sample. Objects observed as part of the ExploreNEOs survey are shown in black. The circles represent ExploreNEOs-only albedos and the squares are an average of ExploreNEOs and NEOWISE albedos. Objects from the MIT Joint Campaign are shown in red. The triangles represent ExploreNEOs-only albedos and the squares are an average of ExploreNEOs and NEOWISE albedos. The errors on the ExploreNEOs albedos of these objects are higher than the ExploreNEOs average since we are targeting the high albedo objects in this band parameter analysis. We place an upper limit on the errors $\sigma_{pV+}$ and $\sigma_{pV-}$ of 0.20. We discard all observations with errors larger than this limit. We find no correlation down to error limits on $\sigma_{pV+}$ and $\sigma_{pV-}$ of 0.05.

Figure 3. Band Area Ratio (BAR) versus Phase Angle ($\alpha$) for (433) Eros, (1036) Ganymed, and (1627) Ivar. For Eros and Ivar, BAR is anti-correlated with phase angle. Ganymed has an opposite correlation with a trend of increasing BAR with increasing phase angle.

Figure 4. Near-infrared spectral slope versus Phase Angle ($\alpha$) for (433) Eros, (1036) Ganymed, and (1627) Ivar. All three objects show evidence of phase reddening. The solid lines indicate the nominal correlations and the dashed lines indicate the correlations calculated without the outlying points.

Figure 5. Our second method of identifying potential ordinary chondrites uses the Dunn ordinary chondrite regions calculated from the ranges of mineralogies determined in Dunn et al. (2010). We calculated the regions in Band I Center-Band Area Ratio space and used the regions to determine potential ordinary chondrite types. Each ordinary chondrite region is labeled with its type. The Gaffey S(III) and S(IV) regions are included for comparison. The L and H regions closely approximate the S(IV) region, but the LL region is much larger than the corresponding subset of the S(IV) region. The Dunn LL region includes most of the Gaffey S(III) regions.

Figure 6. We use the Dunn et al. (2010) equations to derive the mol % of Fayalite (Fa) and Ferrosilite (Fs) and olivine/(olivine+pyroxene) ratio from band parameters of ordinary chondrite-like spectra. The nominal errors on the chemistry calculations are 0.03 for the ol/(ol+px) calculation, 1.3 for the calculation of the mol % of Fa, and 1.4 for the calculation of mol % of Fs. The top panel shows mol % Fayalite versus ol/(ol+px). The bottom panel shows mol % of Ferrosilite versus ol/(ol+px). We determine a potential ordinary chondrite subtype (H, L, LL) based on the location of the objects in each of these compositional regions. The quadratic equations impose an artificial upper limit on Fayalite and Ferrosilite within the LL ordinary chondrite subclass that can be clearly seen.

Figure 7. Band I Center for Sisyphus (top panel) and Cuyo (bottom panel) for our sequences of observations. The x-axis is not directly correlated with time. Each set of observations is connected and presented sequentially. Additional observations for each object are offset in x. We see potential compositional variation on the surface of Sisyphus. Any variation is within the errors, but we set a hard minimum on the errors in order to better include several unquantified sources of error. The differences in BIC exceed the nominal calculated errors for the observations. Cuyo does not show indications of compositional variation on the surface.

Figure 8. We compared the distribution of H magnitudes (as a proxy for size) of our ExploreNEOs spectra to the spectral sample analyzed in Dunn et al. (2013) to determine if we were investigating different size regimes. We have different relative proportions of the ordinary chondrite types within our samples, which cannot be explained by differences in object sizes.

Figure 9. This paper presents the near-infrared spectra observed at the NASA Infrared Telescope Facility (IRTF) using the SpeX instrument as part of the ExploreNEOs spectral survey. We include the spectra of all 125 observations of 92 ExploreNEOs targets. Each spectrum contains the object name and number, the UT date of observation, and, when appropriate, a letter designating the sequence order for all rotationally resolved observations. All spectra are normalized to unity at 1-$\mu$m.

Table 1: Observations of ExploreNEOs targets

| Number | Name | UT | $T_{int}$ | Airmass | Standard | Taxonomic Type |
|---|---|---|---|---|---|---|
| 433 | Eros | 01-Sep-2009 12:06 | 2 | 1.20 | SAO 127422 | Sq/Q |
| 433 | Eros | 01-Nov-2009 05:42 | 18 | 1.05 | L113-276 | Q/Sq |
| 433 | Eros | 06-Mar-2010 04:52 | 8 | 1.32 | SAO 75438 | Q/Sq |
| 433 | Eros | 28-Jan-2012 12:17 | 1.07 | 1.08 | HD 91688 | Q/Sq |
| 1036 | Ganymed | 19-Mar-2010 09:50 | 16 | 1.42 | HD 111017 | Sr/S |
| 1036 | Ganymed | 10-Jun-2011 11:48 | 2 | 1.14 | HD 334883 | Sr |
| 1036 | Ganymed | 04-Jul-2011 10:31 | 2.7 | 1.28 | HD 204570 | Sr |
| 1036 | Ganymed | 17-Sep-2011 14:08 | 3 | 1.46 | Hya 64 | S |
| 1036 | Ganymed | 26-Sep-2011 10:21 | 2.7 | 1.24 | HD 232456 | Sr |
| 1036 | Ganymed | 19-Oct-2011 07:48 | 2.7 | 1.28 | SAO 75105 | Sr/S |
| 1627 | Ivar | 19-Feb-2010 10:26 | 32 | 1.01 | HD 89547 | S |
| 1864 | Daedalus | 07-Mar-2010 14:41 | 32 | 1.42 | HD 132076 | Sq/Q |
| 1865 | Cerberus | 05-Aug-2010 09:08 | 96 | 1.08 | L110-361 & L112-1333 | S |
| 1866 | Sisyphus | 01-May-2011 11:56 | 16 | 1.05 | HD 144684 | S |
| 1866 | Sisyphus A | 10-Jun-2011 06:16 | 16 | 1.18 | HD 125982 | S |
| 1866 | Sisyphus B | 10-Jun-2011 06:52 | 16 | 1.13 | HD 125982 | S |
| 1866 | Sisyphus C | 10-Jun-2011 07:19 | 16 | 1.11 | HD 125982 | S |
| 1866 | Sisyphus D | 10-Jun-2011 08:22 | 16 | 1.14 | HD 125982 | S |
| 1866 | Sisyphus E | 10-Jun-2011 09:07 | 8 | 1.21 | HD 125982 | S |
| 1866 | Sisyphus F | 10-Jun-2011 09:27 | 16 | 1.30 | HD 125982 | S |
| 1866 | Sisyphus A | 24-Jun-2011 06:08 | 16 | 1.16 | HD 123765 | S |
| 1866 | Sisyphus B | 24-Jun-2011 06:40 | 16 | 1.17 | HD 123765 | S |
| 1866 | Sisyphus C | 24-Jun-2011 07:04 | 16 | 1.19 | HD 123765 | S |
| 1917 | Cuyo A | 24-Jun-2011 10:16 | 16 | 1.01 | HD 342904 | Sv |
| 1917 | Cuyo B | 24-Jun-2011 10:46 | 24 | 1.01 | HD 342904 | Sr |
| 1917 | Cuyo C | 24-Jun-2011 11:31 | 24 | 1.02 | HD 342904 | Sv/Sr |
| 1917 | Cuyo D | 24-Jun-2011 12:11 | 24 | 1.05 | HD 342904 | Sr |
| 1917 | Cuyo E | 24-Jun-2011 12:48 | 24 | 1.11 | HD 342904 | Sv/Sr |
| 1917 | Cuyo F | 24-Jun-2011 13:29 | 24 | 1.22 | HD 342904 | Sr |
| 1917 | Cuyo | 14-Aug-2011 06:59 | 48 | 1.01 | L107-684 | Sr |
| 1943 | Anteros | 02-Sep-2009 11:30 | 32 | 1.15 | SAO 107874 | Sq |
| 2102 | Tantalus | 04-Jul-2011 06:36 | 16 | 1.17 | HD 234075 | Sr |
| 3103 | Eger | 24-Jun-2011 14:38 | 40 | 1.06 | HD 206658 | C/X INDET |
| 3122 | Florence | 05-Aug-2010 15:26 | 10 | 1.03 | L115-271 | Sq/Q |
| 3200 | Phaethon | 01-Sep-2009 13:53 | 64 | 1.20 | SAO 3963 | B |
| 3552 | Don Quixote | 18-Oct-2009 05:37 | 20 | 1.54 | L113-276 | D |
| 3554 | Amun | 01-Feb-2010 14:44 | 32 | 1.12 | SAO 97431 | D |
| 4015 | Wilson-Harrington | 02-Sep-2009 05:02 | 48 | 1.40 | HD 148116 | C* |
| 4055 | Magellan | 01-Sep-2010 05:13 | 32 | 1.06 | HD 156486 | V |
| 4283 | Cuno | 19-Oct-2011 12:17 | 32 | 1.07 | HD 16397 | Sq/Q |
| 5143 | Heracles | 27-Sep-2011 12:58 | 32 | 1.04 | HD 275948 | Q |
| 5587 | 1990 SB | 28-Jan-2012 14:25 | 48 | 1.25 | SAO 157011 | S/Sr |
| 5626 | 1991 FE | 02-Sep-2009 06:47 | 32 | 1.25 | HD 190524 | Sq |
| 5646 | 1990 TR | 01-Nov-2009 14:22 | 48 | 1.03 | HD 60913 | Q |
| 5786 | Talos | 05-Aug-2010 11:18 | 72 | 1.02 | L110-361 & L112-1333 | Sq |
| 6239 | Minos | 01-Sep-2010 11:49 | 16 | 1.12 | HD 220685 | Q/Sq |
| 7358 | Oze | 13-Aug-2011 05:44 | 48 | 1.32 | SAO 159708 | S |
| 7358 | Oze | 14-Aug-2011 05:47 | 48 | 1.32 | L107-684 | S |
| 7358 | Oze | 17-Sep-2011 05:02 | 48 | 1.39 | L110-361 | S |
| 7822 | 1991 CS | 06-Mar-2010 05:37 | 48 | 1.09 | HD 288180 | S* |
| 8566 | 1996 EN | 02-Sep-2009 14:58 | 32 | 1.13 | HD 30947 | V |
| 8567 | 1996 HW1 | 26-Sep-2011 15:11 | 16 | 1.02 | HD 247554 | Q |
| 11066 | Sigurd | 19-Feb-2010 12:32 | 48 | 1.10 | HD 129290 | Sr |
| 12711 | Tukmit | 01-Sep-2009 05:07 | 48 | 1.25 | BD+12 2836 | Q |
| 12923 | Zephyr | 04-Apr-2010 12:19 | 64 | 1.19 | HD 125982 | S |
| 16834 | 1997 WU22 | 01-Sep-2009 12:24 | 48 | 1.14 | SAO 127422 | Sq |
| 17274 | 2000 LC16 | 01-Sep-2009 09:07 | 16 | 1.05 | HD 205027 | D |
| 18882 | 1999 YN4 | 27-Dec-2010 07:52 | 32 | 1.33 | SAO 13263 | Sr |
| 19764 | 2000 NF5 | 27-Sep-2010 09:32 | 32 | 1.04 | SAO 109095 | Sq |
| 20086 | 1994 LW | 24-Jun-2011 08:29 | 44 | 1.18 | HD 133600 | C/X INDET |
| 21088 | 1992 BL2 | 19-Feb-2010 08:06 | 64 | 1.20 | HD 70996 | Q |
| 23187 | 2000 PN9 | 01-Sep-2010 09:04 | 80 | 1.04 | HD 206658 | Sq |
| 24761 | Ahau | 06-Jan-2010 15:07 | 16 | 1.20 | HD 96995 | C/X INDET |
| 27346 | 2000 DN8 | 01-Feb-2010 08:50 | 48 | 1.25 | HD 43713 | Q |
| 36284 | 2000 DM8 | 02-Feb-2011 12:26 | 32 | 1.03 | HD 249566 & HD 89179 | Sq |
| 39572 | 1993 DQ1 | 17-Sep-2011 09:40 | 80 | 1.30 | L113-276 | Q |
| 52750 | 1998 KK17 | 19-Mar-2010 14:54 | 32 | 1.27 | HD 175470 | V |
| 53789 | 2000 ED104 | 27-Sep-2010 05:22 | 80 | 1.14 | SA 115-271 & SAO 86827 | Sq |
| 54789 | 2001 MZ7 | 01-Feb-2010 10:51 | 16 | 1.05 | SAO 97431 | C/X INDET |
| 54789 | 2001 MZ7 | 19-Feb-2010 07:21 | 16 | 1.11 | SAO 60533 | C/X INDET |
| 65679 | 1989 UQ | 15-Oct-2010 08:19 | 32 | 1.13 | HD 10488 | C/Cb |
| 68216 | 2001 CV26 | 19-Feb-2010 14:03 | 56 | 1.01 | HD 129290 | Sq |
| 68216 | 2001 CV26 | 07-Mar-2010 11:47 | 32 | 1.06 | HD 123760 | Sq |
| 68216 | 2001 CV26 | 19-Mar-2010 10:55 | 32 | 1.03 | HD 115692 | S/Sq |
| 68350 | 2001 MK3 | 02-Feb-2011 05:25 | 72 | 1.15 | HD 23925 | S/Sr |
| 85839 | 1998 YO4 | 19-Feb-2010 11:32 | 32 | 1.02 | HD 96657 | Q |
| 86067 | 1999 RM28 | 01-Feb-2010 13:22 | 32 | 1.11 | HD 87776 | Q |
| 88254 | 2001 FM129 | 04-Apr-2010 09:22 | 32 | 1.36 | HD 82410 | Q |
| 96590 | 1998 XB | 15-Dec-2010 10:16 | 24 | 1.09 | Hya 64 | Sq |
| 100926 | 1998 MQ | 15-Oct-2010 11:25 | 24 | 1.40 | SAO 40380 | Q |
| 136617 | 1994 CC | 18-Oct-2009 11:49 | 80 | 1.12 | L93-101 & L98-978 | Sa |
| 137032 | 1998 UO1 | 15-Oct-2010 05:06 | 32 | 1.02 | SAO 89822 | Sq |
| 137062 | 1998 WM | 15-Oct-2010 09:23 | 64 | 1.20 | HD 12264 | Sq |
| 137084 | 1998 XS16 | 02-Jan-2011 08:09 | 48 | 1.05 | HD 42160 | S |
| 137084 | 1998 XS16 | 03-Jan-2011 07:10 | 32 | 1.16 | HD 42160 | S |
| 137125 | 1999 CT3 | 02-Feb-2011 13:24 | 84 | 1.25 | HD 92670 | Q |
| 137170 | 1999 HF1 | 01-May-2011 12:32 | 16 | 1.23 | HD 177780 | C/X INDET |
| 137671 | 1999 XP35 | 28-Dec-2011 10:57 | 48 | 1.16 | SA 98-978 | S |
| 138883 | 2000 YL29 | 02-Sep-2009 14:03 | 32 | 1.06 | SAO 55805 | Q |
| 138911 | 2001 AE2 | 28-Jan-2012 09:26 | 92 | 1.05 | SAO 97431 | S |
| 141498 | 2002 EZ16 | 05-Dec-2010 09:40 | 32 | 1.24 | HD 30518 | Sq |
| 141498 | 2002 EZ16 | 12-Dec-2010 06:55 | 68 | 1.08 | L93-101 | Sq |
| 143381 | 2003 BC21 | 27-Sep-2010 08:00 | 48 | 1.05 | SAO 127304 | Sr/S |
| 152558 | 1990 SA | 01-Sep-2010 11:19 | 16 | 1.02 | SAO 90986 | Sq/Q |
| 152563 | 1992 BF | 02-Feb-2011 07:29 | 72 | 1.09 | HD 249566 | K/C/X/ INDET |
| 152931 | 2000 EA107 | 07-Mar-2010 13:13 | 48 | 1.22 | HD 144423 | Q |
| 153814 | 2001 WN5 | 15-Oct-2010 06:24 | 48 | 1.11 | HD 200565 | K/L/Sq INDET |
| 153842 | 2001 XT30 | 05-Dec-2010 10:46 | 80 | 1.07 | HD 288029 | C/X INDET |
| 154029 | 2002 CY46 | 01-Sep-2010 12:23 | 92 | 1.01 | HD 14786 | S |
| 159402 | 1999 AP10 | 01-Sep-2009 09:49 | 16 | 1.32 | HD 216298 | Sq |

| Number | Name | UT | $T_{int}$ | Airmass | Standard | TaxonomicType |
|---|---|---|---|---|---|---|
| 162998 | 2001 SK162 | 19-Feb-2010 05:27 | 52 | 1.10 | HD 285295 | D |
| 163132 | 2002 CU11 | 07-Mar-2010 10:23 | 48 | 1.29 | HD 77066 | C/X INDET |
| 163243 | 2002 FB3 | 04-Apr-2010 05:10 | 64 | 1.18 | HD 296688 | Q |
| 163697 | 2003 EF54 | 02-Sep-2009 10:01 | 48 | 1.13 | SAO 107874 | Q |
| 164121 | 2003 YT1 | 18-Oct-2009 13:55 | 28 | 1.09 | L93-101 & L98-978 | V |
| 198856 | 2005 LR3 | 12-Dec-2010 04:50 | 80 | 1.15 | L115-271 | Sq |
| 198856 | 2005 LR3 | 16-Dec-2010 05:58 | 48 | 1.11 | L115-271 | Q |
| 207945 | 1991 JW | 15-Dec-2010 05:26 | 72 | 1.17 | L115-271 | Q |
| 207945 | 1991 JW | 16-Dec-2010 04:34 | 64 | 1.06 | L115-271 | Q |
| 214088 | 2004 JN13 | 06-Jan-2010 04:51 | 48 | 1.34 | L115-271 | Sq |
| 218863 | 2006 WO127 | 18-Oct-2009 10:23 | 64 | 1.14 | L93-101 & L98-978 | Sq |
| 219071 | 1997 US9 | 05-Dec-2010 07:41 | 64 | 1.34 | HD 16472 | Q |
| 220124 | 2002 TE66 A | 04-Apr-2010 10:44 | 8 | 1.03 | HD 124019 | Sq/Q |
| 220124 | 2002 TE66 B | 04-Apr-2010 14:17 | 16 | 1.30 | HD 124019 | Sq |
| 247517 | 2002 QY6 | 05-Aug-2010 06:25 | 112 | 1.18 | L110-361 & L112-1333 | Sr |
| 265187 | 2003 YS117 | 02-Feb-2011 09:58 | 88 | 1.15 | HD 68972 | Q |
| 275792 | 2001 QH142 | 27-Jan-2012 10:49 | 136 | 1.01 | L102-1081 | C/X INDET |
| 297418 | 2000 SP43 | 17-Sep-2011 14:47 | 36 | 1.04 | Hya 64 | V |
| 347813 | 2002 NP1 | 02-Sep-2009 08:09 | 48 | 1.03 | HD 199268 | Q |
|  | 2000 CO101 | 04-Apr-2010 07:20 | 64 | 1.19 | SAO 117128 | C/X/T INDET |
|  | 2000 TJ1 | 27-Sep-2010 10:38 | 82 | 1.03 | SA 115-271 & HD 5294 | Sq |
|  | 2001 HC | 04-Apr-2010 14:55 | 32 | 1.07 | HD 125982 | Sq |
|  | 2002 OS4 | 02-Jan-2011 06:07 | 64 | 1.21 | SAO 39637 | Sr |
|  | 2002 RQ25 | 02-Sep-2009 12:35 | 48 | 1.36 | HD 5078 | C/X INDET |
|  | 2005 MC | 06-Jan-2010 14:18 | 16 | 1.25 | HD 73668 | Sr |

Table 2: MIT-UH-IRTF Joint Campaign for NEO Spectral Reconnaissance Observations

| Number | Name | File Name | UT Date | Taxonomic Type |
|---|---|---|---|---|
| 433 | Eros | a000433.sp15 | 17-Jun-02 | Sw |
| 433 | Eros | a000433.sp16 | 17-Aug-02 | Sw |
| 433 | Eros | a000433.sp17 | 16-Sep-02 | Sw |
| 433 | Eros | a000433.sp101 | 22-Aug-11 | Sw |
| 433 | Eros | a000433.sp102 | 25-Sep-11 | Sw |
| 433 | Eros | a000433.sp103 | 24-Oct-11 | Sw |
| 433 | Eros | a000433.sp105 | 28-Dec-11 | Sqw |
| 1036 | Ganymed | a001036.sp05 | 28-Mar-01 | Sr |
| 1036 | Ganymed | a001036.sp51 | 03-Jun-06 | Sr |
| 1036 | Ganymed | a001036.sp80 | 27-Apr-09 | Sr |
| 1036 | Ganymed | a001036.sp103 | 24-Oct-11 | S |
| 1036 | Ganymed | a001036.sp104 | 31-Oct-11 | S |
| 1627 | Ivar | a001627.sp09 | 13-Jan-02 | S |
| 1627 | Ivar | a001627.sp73 | 02-Sep-08 | Sqw |
| 1627 | Ivar | a001627.sp74 | 02-Oct-08 | Sw |
| 1627 | Ivar | a001627.sp76 | 03-Dec-08 | S |
| 1627 | Ivar | a001627.sp77 | 27-Jan-09 | Sr |
| 1627 | Ivar | a001627.sp88 | 22-Feb-10 | Sw |
| 1685 | Toro | a001685.sp13 | 08-May-02 | S |
| 1685 | Toro | a001685.sp17[1] | 16-Sep-02 | S |
| 1864 | Daedalus | a001864.sp05 | 29-Mar-01 | Sq |
| 1865 | Cerberus | a001865.sp75 | 31-Oct-08 | S |
| 1866 | Sisyphus | a001866.sp56 | 21-Nov-06 | Sw |
| 1916 | Boreas | a001916.sp07 | 14-Aug-01 | Sw |
| 1917 | Cuyo | a001917.sp72 | 07-Jul-08 | Sv |
| 1943 | Anteros | a001943.sp09 | 12-Jan-02 | Sw |
| 1943 | Anteros | a001943.sp103 | 24-Oct-11 | S |
| 1980 | Tezcatlipoca | a001980.sp55 | 25-Oct-06 | Sw |
| 2212 | Hephaistos | a002212.sp57[1] | 22-Dec-06 | Sq |
| 3102 | Krok | a003102.sp02 | 09-Oct-00 | Sqw |
| 3103 | Eger | a003103.sp06 | 21-Jun-01 | Xe |
| 3199 | Nefertiti | a003199.sp03 | 30-Jan-01 | NO TAX MATCH |
| 3199 | Nefertiti | a003199.sp36 | 03-Mar-05 | K |
| 3200 | Phaethon | a003200.sp19 | 27-Oct-02 | NO TAX MATCH |
| 3200 | Phaethon | a003200.sp34 | 10-Dec-04 | B |
| 3552 | Don Quixote | a003552.sp85 | 25-Oct-09 | D |
| 3554 | Amun | a003554.sp78[1] | 02-Mar-09 | C/X INDET |
| 3554 | Amun | a003554.sp88[1] | 23-Feb-10 | D |
| 3554 | Amun | a003554.sp89[1] | 16-Mar-10 | D |
| 3554 | Amun | a003554.sp98[1] | 05-Apr-11 | D |
| 3671 | Dionysus | a003671.sp91 | 10-May-10 | C/X INDET |
| 3691 | Bede | a003691.sp68 | 10-Mar-08 | Cgh |
| 4015 | Wilson-Harrington | a004015.sp81[1] | 22-Jun-09 | B |
| 4055 | Magellan | a004055.sp38 | 11-Apr-05 | V |
| 4183 | Cuno | a004183.sp103 | 24-Oct-11 | Q |
| 4544 | Xanthus | a004544.sp58[1] | 21-Jan-07 | Sq/Q |
| 4660 | Nereus | a004660.6 | 15-Mar-02 | X* |
| 5011 | Ptah | a005011.sp56[1] | 21-Nov-06 | Q |

| Number | Name | File Name | UT Date | Taxonomic Type |
|---|---|---|---|---|
| 5131 | 1990 BG | a005131.sp67 | 15-Jan-08 | Sq |
| 5131 | 1990 BG | a005131.sp87 | 17-Jan-10 | Sqw |
| 5143 | Heracles | a005143.sp55 | 25-Oct-06 | Q |
| 5332 | 1990 DA | a005332.sp47[1] | 13-Nov-05 | Sr/S |
| 5496 | 1973 NA | a005496.sp96[1] | 06-Jan-11 | C/X INDET |
| 5587 | 1990 SB | a005587.sp05 | 28-Mar-01 | Sr |
| 5604 | 1992 FE | a005604.sp05 | 29-Mar-01 | V |
| 5604 | 1992 FE | a005604.sp80 | 27-Apr-09 | V |
| 5620 | Jasonwheeler | a005620.sp81[1] | 22-Jun-09 | K |
| 5626 | 1991 FE | a005626.sp84 | 20-Sep-09 | S |
| 5645 | 1990 SP | a005645.sp12[1] | 14-Apr-02 | C/X/T INDET |
| 5653 | Camarillo | a005653.sp35[1] | 08-Jan-05 | S/Sr |
| 5693 | 1993 EA | a005693.sp58[1] | 21-Jan-07 | Sq |
| 5786 | Talos | a005786.sp08[1] | 23-Aug-01 | Sq/Q |
| 6047 | 1991 TB1 | a006047.sp45 | 08-Oct-05 | S |
| 6239 | Minos | a006239.sp93 | 07-Sep-10 | Sq |
| 6455 | 1992 HE | a006455.sp11 | 15-Mar-02 | Srw |
| 6455 | 1992 HE | a006455.sp19 | 28-Oct-02 | Srw |
| 6455 | 1992 HE | a006455.sp107 | 21-Apr-12 | Sqw |
| 7350 | 1993 VA | a007350.sp30[1] | 15-Sep-04 | C/X INDET |
| 7350 | 1993 VA | a007350.sp39[1] | 17-Apr-05 | C/X INDET |
| 7358 | Oze | a007358.sp100 | 06-Jun-11 | Sq |
| 8567 | 1996 HW1 | a008567.sp44[1] | 04-Sep-05 | Q |
| 8567 | 1996 HW1 | a008567.sp48[1] | 22-Nov-05 | Q |
| 8567 | 1996 HW1 | a008567.sp73 | 02-Sep-08 | Q |
| 8567 | 1996 HW1 | a008567.sp74[1] | 02-Oct-08 | Sq |
| 8567 | 1996 HW1 | a008567.sp104[1] | 31-Oct-11 | Q |
| 10115 | 1992 SK | a010115.sp55[1] | 24-Oct-06 | Sq/S |
| 10145 | 1994 CK1 | au1994ck1.sp105[1] | 30-Dec-11 | S |
| 11066 | Sigurd | a011066.sp88 | 22-Feb-10 | S |
| 11398 | 1998 YP11 | a011398.sp68 | 10-Mar-08 | Sr |
| 11885 | 1990 SS | a011885.sp98[1] | 05-Apr-11 | CLASSIFIER INDET |
| 12923 | Zephyr | a012923.sp92[1] | 11-Jul-10 | S |
| 14402 | 1991 DB | a014402.sp78 | 02-Mar-09 | Ch |
| 15745 | 1991 PM5 | a015745.sp82 | 07-Aug-09 | S |
| 16834 | 1997 WU22 | a016834.sp77 | 27-Jan-09 | S |
| 16834 | 1997 WU22 | a016834.sp84 | 20-Sep-09 | S |
| 17274 | 2000 LC16 | a017274.sp02 | 10-Oct-00 | D |
| 17274 | 2000 LC16 | a017274.sp83 | 24-Aug-09 | CLASSIFIER INDET[2] |
| 19764 | 2000 NF5 | a019764.sp93[1] | 06-Sep-10 | Sq |
| 20086 | 1994 LW | a020086.sp100[1] | 07-Jun-11 | C/Cb |
| 20790 | 2000 SE45 | a020790.sp03 | 29-Jan-01 | S |
| 21088 | 1992 BL2 | au1992BL2.sp03[1] | 30-Jan-01 | Q |
| 22753 | 1998 WT | a022753.sp19[1] | 27-Oct-02 | S/Sq |
| 22753 | 1998 WT | a022753.sp36[1] | 03-Mar-05 | Sq |
| 23183 | 2000 OY21 | a023183.sp49[1] | 30-Jan-06 | Sq/Q |
| 23187 | 2000 PN9 | a023187.sp05[1] | 28-Mar-01 | S/Sq |
| 23187 | 2000 PN9 | a023187.sp98[1] | 05-Apr-11 | Sq/Q |
| 24761 | Ahau | a024761.sp87[1] | 17-Jan-10 | C/X INDET |

| Number | Name | File Name | UT Date | Taxonomic Type |
|---|---|---|---|---|
| 25143 | Itokawa | a025143.5 | 19-Feb-01 & 28-Mar-01 | Sqw |
| 25916 | 2001 CP44 | a025916.sp55[1] | 24-Oct-06 | Sq/Q |
| 25916 | 2001 CP44 | a025916.sp93[1] | 07-Sep-10 | Sq/Q |
| 31669 | 1999 JT6 | a031669.sp12[1] | 15-Apr-02 | Q |
| 35107 | 1991 VH | a035107.sp20 | 27-Dec-02 | Sq |
| 35107 | 1991 VH | a035107.sp71 | 11-Jun-08 | Sq |
| 36284 | 2000 DM8 | a036284.sp11 | 16-Mar-02 | Sq |
| 36284 | 2000 DM8 | a036284.sp97 | 07-Feb-11 | NO TAX MATCH |
| 39572 | 1993 DQ1 | a039572.sp73[1] | 02-Sep-08 | Q |
| 52387 | 1993 OM7 | au1993OM7.sp20[1] | 27-Dec-02 | Sr |
| 52750 | 1998 KK17 | a052750.sp30[1] | 15-Sep-04 | V |
| 52768 | 1998 OR2 | a052768.sp79[1] | 30-Mar-09 | L |
| 53430 | 1999 TY16 | a053430.sp76[1] | 03-Dec-08 | Sr |
| 53435 | 1999 VM40 | a053435.sp55 | 25-Oct-06 | Srw |
| 53789 | 2000 ED104 | au2000ED104.sp17[1] | 15-Sep-02 | S |
| 54690 | 2001 EB | au2001EB.sp05 | 28-Mar-01 | S |
| 54789 | 2001 MZ7 | a054789.sp21[1] | 16-Mar-03 | C/X INDET |
| 54789 | 2001 MZ7 | a054789.sp87[1] | 17-Jan-10 | C/X INDET |
| 54789 | 2001 MZ7 | a054789.sp89[1] | 16-Mar-10 | C/X INDET |
| 65679 | 1989 UQ | a065679.sp94 | 13-Oct-10 | C |
| 66251 | 1999 GJ2 | a066251.sp44[1] | 05-Sep-05 | Q |
| 66251 | 1999 GJ2 | a066251.sp45[1] | 08-Oct-05 | Q |
| 66251 | 1999 GJ2 | a066251.sp46[1] | 31-Oct-05 | Q |
| 68216 | 2001 CV26 | a068216.sp89[1] | 16-Mar-10 | Sq |
| 68359 | 2001 OZ13 | a068359.sp67[1] | 16-Jan-08 | Sq |
| 69230 | Hermes | a069230.9[1] | 16-Oct-03 | Sq |
| 85709 | 1998 SG36 | a085709.sp50[1] | 30-Apr-06 | S/Sr |
| 85818 | 1998 XM4 | a085818.sp91 | 10-May-10 | Srw |
| 85839 | 1998 YO4 | a085839.sp89[1] | 16-Mar-10 | Sq/Q |
| 85989 | 1999 JD6 | a085989.sp92 | 12-Jul-10 | K |
| 86324 | 1999 WA2 | a086324.sp67[1] | 16-Jan-08 | Sr/S |
| 86667 | 2000 FO10 | a086667.sp107[1] | 21-Apr-12 | Sq |
| 87684 | 2000 SY2 | au2000SY2.sp30[1] | 15-Sep-04 | Q |
| 88254 | 2001 FM129 | a088254.sp89[1] | 17-Mar-10 | Sq |
| 89355 | 2001 VS78 | au2001VS78.sp11[1] | 16-Mar-02 | Sr |
| 89355 | 2001 VS78 | au2001VS78.sp12 | 15-Apr-02 | Sr |
| 90367 | 2003 LC5 | au2003LC5.sp23[1] | 05-Jul-03 | C/X/T INDET |
| 96590 | 1998 XB | a096590.sp96[1] | 06-Jan-11 | Q/Sq |
| 99799 | 2002 LJ3 | a099799.sp105[1] | 30-Dec-11 | Sq |
| 99907 | 1989 VA | au1989VA.sp19 | 27-Oct-02 | Sr |
| 99907 | 1989 VA | a099907.sp65[1] | 11-Nov-07 | S/Sr |
| 99935 | 2002 AV4 | a099935.sp67[1] | 16-Jan-08 | B |
| 100926 | 1998 MQ | a100926.sp94 | 14-Oct-10 | Sqw |
| 108519 | 2001 LF | au2001LF.sp23[1] | 05-Jul-03 | C |
| 136564 | 1977 VA | au1977VA.sp44[1] | 04-Sep-05 | C/X INDET |
| 137032 | 1998 UO1 | au1998UO1.sp19[1] | 27-Oct-02 | S/Sr |
| 137032 | 1998 UO1 | a137032.sp74[1] | 02-Oct-08 | Sq/Q |
| 137062 | 1998 WM | au1998WM.sp19 | 27-Oct-02 | Sr |
| 137170 | 1999 HF1 | au1999HF1.sp13[1] | 07-May-02 | C/X INDET |

| Number | Name | File Name | UT Date | Taxonomic Type |
|---|---|---|---|---|
| 137170 | 1999 HF1 | au1999HF1.sp38[1] | 11-Apr-05 | C/X INDET |
| 137170 | 1999 HF1 | a137170.sp98[1] | 05-Apr-11 | C/X INDET |
| 137671 | 1999 XP35 | au1999xp35.sp105[1] | 30-Dec-11 | S/Sr |
| 137924 | 2000 BD19 | au2000BD19.sp03[1] | 29-Jan-01 | V |
| 138258 | 2000 GD2 | au2000GD2.sp11 | 15-Mar-02 | Sq |
| 138883 | 2000 YL29 | a138883.sp84[1] | 20-Sep-09 | Q |
| 141052 | 2001 XR1 | a141052.sp77 | 27-Jan-09 | Sq |
| 141432 | 2002 CQ11 | a141432.sp99[1] | 30-Apr-11 | S |
| 143651 | 2003 QO104 | a143651.sp79[1] | 30-Mar-09 | Sq/Q |
| 143651 | 2003 QO104 | a143651.sp80 | 27-Apr-09 | Q |
| 144900 | 2004 VG64 | au2004VG64.sp46[1] | 31-Oct-05 | Sq |
| 144922 | 2005 CK38 | au2005CK38.sp37[1] | 08-Mar-05 | Sq |
| 152558 | 1990 SA | a152558.sp93[1] | 07-Sep-10 | Sq |
| 152931 | 2000 EA107 | a152931.sp89 | 16-Mar-10 | Q |
| 153219 | 2000 YM29 | au2000YM29.sp58[1] | 21-Jan-07 | C/X INDET |
| 153814 | 2001 WN5 | a153814.sp94[1] | 12-Oct-10 | NO TAX MATCH[2] |
| 154029 | 2002 CY46 | a154029.sp93[1] | 06-Sep-10 | S |
| 154244 | 2002 KL6 | a154244.sp81[1] | 21-Jun-09 | Q |
| 154715 | 2004 LB6 | a154715.sp96[1] | 06-Jan-11 | Q |
| 159402 | 1999 AP10 | a159402.sp84[1] | 20-Sep-09 | Q |
| 161989 | Cacus | au1978CA.sp21[1] | 16-Mar-03 | Q |
| 161998 | 1988 PA | a161998.sp92[1] | 11-Jul-10 | Q |
| 162483 | 2000 PJ5 | a162483.sp82[1] | 08-Aug-09 | Q |
| 162900 | 2001 HG31 | a162900.sp76[1] | 03-Dec-08 | Q |
| 163001 | 2001 SE170 | au2001SE170.sp33[1] | 24-Oct-04 | Sr |
| 163243 | 2002 FB3 | a163243.dm07[1] | 20-Mar-12 | Sq/Q |
| 163697 | 2003 EF54 | a163697.sp83[1] | 24-Aug-09 | Q |
| 164202 | 2004 EW | au2004EW.sp36[1] | 03-Mar-05 | C/X INDET |
| 164400 | 2005 GN59 | a164400.sp73[1] | 02-Sep-08 | Q |
| 175706 | 1996 FG3 | a175706.sp79 | 30-Mar-09 | C/Ch |
| 175706 | 1996 FG3 | a175706.sp80 | 27-Apr-09 | B |
| 175706 | 1996 FG3 | a175706-obsA.sp105 | 30-Dec-11 | C/Ch[2] |
| 175706 | 1996 FG3 | a175706-obsB.sp105 | 30-Dec-11 | C/Ch[2] |
| 175706 | 1996 FG3 | a175706.dm04 | 1-Dec-11 | NO TAX MATCH[2] |
| 184266 | 2004 VW14 | au2004VW14.sp69[1] | 13-Apr-08 | Q |
| 200840 | 2001 XN254 | au2001XN254.sp12 | 14-Apr-02 | S |
| 207945 | 1991 JW | a207945.sp80 | 27-Apr-09 | Sq/Q |
| 217796 | 2000 TO64 | a217796.sp85[1] | 25-Oct-09 | Sr |
| 217807 | 2000 XK44 | a217807.sp85[1] | 25-Oct-09 | Q |
| 218863 | 2006 WO127 | a218863.sp89[1] | 17-Mar-10 | Sq |
| 219071 | 1997 US9 | a219071.sp104 | 31-Oct-11 | Q |
| 235756 | 2004 VC | au2004VC.sp79[1] | 30-Mar-09 | Sr/S |
| 237442 | 1999 TA10 | a237442.sp91[1] | 10-May-10 | S/Sr |
| 241662 | 2000 KO44 | a241662.sp92[1] | 12-Jul-10 | Sq/Q |
| 248818 | 2006 SZ217 | au2006SZ217.sp76[1] | 03-Dec-08 | Sr/S |
| 256412 | 2007 BT2 | au2007BT2.sp79[1] | 30-Mar-09 | C/Cb |
| 257744 | 2000 AD205 | au2000AD205.sp71[1] | 11-Jun-08 | K |
| 260141 | 2004 QT24 | au2004QT24.sp39[1] | 18-Apr-05 | Q |
| 260141 | 2004 QT24 | a260141.sp99[1] | 30-Apr-11 | Q |
| 267729 | 2003 FC5 | a267729.sp98[1] | 05-Apr-11 | S[3] |

| Number | Name | File Name | UT Date | Taxonomic Type |
|--------|------|-----------|---------|----------------|
| 274138 | 2008 FU6 | au2008FU6.sp98[1] | 05-Apr-11 | Sq |
| 274138 | 2008 FU6 | a274138.sp99[1] | 30-Apr-11 | Q |
| 297418 | 2000 SP43 | au2000SP43.sp02[1] | 09-Oct-00 | V |
| 297418 | 2000 SP43 | au2000sp43.sp102[1] | 25-Sep-11 | V |
| 297418 | 2000 SP43 | a297418.sp103[1] | 23-Oct-11 | V |
| 301844 | 1990 UA | au1990ua.sp103[1] | 24-Oct-11 | Sq |
| 302831 | 2003 FH | au2003fh.sp104[1] | 31-Oct-11 | Q |
| 310442 | 2000 CH59 | au2000CH59.sp67[1] | 16-Jan-08 | Sq |
|  | 2000 CO101 | au2000co101.sp84 | 20-Sep-09 | X |
|  | 2002 GO5 | au2002GO5.sp12[1] | 14-Apr-02 | Sq |
|  | 2002 LV | au2002lv.sp81[1] | 22-Jun-09 | S/Sr |
|  | 2002 LV | au2002lv.sp82[1] | 07-Aug-09 | Sr |
|  | 2005 JE46 | au2005JE46n1.sp47[1] | 12-Nov-05 | NO TAX MATCH[2] |
|  | 2005 JE46 | au2005JE46n2.sp47[1] | 13-Nov-05 | NO TAX MATCH[2] |
|  | 2005 JE46 | au2005JE46.sp48[1] | 22-Nov-05 | C/X/T INDET |
|  | 2005 UL5 | au2005UL5.sp47[1] | 13-Nov-05 | Sq |
|  | 2007 RF5 | au2007RF5.sp64[1] | 02-Oct-07 | Sq |

---

[1] File does not include data at or below $0.75\mu m$. Observations were likely taken with the dichroic in the light path.
[2] Objects can not be formally classified due to the presence of a thermal tail.
[3] Object spectrum is very low resolution, only the complex can be identified. Possible classifications include members of the S-complex that are not S-subtypes.

Table 3: Band Parameters

| Number | Name | $\alpha$ | r (AU) | T(K) | BID | BIC | $\sigma_{BIC}$ | $\delta_{BIC}$ | $\Delta$BIC | BIIC | $\sigma_{BIIC}$ | $\delta_{BIIC}$ | $\Delta$BIIC | BAR | $\sigma_{BAR}$ | $\delta_{BAR}$ | $\Delta$BAR | D | $\sigma_{D+}$ | $\sigma_{D-}$ | p$_V$ | $\sigma_{pV+}$ | $\sigma_{pV-}$ |
|---|---|---|---|---|---|---|---|---|---|---|---|---|---|---|---|---|---|---|---|---|---|---|---|
| 433 | Eros | 9.1 | 1.69 | 206 | 14.3 | 0.984 | 0.010 | 0.000 | 0.983 | 1.931 | 0.010 | 0.019 | 1.949 | 0.473 | 0.100 | -0.075 | 0.398 | 22.30 | | | 0.23 | | |
| 433 | Eros | 37.7 | 1.54 | 216 | 16.5 | 0.995 | 0.010 | 0.000 | 0.995 | 2.015 | 0.010 | 0.017 | 2.032 | 0.392 | 0.100 | -0.068 | 0.324 | 22.30 | | | 0.23 | | |
| 433 | Eros | 45.7 | 1.17 | 247 | 17.2 | 0.962 | 0.010 | 0.000 | 0.962 | 1.963 | 0.010 | 0.011 | 1.974 | 0.390 | 0.100 | -0.045 | 0.345 | 22.30 | | | 0.23 | | |
| 433 | Eros | 29.7 | 1.14 | 251 | 16.9 | 0.976 | 0.010 | 0.000 | 0.976 | 1.987 | 0.010 | 0.010 | 1.996 | 0.550 | 0.100 | -0.042 | 0.507 | 22.30 | | | 0.23 | | |
| 1036 | Ganymed | 6.3 | 3.76 | 137 | 16.2 | 0.921 | 0.010 | -0.001 | 0.919 | 1.853 | 0.010 | 0.033 | 1.885 | 1.114 | 0.100 | -0.125 | 0.989 | | | | | | |
| 1036 | Ganymed | 39.2 | 1.54 | 214 | 18.2 | 0.929 | 0.010 | 0.000 | 0.929 | 1.908 | 0.010 | 0.017 | 1.925 | 1.156 | 0.100 | -0.069 | 1.086 | | | | | | |
| 1036 | Ganymed | 44.7 | 1.40 | 224 | 18.1 | 0.923 | 0.010 | 0.000 | 0.923 | 1.933 | 0.010 | 0.015 | 1.949 | 1.288 | 0.100 | -0.062 | 1.226 | | | | | | |
| 1036 | Ganymed | 46.1 | 1.25 | 237 | 14.1 | 0.925 | 0.010 | 0.000 | 0.925 | 1.950 | 0.010 | 0.013 | 1.963 | 1.575 | 0.100 | -0.053 | 1.522 | | | | | | |
| 1036 | Ganymed | 39.9 | 1.27 | 235 | 17.6 | 0.920 | 0.010 | 0.000 | 0.920 | 1.982 | 0.010 | 0.013 | 1.995 | 1.771 | 0.100 | -0.054 | 1.717 | | | | | | |
| 1036 | Ganymed | 12.8 | 1.35 | 228 | 15.3 | 0.922 | 0.010 | 0.000 | 0.922 | 1.953 | 0.010 | 0.014 | 1.967 | 1.489 | 0.100 | -0.059 | 1.430 | | | | | | |
| 1627 | Ivar | 2.0 | 2.38 | 181 | 12.6 | 1.025 | 0.010 | -0.001 | 1.024 | 1.937 | 0.010 | 0.024 | 1.961 | 0.710 | 0.100 | -0.093 | 0.616 | 10.03 | 3.27 | 2.77 | 0.09 | 0.12 | 0.05 |
| 1864 | Daedalus | 33.4 | 1.61 | 207 | 18.9 | 1.049 | 0.010 | 0.000 | 1.049 | 1.885 | 0.070 | 0.019 | 1.903 | 0.531 | 0.100 | -0.075 | 0.457 | | | | | | |
| 1865 | Cerberus | 40.3 | 1.31 | 209 | 20.0 | 0.930 | 0.010 | 0.000 | 0.930 | 1.945 | 0.010 | 0.018 | 1.963 | 0.605 | 0.100 | -0.073 | 0.533 | 0.82 | 0.15 | 0.13 | 0.50 | 0.29 | 0.21 |
| 1866 | Sisyphus | 13.1 | 2.09 | 194 | 13.3 | 0.929 | 0.010 | 0.000 | 0.929 | 2.011 | 0.015 | 0.021 | 2.032 | 0.633 | 0.100 | -0.084 | 0.549 | 12.94 | 4.57 | 3.66 | 0.07 | 0.09 | 0.04 |
| 1866 | Sisyphus A | 22.2 | 1.83 | 207 | 11.7 | 0.930 | 0.010 | 0.000 | 0.930 | 1.880 | 0.008 | 0.019 | 1.899 | 1.418 | 0.100 | -0.074 | 1.344 | 12.94 | 4.57 | 3.66 | 0.07 | 0.09 | 0.04 |
| 1866 | Sisyphus B | 22.2 | 1.83 | 207 | 11.4 | 0.930 | 0.010 | 0.000 | 0.930 | 1.957 | 0.010 | 0.019 | 1.976 | 2.258 | 0.100 | -0.074 | 2.184 | 12.94 | 4.57 | 3.66 | 0.07 | 0.09 | 0.04 |
| 1866 | Sisyphus C | 22.2 | 1.83 | 207 | 12.8 | 0.920 | 0.010 | 0.000 | 0.919 | 1.905 | 0.011 | 0.019 | 1.923 | 1.262 | 0.100 | -0.074 | 1.188 | 12.94 | 4.57 | 3.66 | 0.07 | 0.09 | 0.04 |
| 1866 | Sisyphus D | 22.2 | 1.83 | 207 | 11.9 | 0.919 | 0.010 | 0.000 | 0.918 | 1.883 | 0.018 | 0.019 | 1.901 | 1.205 | 0.100 | -0.074 | 1.131 | 12.94 | 4.57 | 3.66 | 0.07 | 0.09 | 0.04 |
| 1866 | Sisyphus E | 22.2 | 1.83 | 207 | 11.7 | 0.915 | 0.010 | 0.000 | 0.915 | 1.991 | 0.057 | 0.019 | 2.009 | 1.677 | 0.100 | -0.074 | 1.603 | 12.94 | 4.57 | 3.66 | 0.07 | 0.09 | 0.04 |
| 1866 | Sisyphus F | 22.2 | 1.83 | 207 | 11.8 | 0.919 | 0.010 | 0.000 | 0.918 | 1.858 | 0.013 | 0.019 | 1.876 | 0.844 | 0.100 | -0.074 | 0.770 | 12.94 | 4.57 | 3.66 | 0.07 | 0.09 | 0.04 |
| 1866 | Sisyphus A | 30.7 | 1.73 | 213 | 12.9 | 0.926 | 0.010 | 0.000 | 0.925 | 1.883 | 0.010 | 0.017 | 1.900 | 1.360 | 0.100 | -0.070 | 1.290 | 12.94 | 4.57 | 3.66 | 0.07 | 0.09 | 0.04 |
| 1866 | Sisyphus B | 30.7 | 1.73 | 213 | 18.2 | 0.933 | 0.010 | 0.000 | 0.932 | 1.967 | 0.010 | 0.017 | 1.984 | 1.061 | 0.100 | -0.070 | 0.991 | 12.94 | 4.57 | 3.66 | 0.07 | 0.09 | 0.04 |
| 1866 | Sisyphus C | 30.7 | 1.73 | 213 | 13.1 | 0.926 | 0.010 | 0.000 | 0.925 | 1.946 | 0.010 | 0.017 | 1.963 | 1.736 | 0.100 | -0.070 | 1.667 | 12.94 | 4.57 | 3.66 | 0.07 | 0.09 | 0.04 |
| 1917 | Cuyo A | 24.8 | 1.01 | 263 | 28.3 | 0.922 | 0.010 | 0.001 | 0.922 | 1.889 | 0.010 | 0.007 | 1.897 | 2.458 | 0.100 | -0.033 | 2.425 | 4.36 | 1.16 | 0.93 | 0.27 | 0.21 | 0.14 |
| 1917 | Cuyo B | 24.8 | 1.01 | 263 | 20.1 | 0.922 | 0.010 | 0.001 | 0.922 | 1.932 | 0.010 | 0.007 | 1.939 | 2.003 | 0.100 | -0.033 | 1.969 | 4.36 | 1.16 | 0.93 | 0.27 | 0.21 | 0.14 |
| 1917 | Cuyo C | 24.8 | 1.01 | 263 | 21.7 | 0.922 | 0.010 | 0.001 | 0.922 | 1.925 | 0.010 | 0.007 | 1.932 | 1.355 | 0.100 | -0.033 | 1.322 | 4.36 | 1.16 | 0.93 | 0.27 | 0.21 | 0.14 |
| 1917 | Cuyo D | 24.8 | 1.01 | 263 | 18.1 | 0.925 | 0.010 | 0.001 | 0.926 | 1.914 | 0.010 | 0.007 | 1.921 | 1.618 | 0.100 | -0.033 | 1.585 | 4.36 | 1.16 | 0.93 | 0.27 | 0.21 | 0.14 |
| 1917 | Cuyo E | 24.8 | 1.01 | 263 | 23.6 | 0.922 | 0.010 | 0.001 | 0.922 | 1.907 | 0.033 | 0.007 | 1.914 | 1.429 | 0.100 | -0.033 | 1.396 | 4.36 | 1.16 | 0.93 | 0.27 | 0.21 | 0.14 |
| 1917 | Cuyo F | 24.8 | 1.01 | 263 | 23.9 | 0.922 | 0.010 | 0.001 | 0.922 | 1.928 | 0.010 | 0.007 | 1.935 | 1.746 | 0.100 | -0.033 | 1.713 | 4.36 | 1.16 | 0.93 | 0.27 | 0.21 | 0.14 |
| 1917 | Cuyo | 38.5 | 1.49 | 216 | 19.7 | 0.925 | 0.010 | 0.000 | 0.925 | 1.890 | 0.010 | 0.017 | 1.907 | 0.800 | 0.100 | -0.068 | 0.733 | 4.36 | 1.16 | 0.93 | 0.27 | 0.21 | 0.14 |
| 1943 | Anteros | 16.7 | 1.51 | 223 | 9.8 | 0.987 | 0.010 | 0.000 | 0.987 | 1.993 | 0.010 | 0.015 | 2.008 | 0.694 | 0.100 | -0.063 | 0.631 | 2.40 | 0.67 | 0.56 | 0.17 | 0.15 | 0.08 |
| 2102 | Tantalus | 61.0 | 1.15 | 247 | 23.1 | 0.916 | 0.010 | 0.000 | 0.917 | 1.891 | 0.014 | 0.011 | 1.902 | 1.383 | 0.100 | -0.045 | 1.338 | | | | | | |
| 3122 | Florence | 71.4 | 1.07 | 261 | 35.3 | 1.030 | 0.010 | 0.000 | 1.030 | 2.055 | 0.022 | 0.008 | 2.063 | 0.535 | 0.100 | -0.035 | 0.499 | 4.25 | 1.22 | 0.90 | 0.21 | 0.20 | 0.10 |
| 4055 | Magellan | 54.2 | 1.23 | 230 | 53.2 | 0.935 | 0.010 | 0.003 | 0.938 | 1.951 | 0.010 | 0.010 | 1.962 | 2.289 | 0.100 | 0.000 | 2.289 | 2.46 | 0.53 | 0.41 | 0.36 | 0.27 | 0.17 |
| 4283 | Cuno | 11.7 | 2.11 | 191 | 18.7 | 0.980 | 0.010 | -0.001 | 0.979 | 1.966 | 0.190 | 0.022 | 1.988 | 0.639 | 0.100 | -0.086 | 0.554 | 5.44 | 1.20 | 1.11 | 0.10 | 0.10 | 0.05 |
| 5143 | Heracles | 28.1 | 1.87 | 184 | 23.5 | 1.017 | 0.010 | -0.001 | 1.016 | 1.937 | 0.013 | 0.023 | 1.960 | 0.388 | 0.100 | -0.091 | 0.297 | 3.39 | 0.60 | 0.55 | 0.40 | 0.22 | 0.16 |
| 5587 | 1990 SB | 22.8 | 2.18 | 178 | 16.8 | 0.925 | 0.010 | -0.001 | 0.925 | 1.904 | 0.025 | 0.024 | 1.928 | 1.426 | 0.100 | -0.095 | 1.330 | 3.73 | 2.26 | 1.22 | 0.29 | 0.43 | 0.19 |
| 5626 | 1991 FE | 22.6 | 1.62 | 216 | 12.7 | 0.943 | 0.010 | 0.000 | 0.943 | 1.924 | 0.031 | 0.017 | 1.941 | 0.656 | 0.100 | -0.068 | 0.588 | 3.90 | 1.14 | 0.93 | 0.15 | 0.18 | 0.08 |
| 5646 | 1990 TR | 33.8 | 1.69 | 169 | 28.4 | 1.025 | 0.010 | -0.001 | 1.025 | 2.033 | 0.042 | 0.026 | 2.060 | 0.349 | 0.100 | -0.102 | 0.247 | 2.09 | 0.45 | 0.36 | 0.66 | 0.42 | 0.29 |
| 5786 | Talos | 42.6 | 1.22 | 247 | 30.4 | 0.935 | 0.010 | 0.000 | 0.935 | 1.970 | 0.032 | 0.011 | 1.981 | 0.441 | 0.100 | -0.046 | 0.395 | 1.29 | 0.32 | 0.27 | 0.17 | 0.15 | 0.08 |
| 6239 | Minos | 11.7 | 1.17 | 213 | 16.9 | 0.982 | 0.010 | 0.000 | 0.981 | 1.898 | 0.010 | 0.017 | 1.915 | 0.337 | 0.100 | -0.070 | 0.267 | 0.47 | 0.13 | 0.08 | 0.57 | 0.37 | 0.29 |
| 7358 | Oze | 45.2 | 1.39 | 237 | 12.3 | 0.924 | 0.010 | 0.000 | 0.924 | 1.941 | 0.010 | 0.013 | 1.953 | 1.166 | 0.100 | -0.052 | 1.114 | 6.17 | 2.03 | 1.62 | 0.08 | 0.10 | 0.04 |
| 7358 | Oze | 45.7 | 1.38 | 238 | 11.8 | 0.930 | 0.010 | 0.000 | 0.930 | 1.970 | 0.019 | 0.012 | 1.982 | 2.324 | 0.100 | -0.052 | 2.272 | 6.17 | 2.03 | 1.62 | 0.08 | 0.10 | 0.04 |
| 7358 | Oze | 57.0 | 1.19 | 256 | 8.7 | 0.920 | 0.010 | 0.000 | 0.920 | 2.020 | 0.015 | 0.009 | 2.029 | | | | | 6.17 | 2.03 | 1.62 | 0.08 | 0.10 | 0.04 |
| 7822 | 1991 CS | 60.9 | 1.12 | 249 | 19.1 | 0.908 | 0.010 | 0.000 | 0.909 | 1.977 | 0.029 | 0.010 | 1.987 | 0.982 | 0.100 | -0.044 | 0.938 | 0.84 | 0.16 | 0.13 | 0.28 | 0.18 | 0.12 |
| 8566 | 1996 EN | 79.5 | 1.03 | 258 | 51.0 | 0.950 | 0.010 | 0.002 | 0.952 | 1.993 | 0.010 | 0.006 | 1.999 | 1.754 | 0.100 | 0.000 | 1.754 | 1.25 | 0.26 | 0.22 | 0.30 | 0.22 | 0.14 |
| 8567 | 1996 HW1 | 56.2 | 1.20 | 239 | 17.7 | 1.002 | 0.010 | 0.000 | 1.002 | 2.020 | 0.036 | 0.012 | 2.032 | 0.587 | 0.100 | -0.051 | 0.536 | | | | | | |
| 11066 | Sigurd | 17.4 | 1.65 | 197 | 17.3 | 0.924 | 0.010 | 0.000 | 0.923 | 1.954 | 0.010 | 0.021 | 1.975 | 1.187 | 0.100 | -0.081 | 1.105 | 1.98 | 0.45 | 0.38 | 0.38 | 0.31 | 0.18 |
| 12711 | Tukmit | 80.1 | 0.97 | 275 | 22.1 | 0.978 | 0.010 | 0.001 | 0.979 | 1.965 | 0.224 | 0.005 | 1.970 | 0.332 | 0.100 | -0.025 | 0.307 | 1.95 | 0.40 | 0.34 | 0.19 | 0.14 | 0.09 |
| 12923 | Zephyr | 12.9 | 1.79 | 202 | 12.2 | 0.942 | 0.010 | 0.000 | 0.942 | 1.969 | 0.091 | 0.020 | 1.989 | 1.162 | 0.100 | -0.078 | 1.084 | 1.82 | 0.42 | 0.34 | 0.20 | 0.16 | 0.09 |
| 16834 | 1997 WU22 | 33.6 | 1.48 | 204 | 14.8 | 0.961 | 0.010 | 0.000 | 0.961 | 1.969 | 0.010 | 0.019 | 1.988 | 0.552 | 0.100 | -0.077 | 0.476 | 1.50 | 0.32 | 0.26 | 0.43 | 0.28 | 0.20 |
| 18882 | 1999 YN4 | 22.7 | 1.58 | 216 | 17.9 | 0.911 | 0.010 | 0.000 | 0.911 | 1.987 | 0.224 | 0.017 | 2.004 | 1.813 | 0.100 | -0.068 | 1.746 | 1.74 | 0.47 | 0.36 | 0.18 | 0.15 | 0.09 |
| 19764 | 2000 NF5 | 2.4 | 1.59 | 210 | 15.2 | 0.971 | 0.010 | 0.000 | 0.971 | 1.996 | 0.010 | 0.018 | 2.014 | 0.469 | 0.100 | -0.072 | 0.397 | 1.67 | 0.43 | 0.37 | 0.26 | 0.23 | 0.14 |
| 21088 | 1992 BL2 | 15.7 | 2.11 | 182 | 18.6 | 1.048 | 0.010 | -0.001 | 1.048 | 2.072 | 0.135 | 0.024 | 2.096 | 0.330 | 0.100 | -0.092 | 0.238 | 3.44 | 1.10 | 0.86 | 0.26 | 0.32 | 0.14 |
| 23187 | 2000 PN9 | 12.9 | 1.90 | 196 | 17.2 | 1.032 | 0.010 | 0.000 | 1.031 | 1.933 | 0.228 | 0.021 | 1.954 | 0.728 | 0.100 | -0.082 | 0.646 | 1.84 | 0.37 | 0.35 | 0.20 | 0.15 | 0.10 |
| 27346 | 2000 DN8 | 31.2 | 1.40 | 229 | 13.0 | 1.028 | 0.010 | 0.000 | 1.028 | 1.988 | 0.022 | 0.014 | 2.002 | 0.415 | 0.100 | -0.059 | 0.356 | 1.93 | 0.43 | 0.39 | 0.19 | 0.17 | 0.09 |
| 36284 | 2000 DM8 | 11.0 | 1.78 | 204 | 20.9 | 1.018 | 0.010 | 0.000 | 1.018 | 1.942 | 0.017 | 0.019 | 1.961 | 0.367 | 0.100 | -0.077 | 0.291 | 3.08 | 0.72 | 0.55 | 0.19 | 0.16 | 0.09 |
| 39572 | 1993 DQ1 | 38.7 | 1.34 | 227 | 16.4 | 0.985 | 0.010 | 0.000 | 0.985 | 1.990 | 0.011 | 0.015 | 2.005 | 0.669 | 0.100 | -0.060 | 0.609 | | | | | | |
| 52750 | 1998 KK17 | 85.1 | 0.96 | 250 | 47.1 | 0.939 | 0.010 | 0.002 | 0.941 | 1.928 | 0.050 | 0.007 | 1.935 | 1.339 | 0.100 | 0.000 | 1.339 | 1.03 | 0.17 | 0.16 | 0.46 | 0.25 | 0.20 |
| 53789 | 2000 ED104 | 61.3 | 1.12 | 244 | 12.7 | 0.949 | 0.010 | 0.000 | 0.949 | 2.033 | 0.019 | 0.011 | 2.044 | 0.813 | 0.100 | -0.047 | 0.766 | 0.84 | 0.19 | 0.17 | 0.32 | 0.24 | 0.15 |
| 68216 | 2001 CV26 | 46.3 | 1.24 | 238 | 15.8 | 0.934 | 0.010 | 0.000 | 0.934 | 2.082 | 0.039 | 0.012 | 2.095 | 0.588 | 0.100 | -0.052 | 0.536 | 1.39 | 0.37 | 0.29 | 0.26 | 0.21 | 0.13 |
| 68216 | 2001 CV26 | 30.9 | 1.32 | 231 | 15.6 | 0.960 | 0.010 | 0.000 | 0.960 | 1.885 | 0.010 | 0.014 | 1.899 | 0.484 | 0.100 | -0.057 | 0.427 | 1.39 | 0.37 | 0.29 | 0.26 | 0.21 | 0.13 |
| 68216 | 2001 CV26 | 17.0 | 1.38 | 226 | 14.4 | 0.949 | 0.010 | 0.000 | 0.949 | 1.938 | 0.010 | 0.015 | 1.952 | 0.808 | 0.100 | -0.060 | 0.747 | 1.39 | 0.37 | 0.29 | 0.26 | 0.21 | 0.13 |
| 68350 | 2001 MK3 | 51.1 | 1.26 | 237 | 14.0 | 0.918 | 0.010 | 0.000 | 0.918 | 2.027 | 0.058 | 0.013 | 2.040 | 2.094 | 0.100 | -0.052 | 2.042 | 1.71 | 0.37 | 0.33 | 0.25 | 0.17 | 0.10 |
| 85839 | 1998 YO4 | 16.1 | 1.46 | 226 | 21.4 | 1.001 | 0.010 | 0.000 | 1.001 | 2.049 | 0.012 | 0.015 | 2.064 | 0.332 | 0.100 | -0.061 | 0.271 | 1.81 | 0.41 | 0.40 | 0.17 | 0.16 | 0.08 |
| 86067 | 1999 RM28 | 11.8 | 1.40 | 231 | 18.9 | 0.966 | 0.010 | 0.000 | 0.966 | 1.957 | 0.011 | 0.014 | 1.971 | 0.495 | 0.100 | -0.057 | 0.438 | 1.58 | 0.44 | 0.36 | 0.17 | 0.16 | 0.09 |
| 88254 | 2001 FM129 | 43.0 | 1.20 | 236 | 22.3 | 0.976 | 0.010 | 0.000 | 0.976 | 2.034 | 0.170 | 0.013 | 2.046 | 0.657 | 0.100 | -0.053 | 0.604 | 0.81 | 0.17 | 0.14 | 0.33 | 0.23 | 0.14 |

| Number | Name | $\alpha$ | r (AU) | T(K) | BID | BIC | $\sigma_{BIC}$ | $\delta_{BIC}$ | $\Delta$BIC | BIIC | $\sigma_{BIIC}$ | $\delta_{BIIC}$ | $\Delta$BIIC | BAR | $\sigma_{BAR}$ | $\delta_{BAR}$ | $\Delta$BAR | D | $\sigma_{D+}$ | $\sigma_{D-}$ | $P_V$ | $\sigma_{pV+}$ | $\sigma_{pV-}$ |
|---|---|---|---|---|---|---|---|---|---|---|---|---|---|---|---|---|---|---|---|---|---|---|---|
| 96590 | 1998 XB | 19.2 | 1.21 | 252 | 19.0 | 1.000 | 0.010 | 0.000 | 1.000 | 2.025 | 0.010 | 0.010 | 2.035 | 0.606 | 0.100 | -0.042 | 0.564 | 2.29 | 0.47 | 0.42 | 0.11 | 0.09 | 0.05 |
| 100926 | 1998 MQ | 56.2 | 1.10 | 243 | 21.6 | 0.945 | 0.010 | 0.000 | 0.945 | 2.019 | 0.023 | 0.011 | 2.031 | 0.612 | 0.100 | -0.048 | 0.564 | 1.08 | 0.23 | 0.19 | 0.37 | 0.22 | 0.16 |
| 136617 | 1994 CC | 13.2 | 1.41 | 219 | 29.9 | 1.065 | 0.010 | 0.000 | 1.065 | 2.190 | 0.154 | 0.016 | 2.206 | 0.407 | 0.100 | -0.066 | 0.341 | 0.67 | 0.11 | 0.12 | 0.32 | 0.22 | 0.14 |
| 137032 | 1998 UO1 | 45.3 | 1.16 | 256 | 22.8 | 0.967 | 0.010 | 0.000 | 0.968 | 1.992 | 0.010 | 0.009 | 2.001 | 0.524 | 0.100 | -0.039 | 0.485 | 1.66 | 0.37 | 0.35 | 0.13 | 0.11 | 0.06 |
| 137062 | 1998 WM | 14.8 | 1.58 | 207 | 24.3 | 0.931 | 0.010 | 0.000 | 0.930 | 1.941 | 0.221 | 0.019 | 1.959 | 0.781 | 0.100 | -0.075 | 0.706 | 1.13 | 0.28 | 0.23 | 0.32 | 0.23 | 0.14 |
| 137084 | 1998 XS16 | 5.0 | 1.80 | 195 | 17.3 | 0.909 | 0.010 | 0.000 | 0.908 | 1.872 | 0.186 | 0.021 | 1.893 | 1.096 | 0.125 | -0.083 | 1.013 | 1.36 | 0.28 | 0.29 | 0.30 | 0.23 | 0.14 |
| 137084 | 1998 XS16 | 5.8 | 1.80 | 195 | 19.3 | 0.909 | 0.010 | 0.000 | 0.908 | 1.844 | 0.217 | 0.021 | 1.865 | 1.101 | 0.109 | -0.083 | 1.018 | 1.36 | 0.28 | 0.29 | 0.30 | 0.23 | 0.14 |
| 137125 | 1999 CT3 | 18.2 | 1.40 | 223 | 21.5 | 1.038 | 0.010 | 0.000 | 1.038 | 1.994 | 0.179 | 0.015 | 2.009 | 0.743 | 0.100 | -0.063 | 0.680 | 0.54 | 0.12 | 0.10 | 0.28 | 0.22 | 0.13 |
| 137671 | 1999 XP35 | 23.6 | 1.17 | 252 | 15.9 | 0.936 | 0.010 | 0.000 | 0.937 | 1.977 | 0.191 | 0.010 | 1.987 | 1.826 | 0.100 | -0.041 | 1.785 | 0.60 | 0.17 | 0.13 | 0.17 | 0.16 | 0.08 |
| 138883 | 2000 YL29 | 58.4 | 1.11 | 256 | 23.5 | 0.957 | 0.010 | 0.000 | 0.958 | 1.907 | 0.010 | 0.009 | 1.915 | 0.424 | 0.100 | -0.038 | 0.385 | 1.40 | 0.30 | 0.30 | 0.19 | 0.17 | 0.09 |
| 138911 | 2001 AE2 | 9.3 | 1.34 | 222 | 9.1 | 0.911 | 0.010 | 0.000 | 0.911 | 2.061 | 0.122 | 0.016 | 2.076 | 2.824 | 0.183 | -0.064 | 2.760 | 0.35 | 0.08 | 0.07 | 0.35 | 0.24 | 0.15 |
| 141498 | 2002 EZ16 | 28.1 | 1.15 | 247 | 21.7 | 0.966 | 0.010 | 0.000 | 0.967 | 2.039 | 0.019 | 0.011 | 2.049 | 0.557 | 0.100 | -0.045 | 0.512 | 0.59 | 0.11 | 0.09 | 0.25 | 0.18 | 0.11 |
| 141498 | 2002 EZ16 | 19.5 | 1.20 | 242 | 20.8 | 0.990 | 0.010 | 0.000 | 0.990 | 1.950 | 0.010 | 0.012 | 1.962 | 0.207 | 0.100 | -0.049 | 0.158 | 0.59 | 0.11 | 0.09 | 0.25 | 0.18 | 0.11 |
| 143381 | 2003 BC21 | 20.2 | 1.46 | 212 | 16.9 | 0.924 | 0.010 | 0.000 | 0.924 | 1.924 | 0.010 | 0.018 | 1.941 | 1.008 | 0.100 | -0.070 | 0.937 | 1.50 | 0.33 | 0.25 | 0.35 | 0.23 | 0.15 |
| 152558 | 1990 SA | 27.4 | 1.28 | 232 | 7.9 | 0.970 | 0.010 | 0.000 | 0.970 | 2.005 | 0.011 | 0.014 | 2.019 | 1.169 | 0.100 | -0.056 | 1.113 | | | | | | |
| 152931 | 2000 EA107 | 48.4 | 1.28 | 232 | 22.8 | 1.010 | 0.010 | 0.000 | 1.010 | 1.910 | 0.280 | 0.014 | 1.923 | 0.339 | 0.100 | -0.056 | 0.283 | | | | | | |
| 154029 | 2002 CY46 | 52.9 | 1.10 | 261 | 12.1 | 0.934 | 0.010 | 0.001 | 0.934 | 1.944 | 0.010 | 0.008 | 1.951 | 1.511 | 0.100 | -0.035 | 1.476 | 1.76 | 0.40 | 0.38 | 0.16 | 0.16 | 0.08 |
| 159402 | 1999 AP10 | 9.0 | 1.29 | 226 | 9.7 | 0.950 | 0.010 | 0.000 | 0.950 | 1.907 | 0.010 | 0.015 | 1.921 | 0.723 | 0.100 | -0.060 | 0.663 | 1.22 | 0.25 | 0.22 | 0.35 | 0.24 | 0.16 |
| 163243 | 2002 FB3 | 56.2 | 1.20 | 238 | 26.3 | 0.939 | 0.010 | 0.000 | 0.939 | 1.913 | 0.218 | 0.012 | 1.926 | 0.354 | 0.100 | -0.052 | 0.303 | 1.36 | 0.27 | 0.24 | 0.30 | 0.20 | 0.14 |
| 163697 | 2003 EF54 | 26.7 | 1.12 | 248 | 25.1 | 0.965 | 0.010 | 0.000 | 0.965 | 1.917 | 0.014 | 0.010 | 1.927 | 0.520 | 0.100 | -0.044 | 0.476 | 0.25 | 0.04 | 0.04 | 0.29 | 0.20 | 0.12 |
| 164121 | 2003 YT1 | 68.7 | 1.05 | 252 | 57.1 | 0.935 | 0.010 | 0.002 | 0.937 | 1.950 | 0.010 | 0.007 | 1.957 | 1.594 | 0.100 | 0.000 | 1.594 | 1.37 | 0.28 | 0.24 | 0.34 | 0.23 | 0.15 |
| 198856 | 2005 LR3 | 52.7 | 1.16 | 253 | 21.2 | 0.995 | 0.010 | 0.000 | 0.995 | 1.940 | 0.236 | 0.009 | 1.949 | 0.258 | 0.100 | -0.041 | 0.217 | 1.19 | 0.27 | 0.22 | 0.17 | 0.15 | 0.08 |
| 198856 | 2005 LR3 | 54.4 | 1.15 | 255 | 23.4 | 1.005 | 0.010 | 0.000 | 1.005 | 1.970 | 0.071 | 0.009 | 1.979 | 0.227 | 0.100 | -0.040 | 0.188 | 1.19 | 0.27 | 0.22 | 0.17 | 0.15 | 0.08 |
| 207945 | 1991 JW | 75.3 | 1.01 | 261 | 14.9 | 1.000 | 0.010 | 0.001 | 1.001 | 2.010 | 0.232 | 0.008 | 2.018 | 0.178 | 0.100 | -0.035 | 0.143 | | | | | | |
| 207945 | 1991 JW | 74.5 | 1.01 | 261 | 18.5 | 0.990 | 0.010 | 0.001 | 0.991 | 1.945 | 0.024 | 0.008 | 1.953 | 0.358 | 0.100 | -0.035 | 0.323 | | | | | | |
| 214088 | 2004 JN13 | 62.6 | 1.03 | 261 | 18.8 | 0.980 | 0.010 | 0.001 | 0.981 | 1.955 | 0.161 | 0.008 | 1.963 | 0.601 | 0.100 | -0.035 | 0.566 | 3.01 | 0.76 | 0.65 | 0.26 | 0.20 | 0.13 |
| 218863 | 2006 WO127 | 9.8 | 1.65 | 209 | 18.5 | 0.990 | 0.010 | 0.000 | 0.990 | 1.985 | 0.024 | 0.018 | 2.003 | 0.351 | 0.100 | -0.073 | 0.278 | 1.69 | 0.43 | 0.38 | 0.22 | 0.18 | 0.10 |
| 219071 | 1997 US9 | 39.1 | 1.26 | 228 | 23.7 | 1.024 | 0.010 | 0.000 | 1.024 | 1.985 | 0.159 | 0.014 | 1.999 | 0.306 | 0.100 | -0.059 | 0.248 | 0.79 | 0.15 | 0.13 | 0.35 | 0.22 | 0.15 |
| 220124 | 2002 TE66 A | 30.5 | 1.14 | 235 | 20.1 | 0.976 | 0.010 | 0.000 | 0.976 | 2.020 | 0.176 | 0.013 | 2.033 | 0.339 | 0.100 | -0.054 | 0.286 | 0.48 | 0.09 | 0.08 | 0.40 | 0.24 | 0.17 |
| 220124 | 2002 TE66 B | 30.5 | 1.14 | 235 | 20.0 | 0.957 | 0.010 | 0.000 | 0.957 | 1.993 | 0.093 | 0.013 | 2.006 | 0.519 | 0.100 | -0.054 | 0.465 | 0.48 | 0.09 | 0.08 | 0.40 | 0.24 | 0.17 |
| 247517 | 2002 QY6 | 51.7 | 1.10 | 253 | 24.0 | 0.925 | 0.010 | 0.000 | 0.925 | 1.890 | 0.010 | 0.009 | 1.899 | 1.453 | 0.100 | -0.041 | 1.411 | | | | | | |
| 265187 | 2003 YS117 | 13.8 | 1.41 | 212 | 24.6 | 0.972 | 0.010 | 0.000 | 0.972 | 1.956 | 0.072 | 0.018 | 1.973 | 0.531 | 0.100 | -0.071 | 0.460 | 0.53 | 0.09 | 0.09 | 0.40 | 0.24 | 0.17 |
| 297418 | 2000 SP43 | 70.9 | 1.04 | 244 | 29.0 | 0.935 | 0.010 | 0.003 | 0.938 | 1.835 | 0.013 | 0.008 | 1.843 | 1.126 | 0.100 | 0.000 | 1.126 | | | | | | |
| 347813 | 2002 NP1 | 24.1 | 1.34 | 228 | 17.0 | 0.961 | 0.010 | 0.000 | 0.961 | 1.969 | 0.015 | 0.014 | 1.984 | 0.293 | 0.100 | -0.059 | 0.235 | 0.81 | 0.17 | 0.13 | 0.26 | 0.19 | 0.12 |
| | 2000 TJ1 | 19.5 | 1.21 | 236 | 22.4 | 0.946 | 0.010 | 0.000 | 0.946 | 1.927 | 0.010 | 0.013 | 1.940 | 0.619 | 0.100 | -0.054 | 0.566 | 0.36 | 0.06 | 0.06 | 0.33 | 0.20 | 0.13 |
| | 2001 HC | 60.6 | 1.08 | 252 | 22.1 | 0.968 | 0.010 | 0.000 | 0.968 | 1.986 | 0.166 | 0.010 | 1.996 | 0.287 | 0.100 | -0.041 | 0.246 | | | | | | |
| | 2002 OS4 | 28.9 | 1.26 | 232 | 19.8 | 0.912 | 0.010 | 0.000 | 0.912 | 2.001 | 0.092 | 0.014 | 2.015 | 1.903 | 0.100 | -0.056 | 1.846 | 0.59 | 0.12 | 0.10 | 0.31 | 0.22 | 0.12 |
| | 2005 MC | 20.9 | 1.32 | 236 | 18.4 | 0.915 | 0.010 | 0.000 | 0.916 | 2.005 | 0.036 | 0.013 | 2.017 | 1.876 | 0.100 | -0.054 | 1.822 | 1.63 | 0.50 | 0.41 | 0.19 | 0.21 | 0.09 |

Table 4: Band Parameters MIT

| Number | Name | α | r (AU) | T(K) | BID | BIC | $\sigma_{BIC}$ | $\delta_{BIC}$ | ΔBIC | BIIC | $\sigma_{BIIC}$ | $\delta_{BIIC}$ | ΔBIIC | BAR | $\sigma_{BAR}$ | $\delta_{BAR}$ | ΔBAR | D | $\sigma_{D+}$ | $\sigma_{D-}$ | pV | $\sigma_{pV+}$ | $\sigma_{pV-}$ |
|---|---|---|---|---|---|---|---|---|---|---|---|---|---|---|---|---|---|---|---|---|---|---|---|
| 433 | Eros | 34.9 | 1.77 | 201 | 17.2 | 0.995 | 0.010 | 0.000 | 0.995 | 1.975 | 0.010 | 0.020 | 1.995 | 0.417 | 0.100 | -0.078 | 0.339 | 22.30 | | | 0.23 | | |
| 433 | Eros | 21.6 | 1.69 | 206 | 18.0 | 0.955 | 0.010 | 0.000 | 0.955 | 1.955 | 0.010 | 0.019 | 1.974 | 0.453 | 0.100 | -0.075 | 0.378 | 22.30 | | | 0.23 | | |
| 433 | Eros | 12.8 | 1.62 | 210 | 17.2 | 0.950 | 0.010 | 0.000 | 0.950 | 1.970 | 0.010 | 0.018 | 1.988 | 0.434 | 0.100 | -0.072 | 0.362 | 22.30 | | | 0.23 | | |
| 433 | Eros | 42.8 | 1.49 | 219 | 18.9 | 0.990 | 0.010 | 0.000 | 0.990 | 1.990 | 0.009 | 0.016 | 2.006 | 0.244 | 0.100 | -0.065 | 0.179 | 22.30 | | | 0.23 | | |
| 433 | Eros | 45.5 | 1.38 | 228 | 19.0 | 0.960 | 0.010 | 0.000 | 0.960 | 1.990 | 0.008 | 0.014 | 2.004 | 0.500 | 0.100 | -0.059 | 0.440 | 22.30 | | | 0.23 | | |
| 433 | Eros | 47.2 | 1.29 | 236 | 19.5 | 0.980 | 0.010 | 0.000 | 0.980 | 2.005 | 0.012 | 0.013 | 2.018 | 0.329 | 0.100 | -0.053 | 0.276 | 22.30 | | | 0.23 | | |
| 433 | Eros | 43.7 | 1.14 | 250 | 19.7 | 1.020 | 0.010 | 0.000 | 1.020 | 2.015 | 0.015 | 0.010 | 2.025 | 0.197 | 0.100 | -0.043 | 0.154 | 22.30 | | | 0.23 | | |
| 1036 | Ganymed | 5.0 | 3.97 | 133 | 16.6 | 0.910 | 0.010 | -0.001 | 0.909 | 1.850 | 0.010 | 0.033 | 1.883 | 0.907 | 0.100 | -0.128 | 0.779 | | | | | | |
| 1036 | Ganymed | 15.7 | 3.00 | 153 | 17.9 | 0.915 | 0.010 | -0.001 | 0.914 | 1.890 | 0.010 | 0.029 | 1.919 | 1.049 | 0.100 | -0.114 | 0.936 | | | | | | |
| 1036 | Ganymed | 13.2 | 4.07 | 131 | 18.0 | 0.915 | 0.010 | -0.001 | 0.914 | 1.860 | 0.010 | 0.034 | 1.894 | 0.731 | 0.100 | -0.129 | 0.602 | | | | | | |
| 1036 | Ganymed | 6.1 | 1.37 | 226 | 17.0 | 0.925 | 0.010 | 0.000 | 0.925 | 1.925 | 0.010 | 0.015 | 1.940 | 0.750 | 0.100 | -0.060 | 0.690 | | | | | | |
| 1036 | Ganymed | 2.9 | 1.40 | 224 | 16.6 | 0.930 | 0.010 | 0.000 | 0.930 | 1.930 | 0.010 | 0.015 | 1.945 | 0.932 | 0.100 | -0.062 | 0.869 | | | | | | |
| 1627 | Ivar | 3.7 | 2.60 | 173 | 17.3 | 0.955 | 0.010 | -0.001 | 0.954 | 1.970 | 0.011 | 0.025 | 1.995 | 0.596 | 0.100 | -0.099 | 0.497 | 10.03 | 3.27 | 2.77 | 0.09 | 0.12 | 0.05 |
| 1627 | Ivar | 44.5 | 1.42 | 234 | 20.4 | 0.960 | 0.010 | 0.000 | 0.960 | 1.960 | 0.011 | 0.013 | 1.973 | 0.267 | 0.100 | -0.054 | 0.213 | 10.03 | 3.27 | 2.77 | 0.09 | 0.12 | 0.05 |
| 1627 | Ivar | 31.0 | 1.58 | 222 | 20.6 | 0.950 | 0.010 | 0.000 | 0.950 | 1.970 | 0.007 | 0.016 | 1.986 | 0.262 | 0.100 | -0.063 | 0.199 | 10.03 | 3.27 | 2.77 | 0.09 | 0.12 | 0.05 |
| 1627 | Ivar | 16.3 | 1.90 | 202 | 18.8 | 0.945 | 0.010 | 0.000 | 0.945 | 1.900 | 0.014 | 0.020 | 1.920 | 0.423 | 0.100 | -0.078 | 0.345 | 10.03 | 3.27 | 2.77 | 0.09 | 0.12 | 0.05 |
| 1627 | Ivar | 26.9 | 2.15 | 190 | 21.5 | 0.950 | 0.010 | -0.001 | 0.949 | 1.975 | 0.015 | 0.022 | 1.997 | 0.589 | 0.100 | -0.086 | 0.503 | 10.03 | 3.27 | 2.77 | 0.09 | 0.12 | 0.05 |
| 1627 | Ivar | 2.2 | 2.37 | 181 | 17.2 | 0.950 | 0.010 | -0.001 | 0.949 | 1.780 | 0.009 | 0.024 | 1.804 | 0.283 | 0.100 | -0.093 | 0.190 | 10.03 | 3.27 | 2.77 | 0.09 | 0.12 | 0.05 |
| 1685 | Toro | 8.9 | 1.96 | 181 | 15.9 | 0.940 | 0.010 | -0.001 | 0.939 | 2.045 | 0.090 | 0.024 | 2.069 | 0.324 | 0.100 | -0.093 | 0.232 | 3.09 | 0.72 | 0.59 | 0.38 | 0.33 | 0.18 |
| 1685 | Toro | 31.2 | 1.65 | 198 | 10.2 | 1.015 | 0.010 | 0.000 | 1.015 | 2.050 | 0.092 | 0.020 | 2.070 | | | | | 3.09 | 0.72 | 0.59 | 0.38 | 0.33 | 0.18 |
| 1864 | Daedalus | 4.4 | 1.89 | 191 | 22.0 | 0.960 | 0.010 | -0.001 | 0.959 | 1.915 | 0.033 | 0.022 | 1.937 | 0.238 | 0.100 | -0.086 | 0.152 | | | | | | |
| 1865 | Cerberus | 20.2 | 1.41 | 202 | 17.7 | 0.925 | 0.010 | 0.000 | 0.925 | 1.890 | 0.049 | 0.020 | 1.910 | 0.484 | 0.100 | -0.078 | 0.406 | 0.82 | 0.15 | 0.13 | 0.50 | 0.29 | 0.21 |
| 1866 | Sisyphus | 41.6 | 1.44 | 234 | 15.9 | 0.930 | 0.010 | 0.000 | 0.930 | 1.980 | 0.009 | 0.013 | 1.993 | 0.256 | 0.100 | -0.055 | 0.202 | 12.94 | 4.57 | 3.66 | 0.07 | 0.09 | 0.04 |
| 1916 | Boreas | 38.2 | 1.27 | 240 | 16.3 | 0.955 | 0.010 | 0.000 | 0.955 | 1.985 | 0.010 | 0.012 | 1.997 | 0.461 | 0.100 | -0.050 | 0.411 | 3.24 | 0.78 | 0.71 | 0.20 | 0.17 | 0.11 |
| 1917 | Cuyo | 47.6 | 1.37 | 226 | 21.0 | 0.940 | 0.010 | 0.000 | 0.940 | 1.860 | 0.008 | 0.015 | 1.875 | 1.213 | 0.100 | -0.060 | 1.152 | 4.36 | 1.16 | 0.93 | 0.27 | 0.21 | 0.14 |
| 1943 | Anteros | 14.4 | 1.46 | 226 | 11.3 | 0.940 | 0.010 | 0.000 | 0.940 | 2.025 | 0.032 | 0.015 | 2.040 | 0.424 | 0.100 | -0.060 | 0.363 | 2.40 | 0.67 | 0.56 | 0.17 | 0.15 | 0.08 |
| 1943 | Anteros | 15.9 | 1.80 | 204 | 10.1 | 1.010 | 0.010 | 0.000 | 1.010 | 2.015 | 0.006 | 0.019 | 2.034 | 0.601 | 0.100 | -0.077 | 0.525 | 2.40 | 0.67 | 0.56 | 0.17 | 0.15 | 0.08 |
| 1980 | Tezcatlipoca | 53.8 | 1.17 | 225 | 20.6 | 0.955 | 0.010 | 0.000 | 0.955 | 1.995 | 0.011 | 0.015 | 2.010 | 0.188 | 0.100 | -0.061 | 0.127 | 3.14 | 0.80 | 0.60 | 0.47 | 0.43 | 0.24 |
| 2212 | Hephaistos | 14.2 | 1.81 | 209 | 13.5 | 1.010 | 0.010 | 0.000 | 1.010 | 1.995 | 0.010 | 0.018 | 2.013 | | | | | 9.02 | 3.13 | 2.49 | 0.06 | 0.08 | 0.03 |
| 3102 | Krok | 19.7 | 1.38 | 223 | 20.7 | 1.000 | 0.010 | 0.000 | 1.000 | 2.050 | 0.033 | 0.015 | 2.065 | 0.528 | 0.100 | -0.063 | 0.465 | | | | | | |
| 4055 | Magellan | 17.7 | 1.84 | 188 | 51.2 | 0.925 | 0.010 | 0.005 | 0.930 | 1.915 | 0.010 | 0.017 | 1.932 | 2.182 | 0.100 | 0.000 | 2.182 | 2.46 | 0.53 | 0.41 | 0.36 | 0.27 | 0.17 |
| 4183 | Cuno | 9.4 | 2.07 | 193 | 19.8 | 1.015 | 0.010 | -0.001 | 1.014 | 1.925 | 0.010 | 0.021 | 1.946 | 0.264 | 0.100 | -0.084 | 0.179 | 5.44 | 1.20 | 1.11 | 0.10 | 0.10 | 0.05 |
| 4544 | Xanthus | 50.5 | 1.18 | 234 | 12.4 | 0.970 | 0.010 | 0.000 | 0.970 | 1.940 | 0.162 | 0.013 | 1.953 | | | | | 0.85 | 0.16 | 0.14 | 0.37 | 0.23 | 0.16 |
| 5011 | Ptah | 63.2 | 1.10 | 265 | 18.0 | 0.965 | 0.010 | 0.001 | 0.966 | 1.995 | 0.020 | 0.007 | 2.002 | | | | | 1.54 | 0.42 | 0.39 | 0.11 | 0.11 | 0.05 |
| 5131 | 1990 BG | 13.9 | 1.90 | 95 | 26.2 | 1.020 | 0.010 | -0.002 | 1.018 | 2.020 | 0.032 | 0.041 | 2.061 | 0.278 | 0.100 | -0.156 | 0.122 | 1.31 | 0.29 | 0.19 | 0.96 | 0.38 | 0.39 |
| 5131 | 1990 BG | 22.6 | 1.41 | 110 | 26.6 | 1.040 | 0.010 | -0.002 | 1.038 | 2.000 | 0.152 | 0.038 | 2.038 | 0.119 | 0.100 | -0.144 | -0.026 | 1.31 | 0.29 | 0.19 | 0.96 | 0.38 | 0.39 |
| 5143 | Heracles | 16.2 | 1.70 | 193 | 23.4 | 1.010 | 0.010 | -0.001 | 1.009 | 1.960 | 0.016 | 0.021 | 1.981 | 0.232 | 0.100 | -0.085 | 0.148 | 3.39 | 0.60 | 0.55 | 0.40 | 0.22 | 0.16 |
| 5332 | 1990 DA | 56.1 | 1.18 | 242 | 12.9 | 0.935 | 0.010 | 0.000 | 0.935 | 1.950 | 0.010 | 0.012 | 1.962 | | | | | | | | | | |
| 5587 | 1990 SB | 21.9 | 1.49 | 215 | 16.4 | 0.925 | 0.010 | 0.000 | 0.925 | 1.905 | 0.010 | 0.017 | 1.922 | 0.979 | 0.100 | -0.068 | 0.911 | 3.73 | 2.26 | 1.22 | 0.29 | 0.43 | 0.19 |
| 5604 | 1992 FE | 28.1 | 1.26 | 188 | 48.3 | 0.925 | 0.010 | 0.005 | 0.930 | 1.925 | 0.010 | 0.017 | 1.942 | 1.960 | 0.100 | 0.000 | 1.960 | 0.85 | 0.10 | 0.11 | 0.71 | 0.32 | 0.23 |
| 5604 | 1992 FE | 36.7 | 1.29 | 185 | 51.6 | 0.925 | 0.010 | 0.005 | 0.930 | 1.915 | 0.005 | 0.017 | 1.932 | 2.038 | 0.100 | 0.000 | 2.038 | 0.85 | 0.10 | 0.11 | 0.71 | 0.32 | 0.23 |
| 5626 | 1991 FE | 33.8 | 1.51 | 223 | 15.1 | 0.930 | 0.010 | 0.000 | 0.930 | 1.955 | 0.012 | 0.015 | 1.970 | 0.559 | 0.100 | -0.063 | 0.496 | 3.90 | 1.14 | 0.93 | 0.15 | 0.18 | 0.08 |
| 5653 | Camarillo | 30.3 | 1.30 | 222 | 11.0 | 0.925 | 0.010 | 0.000 | 0.925 | 1.980 | 0.011 | 0.016 | 1.996 | | | | | 1.74 | 0.45 | 0.34 | 0.39 | 0.37 | 0.19 |
| 5693 | 1993 EA | 19.5 | 1.40 | 232 | 13.7 | 1.025 | 0.010 | 0.000 | 1.025 | 2.040 | 0.183 | 0.014 | 2.054 | | | | | 1.39 | 0.28 | 0.24 | 0.15 | 0.11 | 0.07 |
| 5786 | Talos | 16.3 | 1.41 | 230 | 16.1 | 0.960 | 0.010 | 0.000 | 0.960 | 1.990 | 0.023 | 0.014 | 2.004 | | | | | 1.29 | 0.32 | 0.27 | 0.17 | 0.15 | 0.08 |
| 6047 | 1991 TB1 | 20.3 | 1.39 | 162 | 14.7 | 0.930 | 0.010 | -0.001 | 0.929 | 2.005 | 0.040 | 0.028 | 2.033 | 1.889 | 0.130 | -0.107 | 1.782 | 0.55 | 0.10 | 0.08 | 0.80 | 0.44 | 0.35 |
| 6239 | Minos | 0.8 | 1.21 | 210 | 16.8 | 0.985 | 0.010 | 0.000 | 0.985 | 2.020 | 0.113 | 0.018 | 2.038 | 0.268 | 0.100 | -0.072 | 0.196 | 0.47 | 0.13 | 0.08 | 0.57 | 0.37 | 0.29 |
| 6455 | 1992 HE | 24.5 | 1.70 | 200 | 31.4 | 0.935 | 0.010 | 0.000 | 0.935 | 1.980 | 0.054 | 0.020 | 2.000 | 0.727 | 0.100 | -0.079 | 0.648 | 3.96 | 1.72 | 1.10 | 0.31 | 0.42 | 0.18 |
| 6455 | 1992 HE | 14.2 | 1.59 | 207 | 24.4 | 0.940 | 0.010 | 0.000 | 0.940 | 1.955 | 0.011 | 0.019 | 1.974 | 0.482 | 0.100 | -0.075 | 0.407 | 3.96 | 1.72 | 1.10 | 0.31 | 0.42 | 0.18 |
| 6455 | 1992 HE | 23.4 | 1.56 | 209 | 25.4 | 0.980 | 0.010 | 0.000 | 0.980 | 1.930 | 0.010 | 0.018 | 1.948 | 0.373 | 0.100 | -0.073 | 0.300 | 3.96 | 1.72 | 1.10 | 0.31 | 0.42 | 0.18 |
| 7358 | Oze | 1.1 | 1.84 | 206 | 13.7 | 0.940 | 0.010 | 0.000 | 0.940 | 1.950 | 0.018 | 0.019 | 1.969 | 0.312 | 0.100 | -0.075 | 0.237 | 6.17 | 2.03 | 1.62 | 0.08 | 0.10 | 0.04 |
| 8567 | 1996 HW1 | 55.8 | 1.19 | 243 | 14.7 | 1.015 | 0.010 | 0.000 | 1.015 | 2.005 | 0.010 | 0.011 | 2.016 | | | | | | | | | | |
| 8567 | 1996 HW1 | 53.0 | 1.22 | 240 | 13.5 | 1.015 | 0.010 | 0.000 | 1.015 | 2.080 | 0.064 | 0.012 | 2.092 | | | | | | | | | | |
| 8567 | 1996 HW1 | 27.1 | 1.13 | 249 | 17.9 | 0.995 | 0.010 | 0.000 | 0.995 | 1.955 | 0.086 | 0.010 | 1.965 | 0.246 | 0.100 | -0.044 | 0.202 | | | | | | |
| 8567 | 1996 HW1 | 23.9 | 1.15 | 248 | 17.4 | 1.005 | 0.010 | 0.000 | 1.005 | 2.000 | 0.010 | 0.010 | 2.010 | | | | | | | | | | |
| 8567 | 1996 HW1 | 39.6 | 1.36 | 227 | 14.9 | 1.005 | 0.010 | 0.000 | 1.005 | 1.980 | 0.026 | 0.015 | 1.995 | | | | | | | | | | |
| 10145 | 1994 CK1 | 10.2 | 1.35 | 226 | 15.3 | 0.960 | 0.010 | 0.000 | 0.960 | 2.015 | 0.032 | 0.015 | 2.030 | | | | | | | | | | |
| 11066 | Sigurd | 16.2 | 1.66 | 196 | 14.0 | 0.925 | 0.010 | 0.000 | 0.925 | 1.900 | 0.005 | 0.021 | 1.921 | 0.915 | 0.100 | -0.082 | 0.833 | 1.98 | 0.45 | 0.38 | 0.38 | 0.31 | 0.18 |
| 11398 | 1998 YP11 | 32.9 | 1.16 | 252 | 19.0 | 0.925 | 0.010 | 0.000 | 0.925 | 1.940 | 0.003 | 0.010 | 1.950 | 0.645 | 0.100 | -0.042 | 0.603 | 1.73 | 0.47 | 0.39 | 0.18 | 0.17 | 0.09 |
| 12923 | Zephyr | 62.6 | 1.15 | 253 | 15.2 | 0.960 | 0.010 | 0.000 | 0.960 | 1.905 | 0.010 | 0.009 | 1.914 | | | | | 1.82 | 0.42 | 0.34 | 0.20 | 0.16 | 0.09 |
| 15745 | 1991 PM5 | 27.1 | 1.32 | 232 | 11.0 | 0.945 | 0.010 | 0.000 | 0.945 | 1.970 | 0.027 | 0.014 | 1.984 | 0.639 | 0.100 | -0.056 | 0.583 | 0.78 | 0.17 | 0.16 | 0.24 | 0.18 | 0.11 |
| 16834 | 1997 WU22 | 27.8 | 1.35 | 214 | 11.3 | 1.020 | 0.010 | 0.000 | 1.020 | 1.955 | 0.039 | 0.017 | 1.972 | 2.428 | 0.100 | -0.069 | 2.359 | 1.50 | 0.32 | 0.26 | 0.43 | 0.28 | 0.20 |
| 16834 | 1997 WU22 | 24.0 | 1.59 | 197 | 14.2 | 0.960 | 0.010 | 0.000 | 0.960 | 1.975 | 0.016 | 0.021 | 1.996 | 0.487 | 0.100 | -0.082 | 0.406 | 1.50 | 0.32 | 0.26 | 0.43 | 0.28 | 0.20 |
| 19764 | 2000 NF5 | 18.4 | 1.48 | 218 | 17.6 | 1.010 | 0.010 | 0.000 | 1.010 | 1.975 | 0.010 | 0.016 | 1.991 | | | | | 1.67 | 0.43 | 0.37 | 0.26 | 0.23 | 0.14 |
| 20790 | 2000 SE45 | 13.3 | 1.26 | 248 | 12.9 | 0.945 | 0.010 | 0.000 | 0.945 | 1.975 | 0.010 | 0.010 | 1.985 | 1.894 | 0.100 | -0.044 | 1.849 | 2.02 | 0.63 | 0.51 | 0.10 | 0.11 | 0.05 |
| 21088 | 1992 BL2 | 7.6 | 2.10 | 183 | 14.2 | 1.040 | 0.010 | -0.001 | 1.039 | 2.025 | 0.083 | 0.023 | 2.048 | | | | | 3.44 | 1.10 | 0.86 | 0.26 | 0.32 | 0.14 |
| 22753 | 1998 WT | 54.2 | 1.08 | 254 | 14.4 | 0.970 | 0.010 | 0.000 | 0.970 | 2.010 | 0.010 | 0.009 | 2.019 | | | | | 0.74 | 0.14 | 0.12 | 0.27 | 0.17 | 0.11 |

| Number | Name | $\alpha$ | r (AU) | T(K) | BID | BIC | $\sigma_{BIC}$ | $\delta_{BIC}$ | $\Delta$BIC | BIIC | $\sigma_{BIIC}$ | $\delta_{BIIC}$ | $\Delta$BIIC | BAR | $\sigma_{BAR}$ | $\delta_{BAR}$ | $\Delta$BAR | D | $\sigma_{D+}$ | $\sigma_{D-}$ | PV | $\sigma_{pV+}$ | $\sigma_{pV-}$ |
|---|---|---|---|---|---|---|---|---|---|---|---|---|---|---|---|---|---|---|---|---|---|---|---|
| 22753 | 1998 WT | 44.5 | 1.05 | 257 | 15.4 | 1.000 | 0.010 | 0.000 | 1.000 | 1.965 | 0.010 | 0.009 | 1.974 | | | | | 0.74 | 0.14 | 0.12 | 0.27 | 0.17 | 0.11 |
| 23183 | 2000 OY21 | 32.6 | 1.26 | 134 | 20.7 | 1.025 | 0.010 | -0.001 | 1.024 | 2.110 | 0.018 | 0.033 | 2.143 | | | | | 0.74 | 0.11 | 0.10 | 0.92 | 0.39 | 0.39 |
| 23187 | 2000 PN9 | 50.8 | 1.22 | 245 | 14.5 | 1.025 | 0.010 | 0.000 | 1.025 | 2.060 | 0.033 | 0.011 | 2.071 | | | | | 1.84 | 0.37 | 0.35 | 0.20 | 0.15 | 0.10 |
| 23187 | 2000 PN9 | 55.6 | 1.19 | 248 | 18.1 | 1.040 | 0.010 | 0.000 | 1.040 | 1.995 | 0.047 | 0.010 | 2.005 | | | | | 1.84 | 0.37 | 0.35 | 0.20 | 0.15 | 0.10 |
| 25143 | Itokawa | 22.1046-69.3187 | 1.08 | 246 | 19.5 | 0.970 | 0.010 | 0.000 | 0.970 | 2.020 | 0.010 | 0.011 | 2.031 | 0.284 | 0.100 | -0.046 | 0.238 | 0.32 | 0.05 | 0.04 | 0.36 | 0.22 | 0.15 |
| 25916 | 2001 CP44 | 12.4 | 2.22 | 181 | 10.5 | 1.020 | 0.010 | -0.001 | 1.019 | 1.880 | 0.074 | 0.024 | 1.904 | | | | | 6.49 | 1.98 | 1.49 | 0.21 | 0.22 | 0.11 |
| 25916 | 2001 CP44 | 30.7 | 1.70 | 207 | 13.1 | 1.020 | 0.010 | 0.000 | 1.020 | 1.950 | 0.193 | 0.019 | 1.969 | | | | | 6.49 | 1.98 | 1.49 | 0.21 | 0.22 | 0.11 |
| 31669 | 1999 JT6 | 33.3 | 1.28 | 234 | 14.3 | 1.035 | 0.010 | 0.000 | 1.035 | 2.105 | 0.030 | 0.013 | 2.118 | | | | | | | | | | |
| 35107 | 1991 VH | 45.0 | 1.20 | 242 | 19.4 | 0.980 | 0.010 | 0.000 | 0.980 | 2.035 | 0.025 | 0.012 | 2.047 | 0.360 | 0.100 | -0.049 | 0.311 | 1.09 | 0.22 | 0.19 | 0.26 | 0.20 | 0.11 |
| 35107 | 1991 VH | 61.7 | 1.11 | 252 | 19.8 | 1.005 | 0.010 | 0.000 | 1.005 | 1.945 | 0.013 | 0.010 | 1.955 | 0.287 | 0.100 | -0.042 | 0.245 | 1.09 | 0.22 | 0.19 | 0.26 | 0.20 | 0.11 |
| 36284 | 2000 DM8 | 40.3 | 1.37 | 231 | 17.0 | 0.985 | 0.018 | 0.000 | 0.985 | 2.075 | 0.166 | 0.014 | 2.089 | 0.452 | 0.111 | -0.057 | 0.396 | 3.08 | 0.72 | 0.55 | 0.19 | 0.16 | 0.09 |
| 39572 | 1993 DQ1 | 64.8 | 1.08 | 255 | 19.1 | 0.990 | 0.010 | 0.000 | 0.990 | 1.960 | 0.011 | 0.009 | 1.969 | | | | | | | | | | |
| 52387 | 1993 OM7 | 58.8 | 1.07 | 270 | 11.1 | 0.925 | 0.010 | 0.001 | 0.926 | 1.975 | 0.104 | 0.006 | 1.981 | | | | | 1.21 | 0.32 | 0.30 | 0.09 | 0.10 | 0.05 |
| 52750 | 1998 KK17 | 21.0 | 1.36 | 210 | 30.8 | 0.935 | 0.010 | 0.004 | 0.939 | 1.970 | 0.010 | 0.013 | 1.983 | | | | | 1.03 | 0.17 | 0.16 | 0.46 | 0.25 | 0.20 |
| 53430 | 1999 TY16 | 24.0 | 1.29 | 231 | 22.6 | 0.930 | 0.010 | 0.000 | 0.930 | 1.935 | 0.010 | 0.014 | 1.949 | | | | | | | | | | |
| 53435 | 1999 VM40 | 55.0 | 1.21 | 241 | 17.3 | 0.930 | 0.010 | 0.000 | 0.930 | 1.920 | 0.008 | 0.012 | 1.932 | 1.135 | 0.100 | -0.050 | 1.085 | | | | | | |
| 53789 | 2000 ED104 | 58.8 | 1.14 | 243 | 8.9 | 0.925 | 0.045 | 0.000 | 0.925 | 1.890 | 0.257 | 0.011 | 1.901 | | | | | 0.84 | 0.19 | 0.17 | 0.32 | 0.24 | 0.15 |
| 54690 | 2001 EB | 33.0 | 1.36 | 229 | 6.9 | 0.940 | 0.035 | 0.000 | 0.940 | 1.975 | 0.279 | 0.014 | 1.989 | 0.567 | 0.325 | -0.058 | 0.509 | 0.96 | 0.25 | 0.20 | 0.24 | 0.21 | 0.11 |
| 66251 | 1999 GJ2 | 37.2 | 1.24 | 229 | 11.6 | 0.955 | 0.010 | 0.000 | 0.955 | 2.015 | 0.012 | 0.014 | 2.029 | | | | | 0.88 | 0.29 | 0.17 | 0.37 | 0.32 | 0.21 |
| 66251 | 1999 GJ2 | 7.4 | 1.28 | 226 | 11.0 | 0.945 | 0.010 | 0.000 | 0.945 | 2.005 | 0.011 | 0.015 | 2.020 | | | | | 0.88 | 0.29 | 0.17 | 0.37 | 0.32 | 0.21 |
| 66251 | 1999 GJ2 | 16.6 | 1.32 | 222 | 9.8 | 0.945 | 0.010 | 0.000 | 0.945 | 1.925 | 0.017 | 0.016 | 1.941 | | | | | 0.88 | 0.29 | 0.17 | 0.37 | 0.32 | 0.21 |
| 68216 | 2001 CV26 | 20.5 | 1.36 | 227 | 11.3 | 0.980 | 0.010 | 0.000 | 0.980 | 2.000 | 0.010 | 0.015 | 2.015 | | | | | 1.39 | 0.37 | 0.29 | 0.26 | 0.21 | 0.13 |
| 68359 | 2001 OZ13 | 21.5 | 1.26 | 217 | 18.0 | 0.995 | 0.010 | 0.000 | 0.995 | 1.845 | 0.308 | 0.017 | 1.862 | | | | | 0.61 | 0.14 | 0.11 | 0.47 | 0.32 | 0.22 |
| 69230 | Hermes | 6.5 | 1.19 | 254 | 14.2 | 0.975 | 0.010 | 0.000 | 0.975 | 1.965 | 0.010 | 0.009 | 1.974 | | | | | 1.25 | 0.27 | 0.21 | 0.12 | 0.10 | 0.05 |
| 85709 | 1998 SG36 | 66.1 | 1.10 | 248 | 11.9 | 0.930 | 0.010 | 0.000 | 0.930 | 1.960 | 0.010 | 0.010 | 1.970 | | | | | 1.53 | 0.37 | 0.30 | 0.32 | 0.27 | 0.15 |
| 85818 | 1998 XM4 | 48.5 | 1.18 | 241 | 18.1 | 0.935 | 0.010 | 0.000 | 0.935 | 1.985 | 0.010 | 0.012 | 1.997 | 1.401 | 0.100 | -0.050 | 1.351 | 2.00 | 0.50 | 0.40 | 0.30 | 0.23 | 0.15 |
| 85839 | 1998 YO4 | 22.2 | 1.38 | 232 | 21.4 | 1.025 | 0.010 | 0.000 | 1.025 | 2.035 | 0.034 | 0.014 | 2.049 | | | | | 1.81 | 0.41 | 0.40 | 0.17 | 0.16 | 0.08 |
| 86324 | 1999 WA2 | 37.4 | 1.47 | 217 | 16.3 | 0.935 | 0.010 | 0.000 | 0.935 | 1.940 | 0.112 | 0.017 | 1.957 | | | | | | | | | | |
| 86667 | 2000 FO10 | 3.3 | 1.34 | 238 | 15.2 | 1.035 | 0.010 | 0.000 | 1.035 | 1.945 | 0.012 | 0.012 | 1.957 | | | | | 1.22 | 0.29 | 0.25 | 0.13 | 0.11 | 0.06 |
| 87684 | 2000 SY2 | 44.9 | 1.18 | 244 | 19.6 | 0.965 | 0.010 | 0.000 | 0.965 | 1.950 | 0.010 | 0.011 | 1.961 | | | | | | | | | | |
| 88254 | 2001 FM129 | 79.5 | 1.01 | 257 | 21.6 | 0.985 | 0.010 | 0.000 | 0.985 | 2.005 | 0.035 | 0.009 | 2.014 | | | | | 0.81 | 0.17 | 0.14 | 0.33 | 0.23 | 0.14 |
| 89355 | 2001 VS78 | 51.9 | 1.24 | 236 | 12.0 | 0.945 | 0.010 | 0.000 | 0.945 | 1.950 | 0.042 | 0.013 | 1.963 | | | | | | | | | | |
| 89355 | 2001 VS78 | 49.0 | 1.27 | 233 | 18.5 | 0.925 | 0.010 | 0.000 | 0.925 | 2.065 | 0.015 | 0.013 | 2.078 | 0.907 | 0.100 | -0.055 | 0.852 | | | | | | |
| 96590 | 1994 XB | 28.6 | 1.23 | 251 | 17.4 | 1.010 | 0.010 | 0.000 | 1.010 | 1.995 | 0.010 | 0.010 | 2.005 | | | | | 2.29 | 0.47 | 0.42 | 0.11 | 0.09 | 0.05 |
| 99799 | 2002 LJ3 | 16.0 | 1.36 | 214 | 20.4 | 1.005 | 0.010 | 0.000 | 1.005 | 1.875 | 0.234 | 0.017 | 1.892 | | | | | 0.50 | 0.09 | 0.09 | 0.42 | 0.26 | 0.17 |
| 99907 | 1989 VA | 62.8 | 1.06 | 259 | 24.5 | 0.905 | 0.010 | 0.000 | 0.905 | 1.965 | 0.008 | 0.008 | 1.973 | 0.589 | 0.100 | -0.036 | 0.552 | 0.74 | 0.14 | 0.14 | 0.24 | 0.19 | 0.11 |
| 99907 | 1989 VA | 36.7 | 1.15 | 249 | 21.5 | 0.975 | 0.010 | 0.000 | 0.975 | 1.965 | 0.010 | 0.010 | 1.975 | | | | | 0.74 | 0.14 | 0.14 | 0.24 | 0.19 | 0.11 |
| 100926 | 1998 MQ | 57.4 | 1.10 | 244 | 23.2 | 0.945 | 0.010 | 0.000 | 0.945 | 2.030 | 0.012 | 0.011 | 2.041 | 0.311 | 0.100 | -0.048 | 0.263 | 1.08 | 0.23 | 0.19 | 0.37 | 0.22 | 0.16 |
| 137032 | 1998 UO1 | 22.2 | 1.57 | 220 | 14.3 | 0.925 | 0.010 | 0.000 | 0.925 | 1.890 | 0.019 | 0.016 | 1.906 | | | | | 1.66 | 0.37 | 0.35 | 0.13 | 0.11 | 0.06 |
| 137032 | 1998 UO1 | 52.0 | 1.07 | 267 | 21.0 | 0.985 | 0.010 | 0.001 | 0.986 | 2.010 | 0.010 | 0.007 | 2.017 | | | | | 1.66 | 0.37 | 0.35 | 0.13 | 0.11 | 0.06 |
| 137062 | 1998 WM | 18.1 | 1.60 | 205 | 22.7 | 0.915 | 0.010 | 0.000 | 0.915 | 1.990 | 0.010 | 0.019 | 2.009 | 0.778 | 0.100 | -0.076 | 0.702 | 1.13 | 0.28 | 0.23 | 0.32 | 0.23 | 0.14 |
| 137671 | 1999 XP35 | 26.1 | 1.17 | 252 | 16.1 | 0.925 | 0.010 | 0.000 | 0.925 | 2.020 | 0.012 | 0.010 | 2.030 | | | | | 0.60 | 0.17 | 0.13 | 0.17 | 0.16 | 0.08 |
| 137924 | 2000 BD19 | 19.8 | 1.50 | 226 | 55.8 | 0.950 | 0.010 | 0.004 | 0.954 | 1.965 | 0.010 | 0.011 | 1.976 | | | | | 1.46 | 0.31 | 0.29 | 0.11 | 0.09 | 0.05 |
| 138258 | 2000 GD2 | 26.5 | 1.06 | 257 | 21.0 | 1.010 | 0.010 | 0.000 | 1.010 | 1.950 | 0.038 | 0.009 | 1.959 | 0.499 | 0.100 | -0.038 | 0.461 | | | | | | |
| 138883 | 2000 YL29 | 34.8 | 1.19 | 248 | 26.2 | 0.995 | 0.010 | 0.000 | 0.995 | | | | | | | | | 1.40 | 0.30 | 0.30 | 0.19 | 0.17 | 0.09 |
| 141052 | 2001 XR1 | 39.9 | 1.12 | 255 | 16.8 | 0.960 | 0.010 | 0.000 | 0.960 | 1.965 | 0.047 | 0.009 | 1.974 | 0.395 | 0.100 | -0.039 | 0.356 | 0.94 | 0.16 | 0.14 | 0.20 | 0.16 | 0.09 |
| 141052 | 2001 XR1 | | | | | | | | | | | | | | | | | 0.91 | 0.16 | 0.16 | 0.22 | 0.16 | 0.09 |
| 141052 | 2001 XR1 | | | | | | | | | | | | | | | | | 0.82 | 0.14 | 0.12 | 0.27 | 0.18 | 0.12 |
| 141052 | 2001 XR1 | | | | | | | | | | | | | | | | | 0.96 | 0.22 | 0.19 | 0.20 | 0.15 | 0.09 |
| 141432 | 2002 CQ11 | 12.4 | 1.15 | 245 | 17.2 | 0.925 | 0.010 | 0.000 | 0.925 | 2.005 | 0.058 | 0.011 | 2.016 | | | | | | | | | | |
| 143651 | 2003 QO104 | 40.8 | 1.24 | 247 | 21.4 | 0.985 | 0.010 | 0.000 | 0.985 | 2.060 | 0.010 | 0.011 | 2.071 | | | | | 2.30 | 0.56 | 0.49 | 0.14 | 0.12 | 0.07 |
| 143651 | 2003 QO104 | 64.3 | 1.09 | 264 | 27.1 | 0.995 | 0.010 | 0.001 | 0.996 | 1.905 | 0.014 | 0.007 | 1.912 | 0.329 | 0.100 | -0.033 | 0.296 | 2.30 | 0.56 | 0.49 | 0.14 | 0.12 | 0.07 |
| 144900 | 2004 VG64 | 44.2 | 1.10 | 249 | 10.7 | 0.980 | 0.010 | 0.000 | 0.980 | 2.020 | 0.021 | 0.010 | 2.030 | | | | | 0.55 | 0.11 | 0.09 | 0.30 | 0.20 | 0.13 |
| 144922 | 2005 CK38 | 11.4 | 1.40 | 230 | 13.1 | 1.000 | 0.010 | 0.000 | 1.000 | 1.990 | 0.013 | 0.014 | 2.004 | | | | | 1.15 | 0.28 | 0.26 | 0.17 | 0.16 | 0.08 |
| 152558 | 1990 SA | 15.7 | 1.31 | 232 | 5.7 | 0.950 | 0.010 | 0.000 | 0.950 | 2.025 | 0.019 | 0.014 | 2.039 | | | | | | | | | | |
| 152931 | 2000 EA107 | 50.2 | 1.24 | 238 | 27.0 | 1.005 | 0.010 | 0.000 | 1.005 | 1.950 | 0.025 | 0.012 | 1.962 | 0.326 | 0.100 | -0.052 | 0.274 | | | | | | |
| 154029 | 2002 CY46 | 45.5 | 1.12 | 258 | 11.0 | 0.935 | 0.010 | 0.000 | 0.935 | 1.980 | 0.010 | 0.008 | 1.988 | | | | | 1.76 | 0.40 | 0.38 | 0.16 | 0.16 | 0.08 |
| 154244 | 2002 KL6 | 48.7 | 1.11 | 251 | 22.5 | 1.000 | 0.010 | 0.000 | 1.000 | 1.920 | 0.127 | 0.010 | 1.930 | | | | | | | | | | |
| 154715 | 2004 LB6 | 50.9 | 1.13 | 249 | 22.0 | 1.015 | 0.010 | 0.000 | 1.015 | 1.965 | 0.045 | 0.010 | 1.975 | | | | | | | | | | |
| 159402 | 1999 AP10 | 19.3 | 1.16 | 238 | 9.8 | 0.970 | 0.010 | 0.000 | 0.970 | 1.955 | 0.013 | 0.012 | 1.967 | | | | | 1.22 | 0.25 | 0.22 | 0.35 | 0.24 | 0.16 |
| 161989 | Cacus | 53.6 | 1.10 | 250 | 14.8 | 1.010 | 0.010 | 0.000 | 1.010 | 1.990 | 0.012 | 0.010 | 2.000 | | | | | 0.95 | 0.19 | 0.16 | 0.29 | 0.21 | 0.13 |
| 161998 | 1988 PA | 25.7 | 1.29 | 241 | 16.8 | 0.980 | 0.010 | 0.000 | 0.980 | 1.935 | 0.151 | 0.012 | 1.947 | | | | | 1.10 | 0.29 | 0.23 | 0.16 | 0.14 | 0.08 |
| 162483 | 2000 PJ5 | 52.8 | 1.18 | 248 | 26.3 | 0.965 | 0.010 | 0.000 | 0.965 | 1.915 | 0.265 | 0.010 | 1.925 | | | | | 0.93 | 0.21 | 0.17 | 0.22 | 0.17 | 0.10 |
| 162900 | 2001 HG31 | 16.4 | 1.31 | 232 | 17.4 | 1.065 | 0.010 | 0.000 | 1.065 | 1.945 | 0.011 | 0.014 | 1.959 | | | | | | | | | | |
| 163001 | 2001 SE170 | 28.0 | 1.26 | 248 | 10.3 | 0.925 | 0.010 | 0.000 | 0.925 | 1.980 | 0.028 | 0.010 | 1.990 | | | | | 1.02 | 0.30 | 0.23 | 0.09 | 0.10 | 0.05 |
| 163243 | 2002 FB3 | 57.6 | 1.17 | 242 | 15.2 | 1.015 | 0.010 | 0.000 | 1.015 | 1.943 | 0.015 | 0.012 | 1.954 | | | | | 1.36 | 0.27 | 0.24 | 0.30 | 0.20 | 0.14 |
| 163697 | 2003 EF54 | 46.4 | 1.06 | 255 | 20.8 | 0.990 | 0.010 | 0.000 | 0.990 | 1.965 | 0.227 | 0.009 | 1.974 | | | | | 0.25 | 0.04 | 0.04 | 0.29 | 0.20 | 0.12 |

| Number | Name | $\alpha$ | r (AU) | T(K) | BID | BIC | $\sigma_{BIC}$ | $\delta_{BIC}$ | $\Delta$BIC | BIIC | $\sigma_{BIIC}$ | $\delta_{BIIC}$ | $\Delta$BIIC | BAR | $\sigma_{BAR}$ | $\delta_{BAR}$ | $\Delta$BAR | D | $\sigma_{D+}$ | $\sigma_{D-}$ | PV | $\sigma_{pV+}$ | $\sigma_{pV-}$ |
|---|---|---|---|---|---|---|---|---|---|---|---|---|---|---|---|---|---|---|---|---|---|---|---|
| 164400 | 2005 GN59 | 27.0 | 1.20 | 252 | 18.1 | 0.995 | 0.010 | 0.000 | 0.995 | 1.940 | 0.016 | 0.010 | 1.950 | | | | | 1.30 | 0.29 | 0.26 | 0.13 | 0.11 | 0.05 |
| 184266 | 2004 VW14 | 50.8 | 1.06 | 265 | 24.0 | 1.025 | 0.010 | 0.001 | 1.026 | 2.020 | 0.015 | 0.007 | 2.027 | | | | | 0.44 | 0.10 | 0.09 | 0.17 | 0.14 | 0.08 |
| 200840 | 2001 XN254 | 63.8 | 1.03 | 258 | 17.1 | 0.950 | 0.010 | 0.000 | 0.950 | 2.000 | 0.010 | 0.008 | 2.008 | 0.732 | 0.100 | -0.037 | 0.695 | | | | | | |
| 207945 | 1991 JW | 23.1 | 1.12 | 250 | 19.9 | 0.970 | 0.010 | 0.000 | 0.970 | 1.955 | 0.087 | 0.010 | 1.965 | 0.318 | 0.100 | -0.043 | 0.275 | | | | | | |
| 217796 | 2000 TO64 | 64.9 | 1.07 | 255 | 26.4 | 0.935 | 0.010 | 0.000 | 0.935 | 1.890 | 0.028 | 0.009 | 1.899 | | | | | 1.14 | 0.25 | 0.21 | 0.28 | 0.21 | 0.12 |
| 217807 | 2000 XK44 | 38.8 | 1.06 | 255 | 16.5 | 1.000 | 0.010 | 0.000 | 1.000 | 1.980 | 0.019 | 0.009 | 1.989 | | | | | 0.73 | 0.15 | 0.13 | 0.29 | 0.19 | 0.13 |
| 218863 | 2006 WO127 | 66.1 | 1.07 | 259 | 19.1 | 0.970 | 0.010 | 0.000 | 0.970 | 1.995 | 0.044 | 0.008 | 2.003 | | | | | 1.69 | 0.43 | 0.38 | 0.22 | 0.18 | 0.10 |
| 219071 | 1997 US9 | 23.5 | 1.35 | 221 | 23.9 | 1.030 | 0.010 | 0.000 | 1.030 | 1.965 | 0.131 | 0.016 | 1.981 | 0.265 | 0.100 | -0.064 | 0.201 | 0.79 | 0.15 | 0.13 | 0.35 | 0.22 | 0.15 |
| 235756 | 2004 VC | 59.2 | 1.06 | 261 | 18.7 | 0.915 | 0.010 | 0.001 | 0.916 | 1.875 | 0.026 | 0.008 | 1.883 | | | | | 0.52 | 0.10 | 0.09 | 0.21 | 0.16 | 0.09 |
| 237442 | 1999 TA10 | 20.8 | 1.30 | 233 | 17.4 | 0.930 | 0.010 | 0.000 | 0.930 | 1.960 | 0.017 | 0.013 | 1.973 | | | | | 0.75 | 0.16 | 0.13 | 0.25 | 0.18 | 0.11 |
| 241662 | 2000 KO44 | 24.0 | 1.26 | 236 | 16.3 | 1.005 | 0.010 | 0.000 | 1.005 | 1.945 | 0.012 | 0.013 | 1.958 | | | | | | | | | | |
| 248818 | 2006 SZ217 | 21.7 | 1.23 | 246 | 16.1 | 0.925 | 0.010 | 0.000 | 0.925 | 1.910 | 0.010 | 0.011 | 1.921 | | | | | 1.08 | 0.26 | 0.20 | 0.16 | 0.14 | 0.07 |
| 260141 | 2004 QT24 | 27.4 | 1.10 | 238 | 14.1 | 0.995 | 0.010 | 0.000 | 0.995 | 2.000 | 0.010 | 0.012 | 2.012 | | | | | 0.46 | 0.08 | 0.07 | 0.42 | 0.24 | 0.17 |
| 260141 | 2004 QT24 | 51.4 | 1.10 | 238 | 17.7 | 0.980 | 0.010 | 0.000 | 0.980 | 1.985 | 0.019 | 0.012 | 1.997 | | | | | 0.46 | 0.08 | 0.07 | 0.42 | 0.24 | 0.17 |
| 274138 | 2008 FU6 | 4.0 | 1.34 | 229 | 18.2 | 1.045 | 0.010 | 0.000 | 1.045 | 1.960 | 0.227 | 0.014 | 1.974 | | | | | | | | | | |
| 274138 | 2008 FU6 | 53.2 | 1.11 | 251 | 20.8 | 1.025 | 0.010 | 0.000 | 1.025 | 1.965 | 0.148 | 0.010 | 1.975 | | | | | | | | | | |
| 297418 | 2000 SP43 | 56.6 | 1.09 | 239 | 26.3 | 0.940 | 0.010 | 0.003 | 0.943 | 2.000 | 0.010 | 0.009 | 2.009 | | | | | | | | | | |
| 297418 | 2000 SP43 | 51.3 | 1.08 | 239 | 30.5 | 0.945 | 0.010 | 0.003 | 0.948 | 1.945 | 0.010 | 0.009 | 1.954 | | | | | | | | | | |
| 297418 | 2000 SP43 | 35.6 | 1.17 | 230 | 29.2 | 0.945 | 0.010 | 0.003 | 0.948 | 1.985 | 0.010 | 0.010 | 1.995 | | | | | | | | | | |
| 301844 | 1990 UA | 44.1 | 1.08 | 236 | 16.1 | 0.990 | 0.010 | 0.000 | 0.990 | 1.965 | 0.106 | 0.013 | 1.978 | | | | | 0.24 | 0.04 | 0.04 | 0.46 | 0.25 | 0.19 |
| 302831 | 2003 FH | 64.2 | 1.04 | 258 | 24.9 | 1.000 | 0.010 | 0.000 | 1.000 | 2.065 | 0.094 | 0.008 | 2.073 | | | | | 0.40 | 0.08 | 0.07 | 0.28 | 0.19 | 0.13 |
| 310442 | 2000 CH59 | 21.2 | 1.14 | 229 | 19.7 | 0.985 | 0.010 | 0.000 | 0.985 | 2.100 | 0.143 | 0.014 | 2.114 | | | | | 0.23 | 0.03 | 0.03 | 0.46 | 0.23 | 0.19 |
| | 2002 GO5 | 11.2 | 1.15 | 249 | 14.5 | 1.000 | 0.010 | 0.000 | 1.000 | 1.990 | 0.010 | 0.010 | 2.000 | | | | | 0.76 | 0.16 | 0.14 | 0.25 | 0.20 | 0.11 |
| | 2002 LV | 34.8 | 1.47 | 227 | 22.1 | 0.960 | 0.010 | 0.000 | 0.960 | 1.995 | 0.284 | 0.015 | 2.010 | | | | | 1.73 | 0.36 | 0.30 | 0.15 | 0.13 | 0.07 |
| | 2002 LV | 65.3 | 1.09 | 263 | 13.2 | 0.930 | 0.010 | 0.001 | 0.931 | 1.890 | 0.013 | 0.007 | 1.897 | | | | | 1.73 | 0.36 | 0.30 | 0.15 | 0.13 | 0.07 |
| | 2005 UL5 | 27.6 | 1.10 | 253 | 18.5 | 0.975 | 0.010 | 0.000 | 0.975 | 2.000 | 0.010 | 0.009 | 2.009 | | | | | | | | | | |
| | 2007 RF5 | 16.7 | 1.28 | 238 | 14.6 | 1.035 | 0.010 | 0.000 | 1.035 | 1.955 | 0.010 | 0.012 | 1.967 | | | | | 0.62 | 0.12 | 0.12 | 0.21 | 0.18 | 0.09 |

Table 5: Mineralogies and ordinary chondrite analogues for potential ordinary chondrites in the ExploreNEOs survey

| Number | Name | Type | ΔBIC | ΔBAR | ol/(ol+px) | Fa | Fs | OC Type Fa | OC Type Fs | OC Type |
|---|---|---|---|---|---|---|---|---|---|---|
| S(IV) | | | | | | | | | | |
| 433 | Eros | Q/Sq | 0.962 | 0.345 | 0.64 | 24.2 | 20.3 | L | L | L |
| 433 | Eros | Q/Sq | 0.976 | 0.507 | 0.61 | 26.5 | 22.0 | L/LL | L/LL | L/LL |
| 433 | Eros | Sq/Q | 0.983 | 0.398 | 0.63 | 27.5 | 22.7 | L/LL | L/LL | L/LL |
| 433 | Eros | Q/Sq | 0.995 | 0.324 | 0.65 | 28.8 | 23.7 | LL | LL | LL |
| 1036 | Ganymed | Sr/S | 0.919 | 0.989 | 0.49 | 13.9 | 13.0 | H? | H | H? |
| 1036 | Ganymed | Sr | 0.929 | 1.086 | 0.47 | 16.7 | 15.0 | H | H | H |
| 1865 | Cerberus | S | 0.930 | 0.533 | 0.60 | 16.9 | 15.1 | H | H | H |
| 1866 | Sisyphus | S | 0.929 | 0.549 | 0.60 | 16.6 | 14.9 | H | H | H |
| 1866 | Sisyphus B | S | 0.932 | 0.991 | 0.49 | 17.6 | 15.6 | H | H | H |
| 1866 | Sisyphus D | S | 0.918 | 1.131 | 0.45 | 13.7 | 12.8 | H? | H? | H? |
| 1866 | Sisyphus F | S | 0.918 | 0.770 | 0.54 | 13.7 | 12.8 | H? | H? | H? |
| 1917 | Cuyo | Sr | 0.925 | 0.733 | 0.55 | 15.5 | 14.1 | H | H | H |
| 4283 | Cuno | Sq/Q | 0.979 | 0.554 | 0.59 | 26.9 | 22.3 | L/LL | L/LL | L/LL |
| 5143 | Heracles | Q | 1.016 | 0.297 | 0.66 | 30.4 | 24.9 | LL | LL | LL |
| 5626 | 1991 FE | Sq | 0.943 | 0.588 | 0.59 | 20.1 | 17.4 | H | H | H |
| 5646 | 1990 TR | Q | 1.025 | 0.247 | 0.67 | 30.7 | 25.2 | LL | LL | LL |
| 7358 | Oze | S | 0.924 | 1.114 | 0.46 | 15.3 | 14.0 | H | H | H |
| 11066 | Sigurd | Sr | 0.923 | 1.105 | 0.46 | 15.0 | 13.8 | H | H | H |
| 12711 | Tukmit | Q | 0.979 | 0.307 | 0.65 | 26.9 | 22.3 | LL | LL | LL |
| 12923 | Zephyr | S | 0.942 | 1.084 | 0.47 | 19.9 | 17.3 | H | H | H |
| 16834 | 1997 WU22 | Sq | 0.961 | 0.476 | 0.61 | 23.9 | 20.1 | L | L | L |
| 19764 | 2000 NF5 | Sq | 0.971 | 0.397 | 0.63 | 25.7 | 21.4 | L/LL | L/LL | L/LL |
| 27346 | 2000 DN8 | Q | 1.028 | 0.356 | 0.64 | 30.7 | 25.3 | LL | LL | LL |
| 36284 | 2000 DM8 | Sq | 1.018 | 0.291 | 0.66 | 30.4 | 25.0 | LL | LL | LL |
| 53789 | 2000 ED104 | Sq | 0.949 | 0.766 | 0.54 | 21.6 | 18.5 | L | H/L | H/L |
| 68216 | 2001 CV26 | Sq | 0.934 | 0.536 | 0.60 | 18.1 | 15.9 | H | H | H |
| 68216 | 2001 CV26 | S/Sq | 0.949 | 0.747 | 0.55 | 21.6 | 18.4 | L | H/L | H/L |
| 68216 | 2001 CV26 | Sq | 0.960 | 0.427 | 0.62 | 23.8 | 20.1 | L | L | L |
| 85839 | 1998 YO4 | Q | 1.001 | 0.271 | 0.66 | 29.4 | 24.2 | LL | LL | LL |
| 86067 | 1999 RM28 | Q | 0.966 | 0.438 | 0.62 | 24.9 | 20.9 | L/LL | L/LL | L/LL |
| 88254 | 2001 FM129 | Q | 0.976 | 0.604 | 0.58 | 26.4 | 22.0 | L/LL | L/LL | L/LL |
| 100926 | 1998 MQ | Q | 0.945 | 0.564 | 0.59 | 20.7 | 17.8 | H | H/L | H/L |
| 137032 | 1998 UO1 | Sq | 0.968 | 0.485 | 0.61 | 25.1 | 21.0 | L/LL | L/LL | L/LL |
| 137062 | 1998 WM | Sq | 0.930 | 0.706 | 0.56 | 17.0 | 15.2 | H | H | H |
| 138883 | 2000 YL29 | Q | 0.958 | 0.385 | 0.63 | 23.3 | 19.7 | L | L | L |
| 141498 | 2002 EZ16 | Sq | 0.967 | 0.512 | 0.60 | 25.0 | 20.9 | L/LL | L/LL | L/LL |
| 143381 | 2003 BC21 | Sr/S | 0.924 | 0.937 | 0.50 | 15.3 | 13.9 | H | H | H |
| 152931 | 2000 EA107 | Q | 1.010 | 0.283 | 0.66 | 30.0 | 24.7 | LL | LL | LL |
| 159402 | 1999 AP10 | Sq | 0.950 | 0.663 | 0.57 | 21.8 | 18.6 | L | H/L | H/L |
| 163697 | 2003 EF54 | Q | 0.965 | 0.476 | 0.61 | 24.7 | 20.7 | L/LL | L/LL | L/LL |
| 207945 | 1991 JW | Q | 0.991 | 0.323 | 0.65 | 28.4 | 23.4 | LL | LL | LL |
| 214088 | 2004 JN13 | Sq | 0.981 | 0.566 | 0.59 | 27.1 | 22.5 | L/LL | L/LL | L/LL |
| 218863 | 2006 WO127 | Sq | 0.990 | 0.278 | 0.66 | 28.3 | 23.3 | LL | LL | LL |
| 219071 | 1997 US9 | Q | 1.024 | 0.248 | 0.67 | 30.6 | 25.2 | LL | LL | LL |

| Number | Name | Type | $\Delta$BIC | $\Delta$BAR | ol/(ol+px) | Fa | Fs | OC Type Fa | OC Type Fs | OC Type |
|---|---|---|---|---|---|---|---|---|---|---|
| 220124 | 2002 TE66 B | Sq | 0.957 | 0.465 | 0.62 | 23.2 | 19.6 | L | L | L |
| 265187 | 2003 YS117 | Q | 0.972 | 0.460 | 0.62 | 25.9 | 21.6 | L/LL | L/LL | L/LL |
| 297418 | 2000 SP43 | V | 0.936 | 1.126 | 0.46 | 18.5 | 16.2 | H | H | H |
|  | 2000 TJ1 | Sq | 0.946 | 0.566 | 0.59 | 20.8 | 17.9 | H | H/L | H/L |
| Dunn regions | | | | | | | | | | |
| 1036 | Ganymed | Sr | 0.923 | 1.226 | 0.43 | 15.0 | 13.8 | H | H | H |
| 1627 | Ivar | S | 1.024 | 0.616 | 0.58 | 30.6 | 25.2 | LL | LL | LL |
| 1943 | Anteros | Sq | 0.987 | 0.631 | 0.58 | 28.0 | 23.1 | LL | L/LL | L/LL |
| 3122 | Florence | Sq/Q | 1.030 | 0.499 | 0.61 | 30.7 | 25.3 | LL | LL | LL |
| 6239 | Minos | Q/Sq | 0.981 | 0.267 | 0.66 | 27.2 | 22.5 | LL | LL | LL |
| 8567 | 1996 HW1 | Q | 1.002 | 0.536 | 0.60 | 29.5 | 24.2 | LL | LL | LL |
| 23187 | 2000 PN9 | Sq | 1.031 | 0.646 | 0.57 | 30.8 | 25.3 | LL | LL | LL |
| 39572 | 1993 DQ1 | Q | 0.985 | 0.609 | 0.58 | 27.7 | 22.9 | LL | L/LL | L/LL |
| 96590 | 1998 XB | Sq | 1.000 | 0.564 | 0.59 | 29.3 | 24.1 | LL | LL | LL |
| 137125 | 1999 CT3 | Q | 1.038 | 0.680 | 0.56 | 30.7 | 25.4 | LL | LL | LL |
| 141498 | 2002 EZ16 | Sq | 0.990 | 0.158 | 0.69 | 28.3 | 23.3 | LL | LL | LL |
| 198856 | 2005 LR3 | Sq | 0.995 | 0.217 | 0.68 | 28.8 | 23.7 | LL | LL | LL |
| 198856 | 2005 LR3 | Q | 1.005 | 0.188 | 0.68 | 29.7 | 24.4 | LL | LL | LL |
| 207945 | 1991 JW | Q | 1.001 | 0.143 | 0.69 | 29.4 | 24.2 | LL | LL | LL |
| 220124 | 2002 TE66 A | Sq/Q | 0.976 | 0.286 | 0.66 | 26.5 | 22.0 | LL | LL | LL |
| 347813 | 2002 NP1 | Q | 0.961 | 0.235 | 0.67 | 24.0 | 20.2 | LL? | LL? | LL? |
|  | 2001 HC | Sq | 0.968 | 0.246 | 0.67 | 25.2 | 21.1 | LL | LL | LL |

Table 6: Mineralogies and ordinary chondrite analogues for potential ordinary chondrites in the MIT-UH-IRTF Joint Campaign

| Number | Name | Type | $\Delta$BIC | $\Delta$BAR | ol/(ol+px) | Fa | Fs | OC Type Fa | OC Type Fs | OC Type |
|---|---|---|---|---|---|---|---|---|---|---|
| S(IV) | | | | | | | | | | |
| 433 | Eros | Sw | 0.950 | 0.362 | 0.64 | 21.7 | 18.5 | L | L | L |
| 433 | Eros | Sw | 0.955 | 0.378 | 0.64 | 22.7 | 19.3 | L | L | L |
| 433 | Eros | Sw | 0.960 | 0.440 | 0.62 | 23.8 | 20.0 | L | L | L |
| 433 | Eros | Sw | 0.980 | 0.276 | 0.66 | 27.1 | 22.4 | LL | LL | LL |
| 433 | Eros | Sw | 0.990 | 0.179 | 0.68 | 28.3 | 23.3 | LL | LL | LL |
| 433 | Eros | Sw | 0.995 | 0.339 | 0.65 | 28.8 | 23.7 | LL | LL | LL |
| 433 | Eros | Sqw | 1.020 | 0.154 | 0.69 | 30.5 | 25.1 | LL | LL | LL |
| 1036 | Ganymed | S | 0.925 | 0.690 | 0.56 | 15.6 | 14.1 | H | H | H |
| 1036 | Ganymed | S | 0.930 | 0.869 | 0.52 | 16.9 | 15.1 | H | H | H |
| 1627 | Ivar | Sr | 0.949 | 0.503 | 0.61 | 21.6 | 18.5 | H/L | H/L | H/L |
| 1627 | Ivar | S | 0.954 | 0.497 | 0.61 | 22.6 | 19.2 | L | H/L | H/L |
| 1916 | Boreas | Sw | 0.955 | 0.411 | 0.63 | 22.8 | 19.3 | L | L | L |
| 1917 | Cuyo | Sv | 0.940 | 1.152 | 0.45 | 19.5 | 16.9 | H | H | H |
| 3102 | Krok | Sqw | 1.000 | 0.465 | 0.62 | 29.3 | 24.1 | LL | LL | LL |
| 5587 | 1990 SB | Sr | 0.925 | 0.911 | 0.51 | 15.5 | 14.1 | H | H | H |
| 5626 | 1991 FE | S | 0.930 | 0.496 | 0.61 | 16.9 | 15.1 | H | H | H |
| 6455 | 1992 HE | Srw | 0.935 | 0.648 | 0.57 | 18.1 | 16.0 | H | H | H |
| 6455 | 1992 HE | Sqw | 0.980 | 0.300 | 0.66 | 27.0 | 22.4 | LL | LL | LL |
| 11066 | Sigurd | S | 0.925 | 0.833 | 0.53 | 15.4 | 14.1 | H | H | H |
| 11398 | 1998 YP11 | Sr | 0.925 | 0.603 | 0.58 | 15.7 | 14.2 | H | H | H |
| 15745 | 1991 PM5 | S | 0.945 | 0.583 | 0.59 | 20.7 | 17.8 | H | H/L | H/L |
| 16834 | 1997 WU22 | S | 0.960 | 0.406 | 0.63 | 23.7 | 20.0 | L | L | L |
| 25143 | Itokawa | Sqw | 0.970 | 0.238 | 0.67 | 25.6 | 21.3 | LL | LL | LL |
| 35107 | 1991 VH | Sq | 0.980 | 0.311 | 0.65 | 27.1 | 22.4 | LL | LL | LL |
| 36284 | 2000 DM8 | Sq | 0.985 | 0.396 | 0.63 | 27.7 | 22.9 | LL | L/LL | L/LL |
| 53435 | 1999 VM40 | Srw | 0.930 | 1.085 | 0.47 | 17.0 | 15.2 | H | H | H |
| 54690 | 2001 EB | S | 0.940 | 0.509 | 0.60 | 19.5 | 16.9 | H | H | H |
| 89355 | 2001 VS78 | Sr | 0.925 | 0.852 | 0.52 | 15.6 | 14.2 | H | H | H |
| 137062 | 1998 WM | Sr | 0.915 | 0.702 | 0.56 | 12.5 | 12.0 | H? | H? | H? |
| 138258 | 2000 GD2 | Sq | 1.010 | 0.461 | 0.62 | 30.1 | 24.7 | LL | LL | LL |
| 141052 | 2001 XR1 | Sq | 0.960 | 0.356 | 0.64 | 23.9 | 20.1 | L | L | L |
| 143651 | 2003 QO104 | Q | 0.996 | 0.296 | 0.66 | 28.9 | 23.8 | LL | LL | LL |
| 152931 | 2000 EA107 | Q | 1.005 | 0.274 | 0.66 | 29.7 | 24.4 | LL | LL | LL |
| 200840 | 2001 XN254 | S | 0.950 | 0.695 | 0.56 | 21.9 | 18.6 | L | H/L | H/L |
| Dunn regions | | | | | | | | | | |
| 1627 | Ivar | Sqw | 0.960 | 0.213 | 0.68 | 23.8 | 20.0 | LL? | LL? | LL? |
| 1943 | Anteros | S | 1.010 | 0.525 | 0.60 | 30.0 | 24.7 | LL | LL | LL |
| 4183 | Cuno | Q | 1.014 | 0.179 | 0.68 | 30.3 | 24.9 | LL | LL | LL |
| 5131 | 1990 BG | Sq | 1.018 | 0.122 | 0.70 | 30.4 | 25.0 | LL | LL | LL |
| 5143 | Heracles | Q | 1.009 | 0.148 | 0.69 | 30.0 | 24.6 | LL | LL | LL |
| 6239 | Minos | Sq | 0.985 | 0.196 | 0.68 | 27.7 | 22.9 | LL | LL | LL |
| 8567 | 1996 HW1 | Q | 0.995 | 0.202 | 0.68 | 28.8 | 23.7 | LL | LL | LL |
| 35107 | 1991 VH | Sq | 1.005 | 0.245 | 0.67 | 29.7 | 24.4 | LL | LL | LL |
| 207945 | 1991 JW | Sq/Q | 0.970 | 0.275 | 0.66 | 25.5 | 21.3 | LL | LL | LL |

| Number | Name | Type | $\Delta$BIC | $\Delta$BAR | ol/(ol+px) | Fa | Fs | OC Type Fa | OC Type Fs | OC Type |
|---|---|---|---|---|---|---|---|---|---|---|
| 219071 | 1997 US9 | Q | 1.030 | 0.201 | 0.68 | 30.7 | 25.3 | LL | LL | LL |

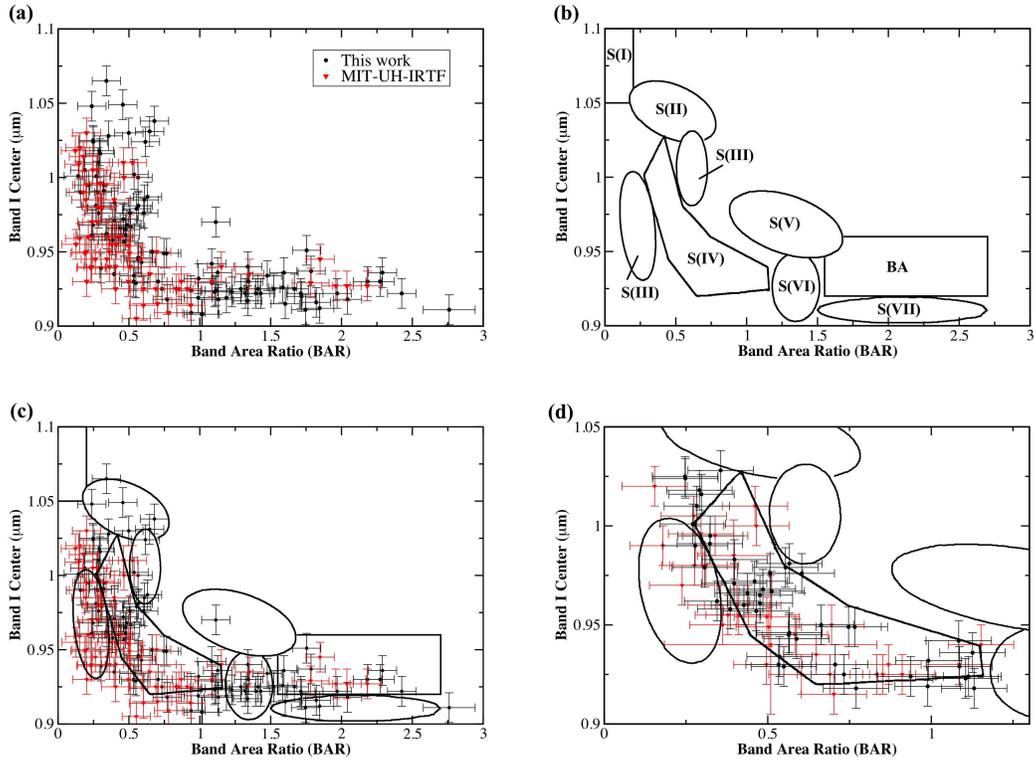

Figure 1: (a) Band I Centers (BICs) versus Band Area Ratios (BARs) for all S-, Q-, and V-complex objects in our full spectral sample. Objects observed as part of the ExploreNEOs survey are displayed using black circles and those from the MIT Joint Campaign are shown using red triangles. (b) The Gaffey et al. (1993) S-subclasses and basaltic achondrites (BA) in BIC-BAR space. (c) The comparison of our spectral band parameters with the Gaffey et al. (1993) S-subclasses. The shape of the distribution of S-subclasses matches the distribution of objects in our analysis. (d) We use the S(IV) region as one method of identifying potential ordinary chondrites in our sample. This panel shows the S(IV) region with objects classified as S(IV) in our analysis. We include Eros and Itokawa because of their confirmed ordinary chondritic nature.

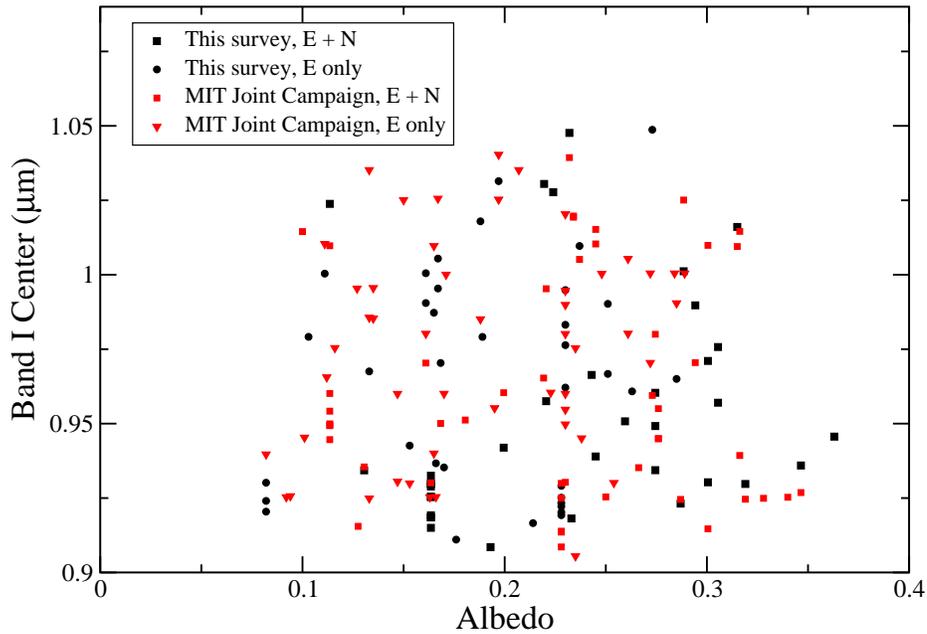

Figure 2: We find no correlation between ExploreNEOs (E) and NEOWISE (N) albedos and Band I Center for our spectral sample. Objects observed as part of the ExploreNEOs survey are shown in black. The circles represent ExploreNEOs-only albedos and the squares are an average of ExploreNEOs and NEOWISE albedos. Objects from the MIT Joint Campaign are shown in red. The triangles represent ExploreNEOs-only albedos and the squares are an average of ExploreNEOs and NEOWISE albedos. The errors on the ExploreNEOs albedos of these objects are higher than the ExploreNEOs average since we are targeting the high albedo objects in this band parameter analysis. We place an upper limit on the errors $\sigma_{pV+}$ and $\sigma_{pV-}$ of 0.20. We discard all observations with errors larger than this limit. We find no correlation down to error limits on $\sigma_{pV+}$ and $\sigma_{pV-}$ of 0.05.

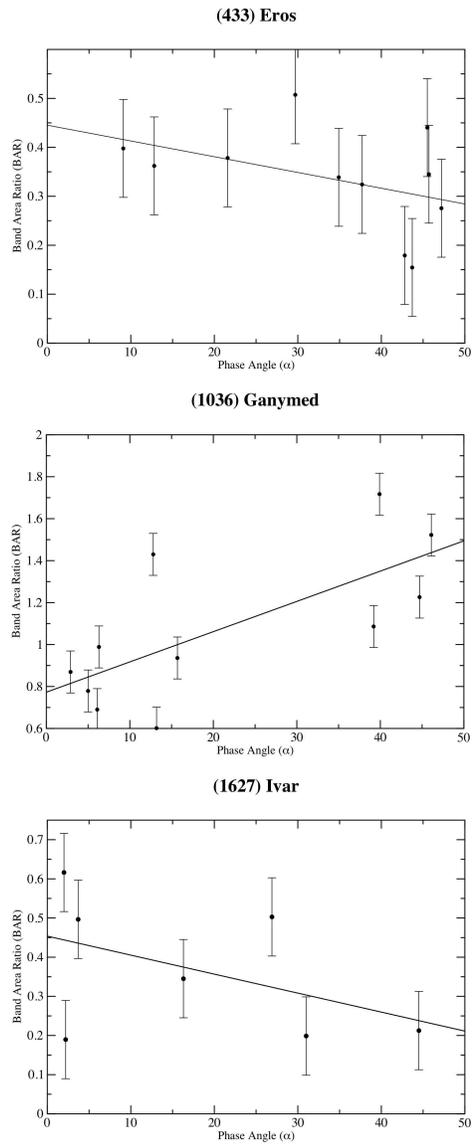

Figure 3: Band Area Ratio (BAR) versus Phase Angle ($\alpha$) for (433) Eros, (1036) Ganymed, and (1627) Ivar. For Eros and Ivar, BAR is anti-correlated with phase angle. Ganymed has an opposite correlation with a trend of increasing BAR with increasing phase angle.

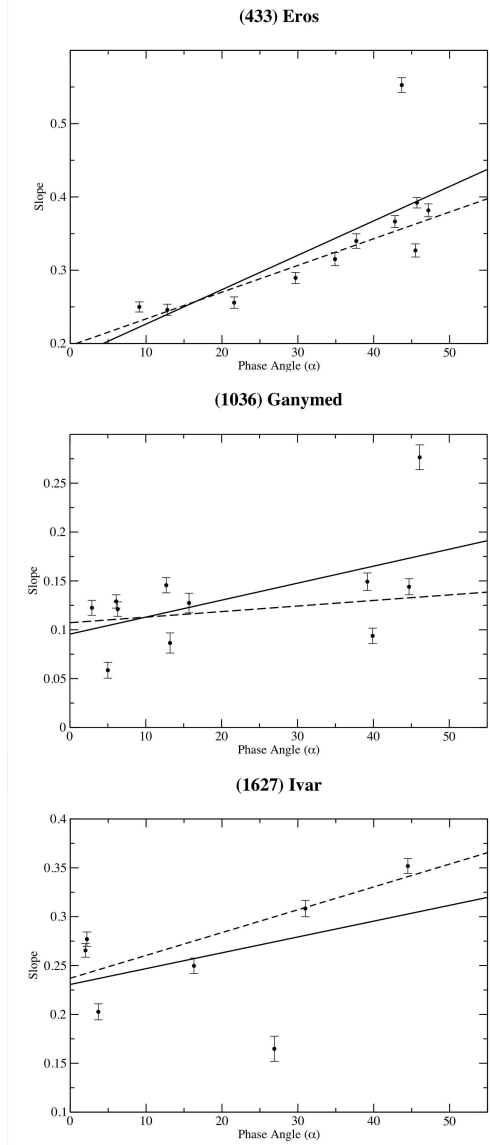

Figure 4: Near-infrared spectral slope versus Phase Angle ($\alpha$) for (433) Eros, (1036) Ganymed, and (1627) Ivar. All three objects show evidence of phase reddening. The solid lines indicate the nominal correlations and the dashed lines indicate the correlations calculated without the outlying points.

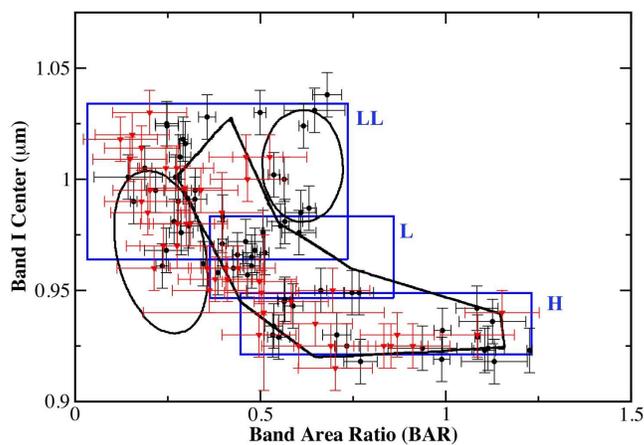

Figure 5: Our second method of identifying potential ordinary chondrites uses the Dunn ordinary chondrite regions calculated from the ranges of mineralogies determined in Dunn et al. (2010). We calculated the regions in Band I Center-Band Area Ratio space and used the regions to determine potential ordinary chondrite types. Each ordinary chondrite region is labeled with its type. The Gaffey S(III) and S(IV) regions are included for comparison. The L and H regions closely approximate the S(IV) region, but the LL region is much larger than the corresponding subset of the S(IV) region. The Dunn LL region includes most of the Gaffey S(III) regions.

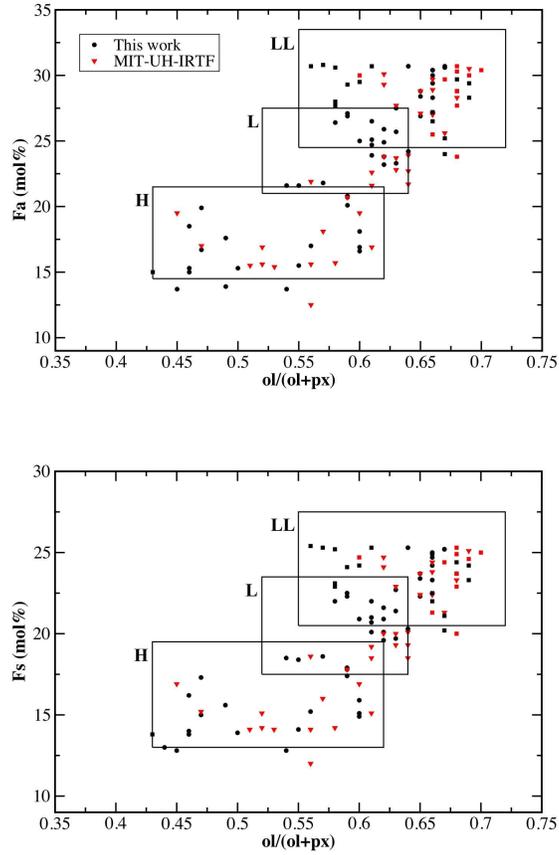

Figure 6: We use the Dunn et al. (2010) equations to derive the mol % of Fayalite (Fa) and Ferrosilite (Fs) and olivine/(olivine+pyroxene) ratio from band parameters of ordinary chondrite-like spectra. The nominal errors on the chemistry calculations are 0.03 for the ol/(ol+px) calculation, 1.3 for the calculation of the mol % of Fa, and 1.4 for the calculation of mol % of Fs. The top panel shows mol % Fayalite versus ol/(ol+px). The bottom panel shows mol % of Ferrosilite versus ol/(ol+px). We determine a potential ordinary chondrite subtype (H, L, LL) based on the location of the objects in each of these compositional regions. The quadratic equations impose an artificial upper limit on Fayalite and Ferrosilite within the LL ordinary chondrite subclass that can be clearly seen.

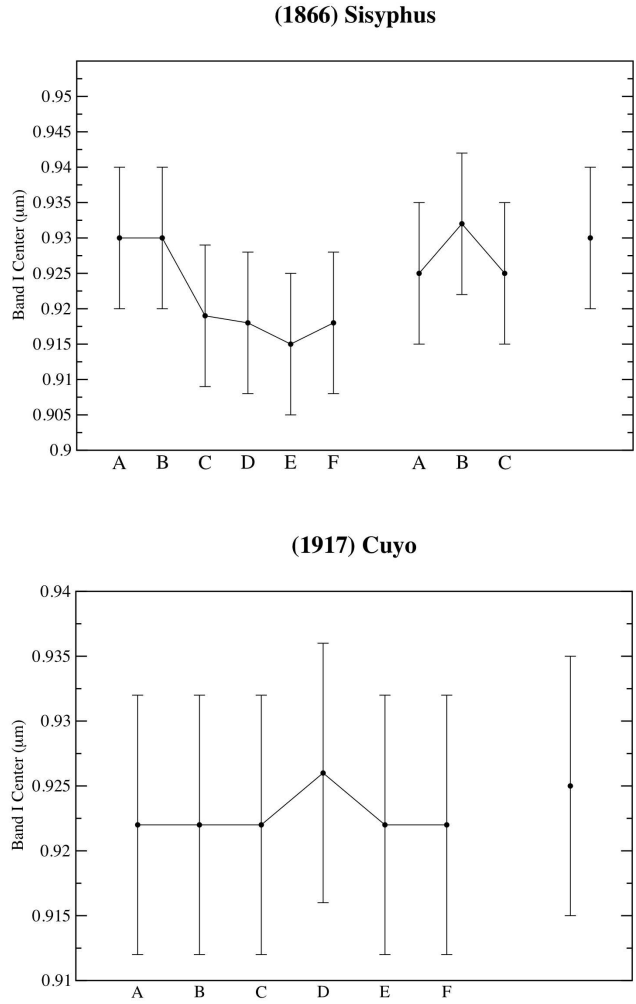

Figure 7: Band I Center for Sisyphus (top panel) and Cuyo (bottom panel) for our sequences of observations. The x-axis is not directly correlated with time. Each set of observations is connected and presented sequentially. Additional observations for each object are offset in x. We see potential compositional variation on the surface of Sisyphus. Any variation is within the errors, but we set a hard minimum on the errors in order to better include several unquantified sources of error. The differences in BIC exceed the nominal calculated errors for the observations. Cuyo does not show indications of compositional variation on the surface.

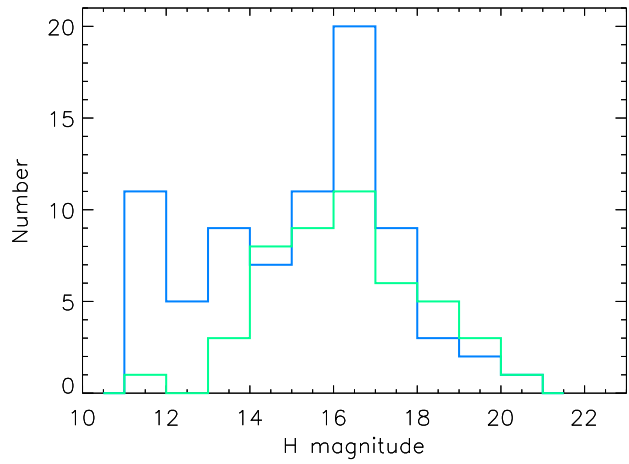

Figure 8: We compared the distribution of H magnitudes (as a proxy for size) of our ExploreNEOs spectra to the spectral sample analyzed in Dunn et al. (2013) to determine if we were investigating different size regimes. We have different relative proportions of the ordinary chondrite types within our samples, which cannot be explained by differences in object sizes.

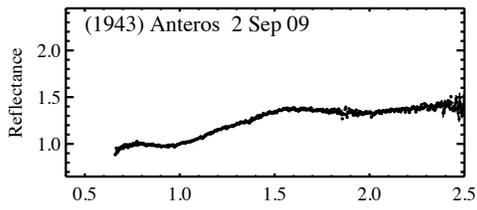
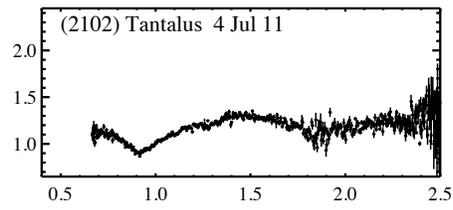
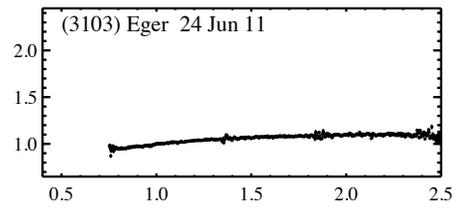
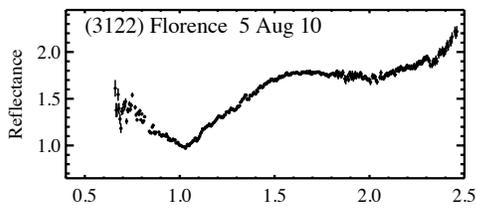
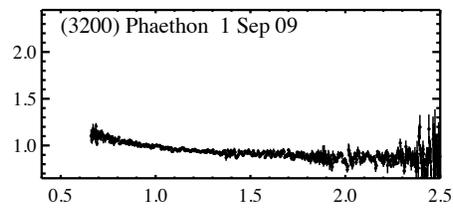
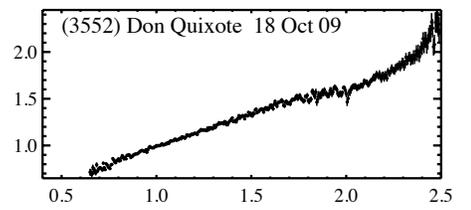
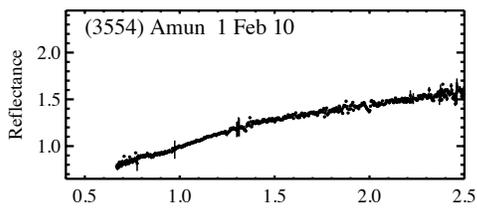
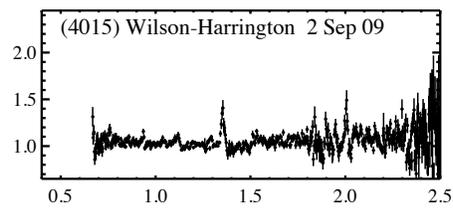
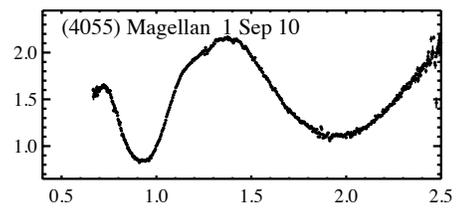
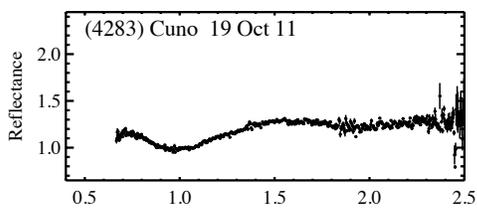
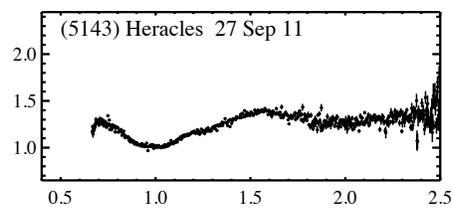
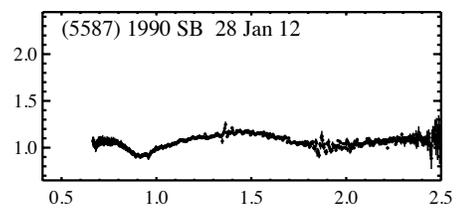
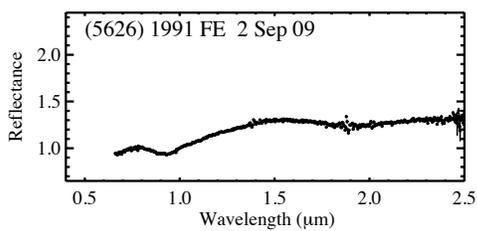
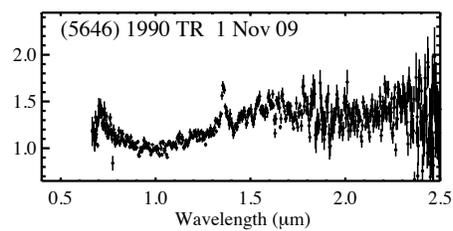
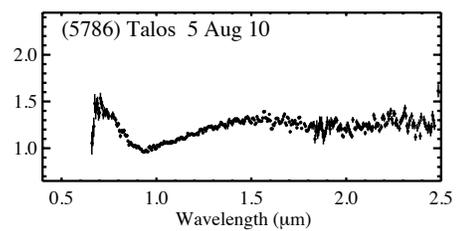

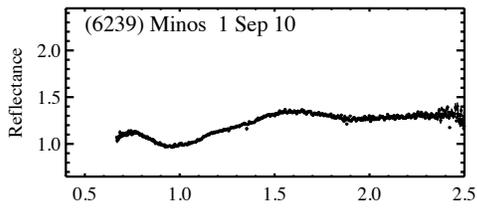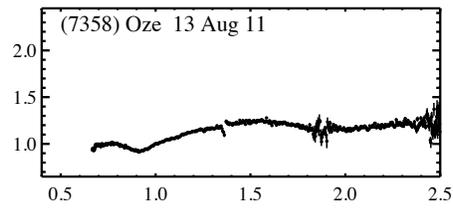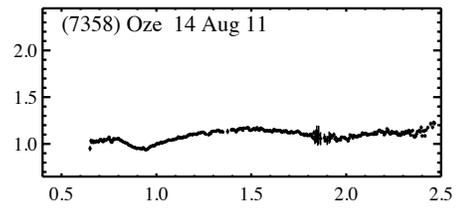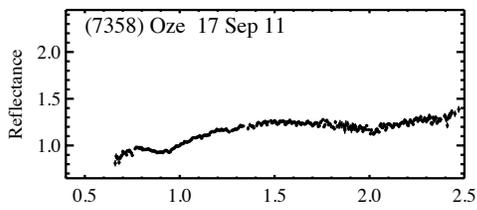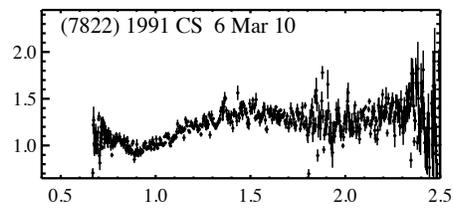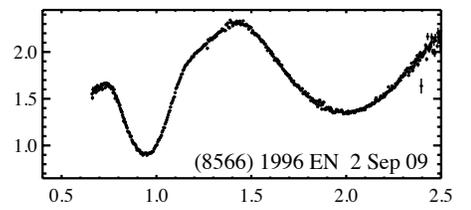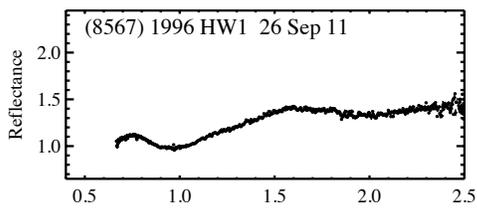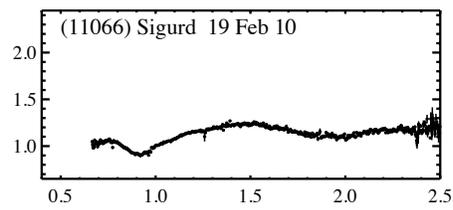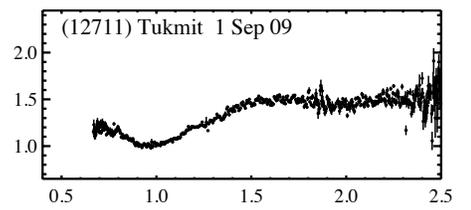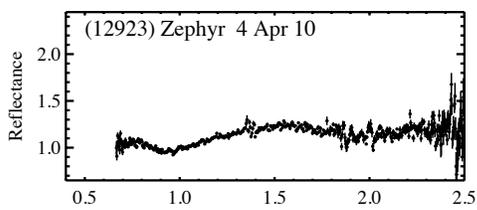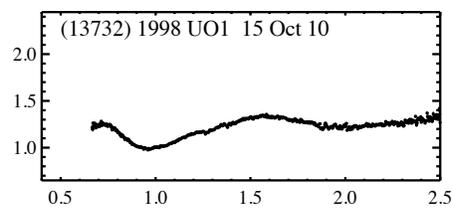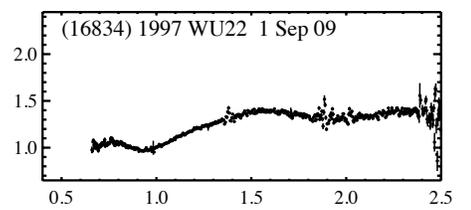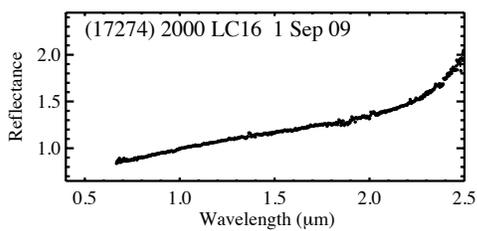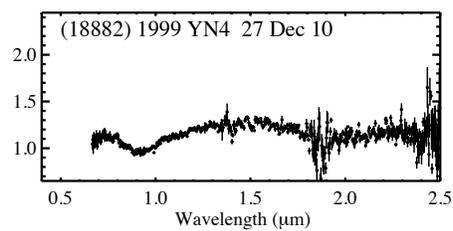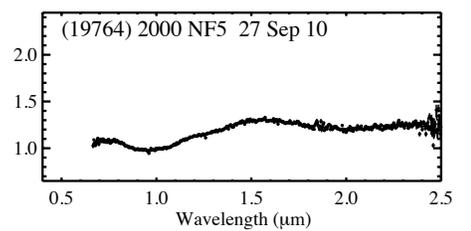

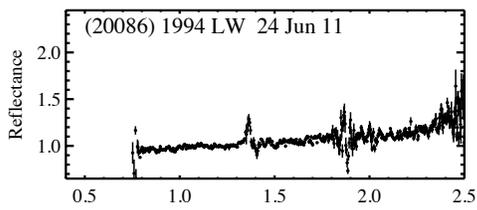
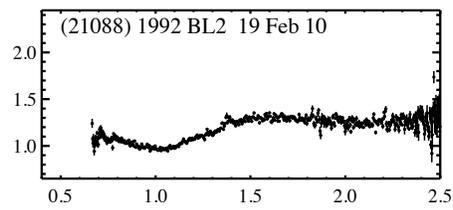
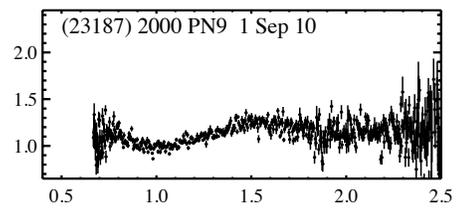
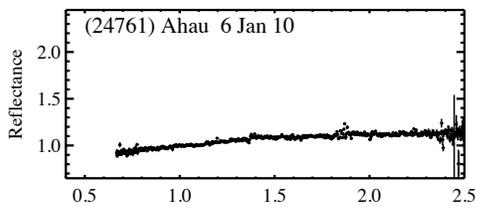
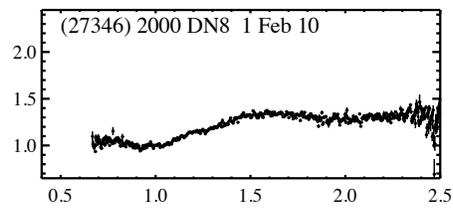
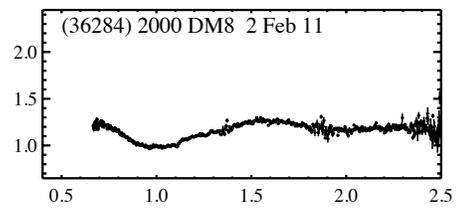
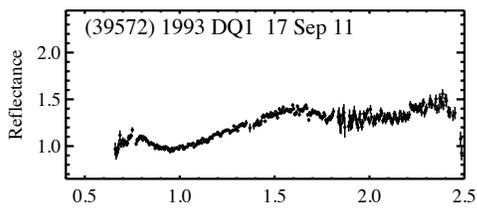
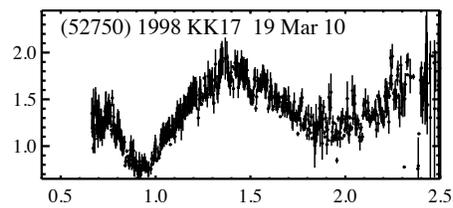
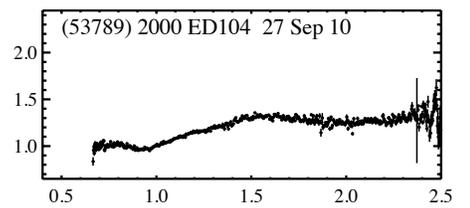
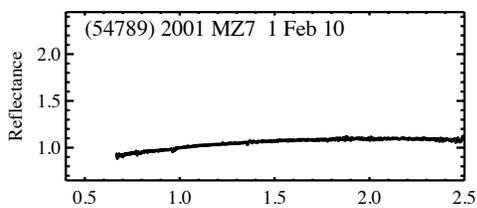
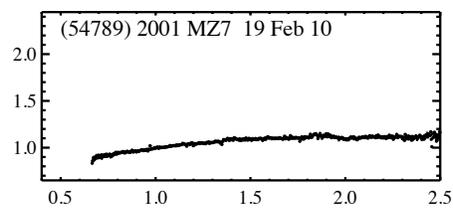
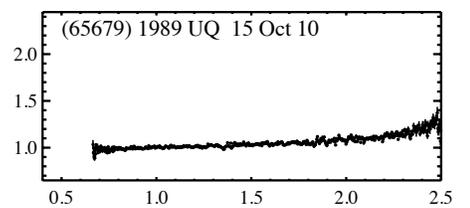
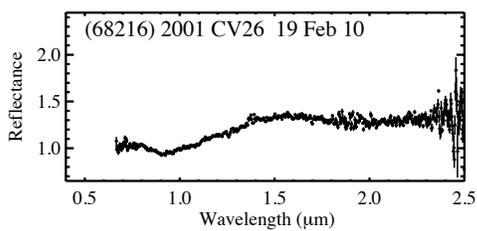
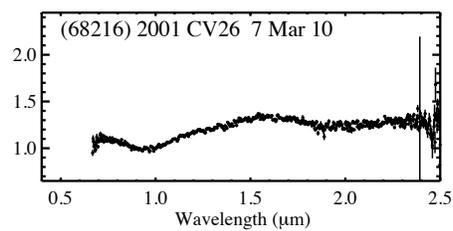
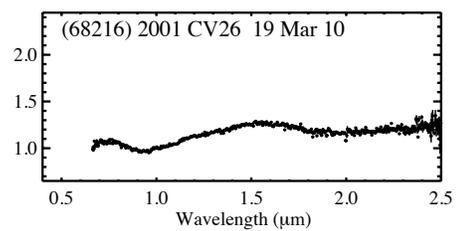

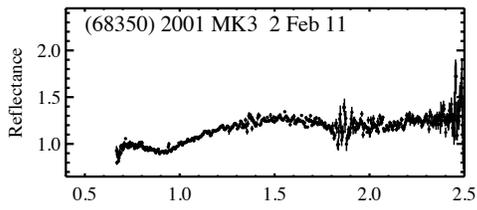
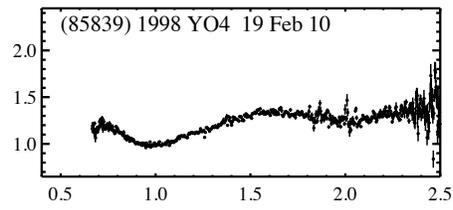
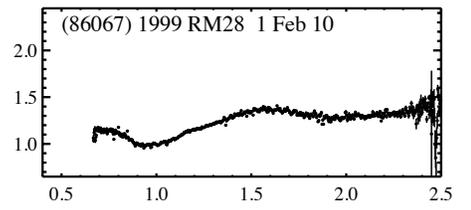
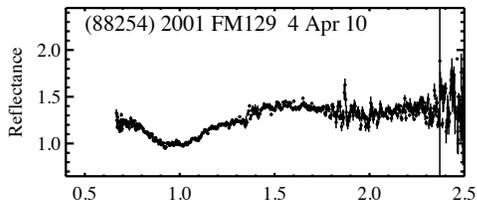
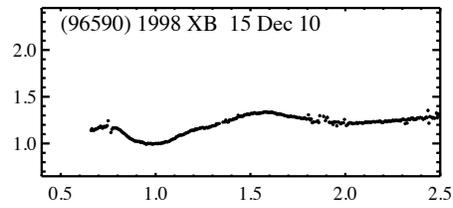
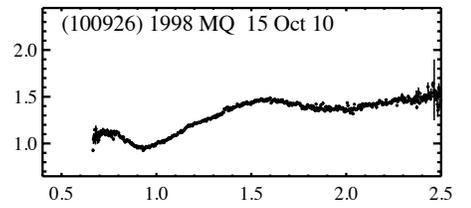
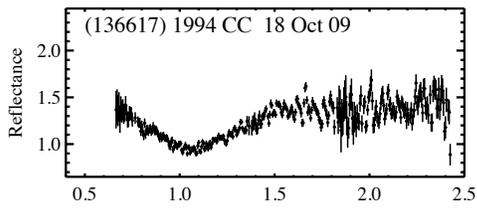
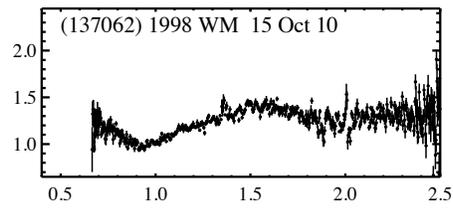
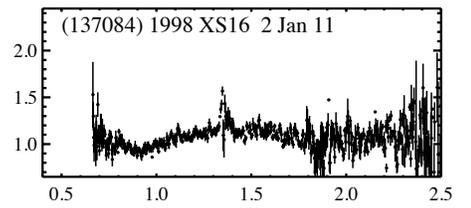
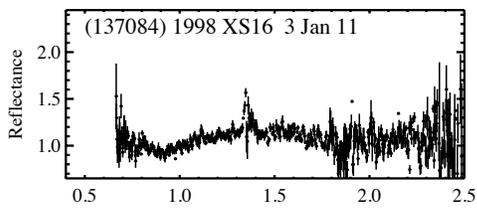
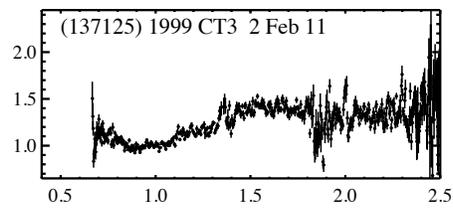
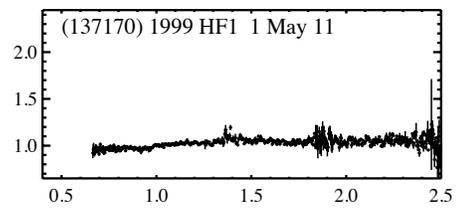
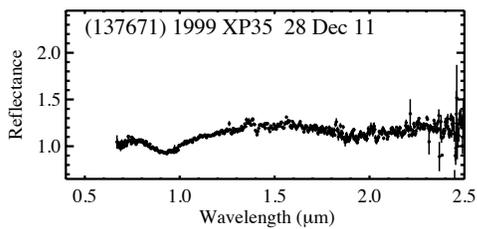
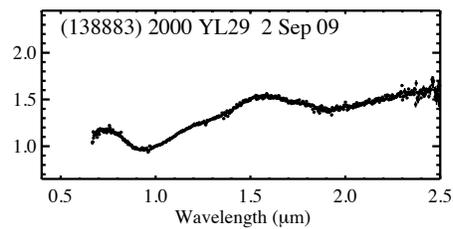
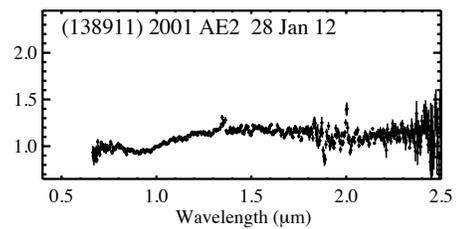

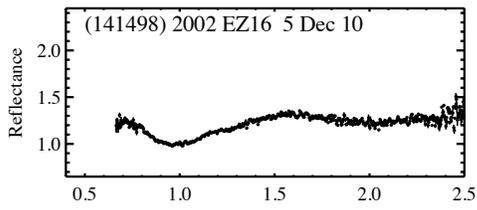
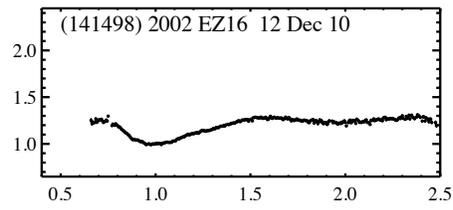
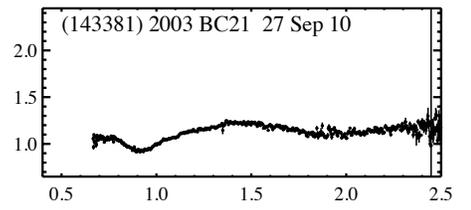
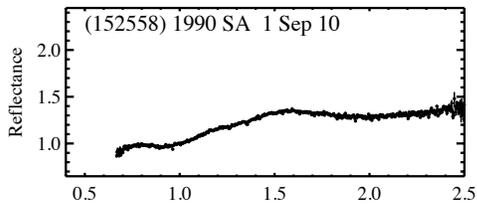
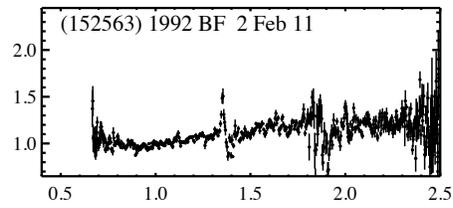
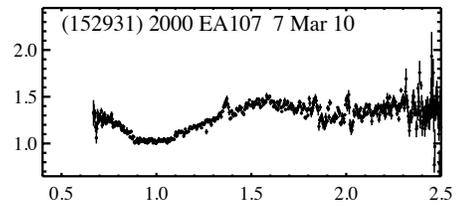
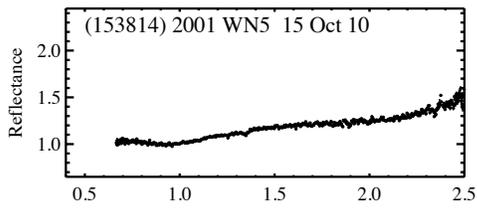
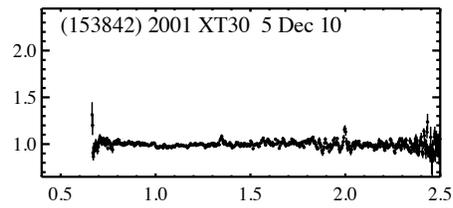
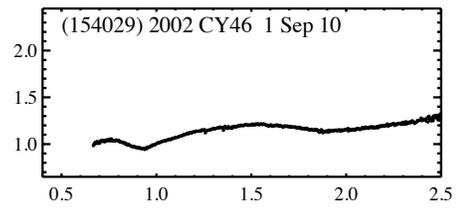
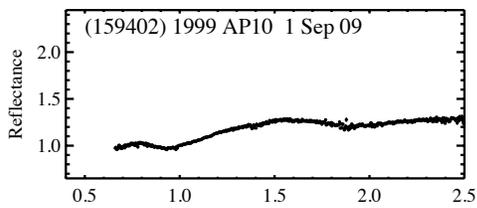
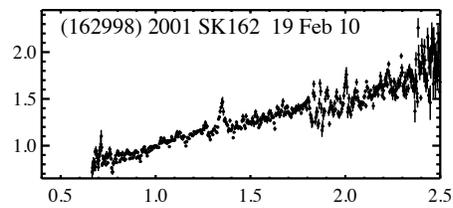
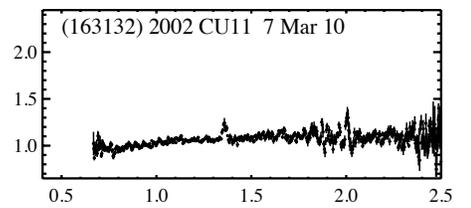
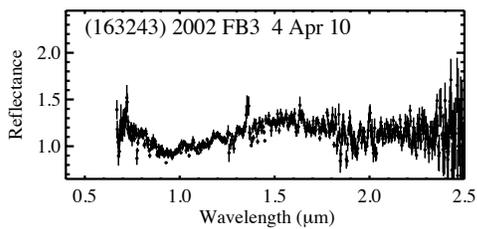
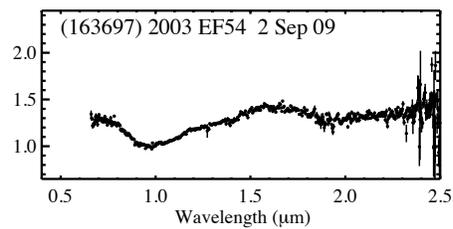
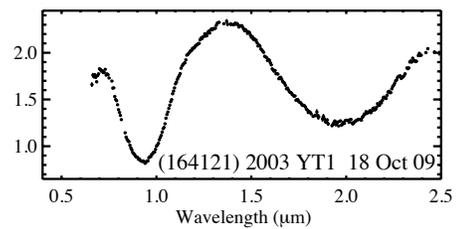

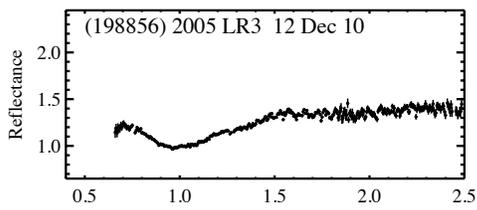
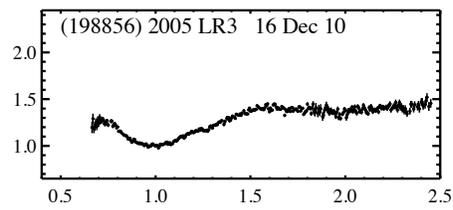
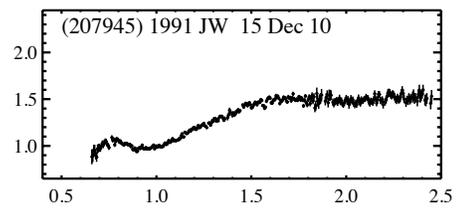
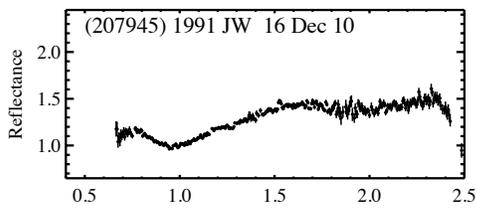
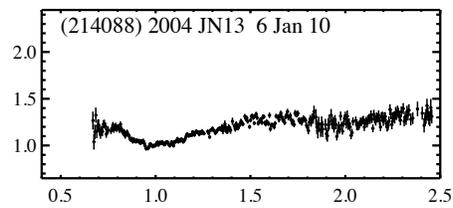
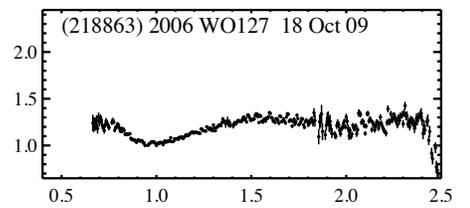
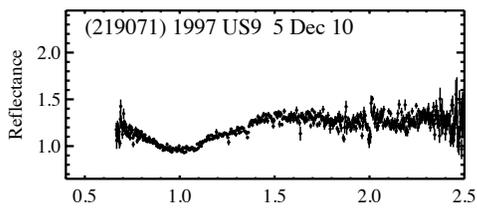
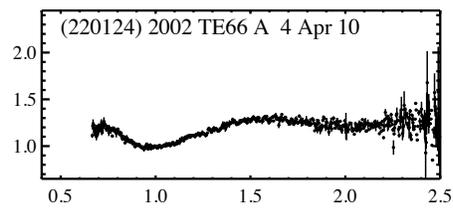
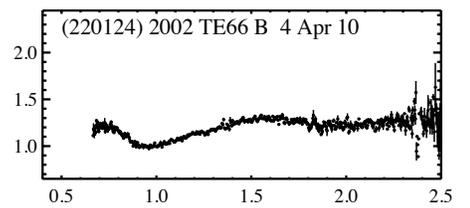
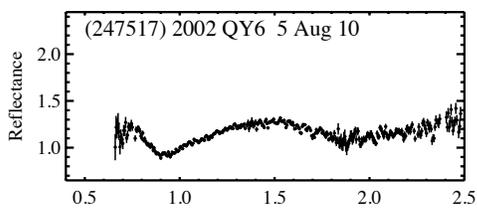
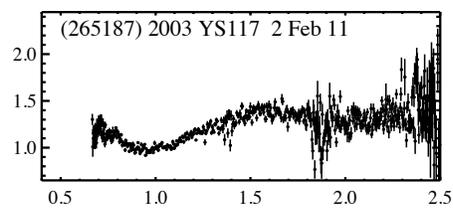
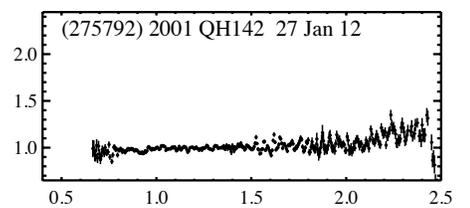
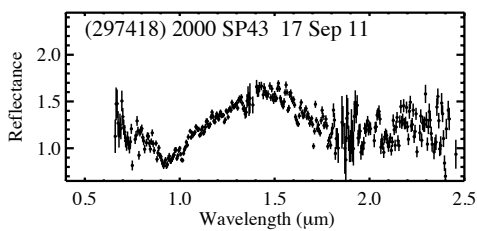
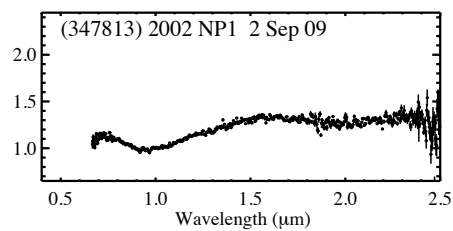
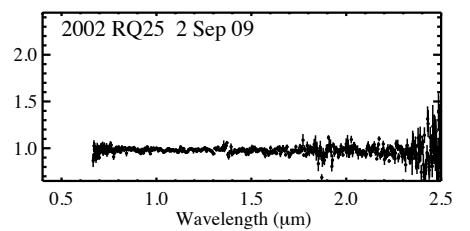

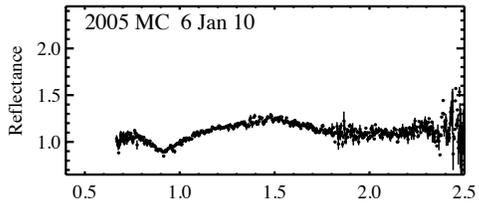
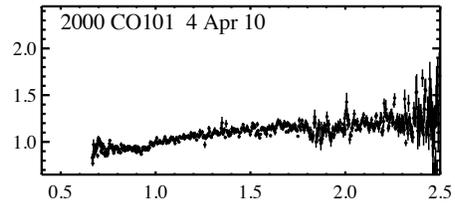
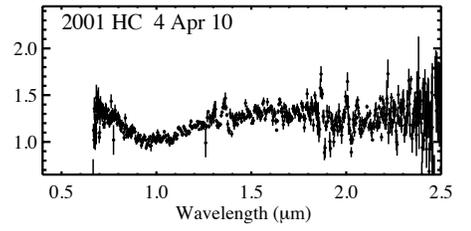
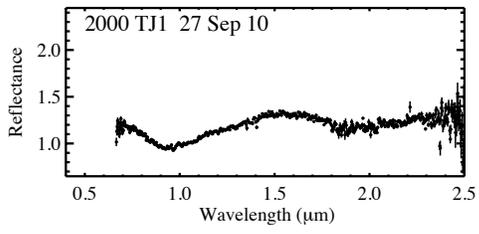
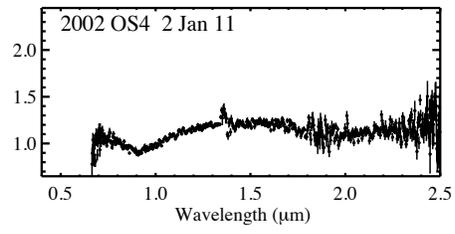